\def\spacingNumerator{5}
\def\spacingDenominator{4}

\def\ifundefined#1{\expandafter\ifx\csname#1\endcsname\relax}
\ifundefined{ftmagnification}  \def\ftmagnification{1200} \fi
\ifundefined{spacingNumerator}  \def\spacingNumerator{5} \fi
\ifundefined{spacingDenominator}  \def\spacingDenominator{4} \fi


\magnification\ftmagnification
\tolerance=10000
\hsize=17truecm\vsize=23truecm

\parindent=40pt
\mathsurround=0pt
     \multiply\baselineskip by \spacingNumerator
     \divide \baselineskip by \spacingDenominator 

%
%
\def\today{\ifcase\month\or January\or February\or March\or April\or
     May\or June\or July\or August\or September\or October\or November\or
     December\fi\space\number\day, \number\year}
%
%
\def\dst{\displaystyle}
\def\sst{\scriptstyle}
\def\tst{\textstyle}
\def\ssst{\scriptscriptstyle}
%
%
\def\frac#1#2{\dst {#1\over#2}}     
\def\sfrac#1#2{{\tst{#1\over#2}}}   

\def\deqalign#1{\vcenter{\openup1\jot \mathsurround=0pt \ialign{
                \strut\hfil$\displaystyle{##}$&&$\displaystyle{{}##}$\hfil
                \crcr
                #1\crcr}}}         

\def\meqalign#1{\vcenter{\openup1\jot \mathsurround=0pt \ialign{
                &\strut\hfil$\displaystyle{##}$&$\displaystyle{{}##}$\hfil&
                \quad$##$\crcr
                #1\crcr}}}         

%
%
\def\al{\alpha}

\def\ga{\gamma}
\def\de{\delta}

\def\ze{\zeta}
\def\et{\eta}

\def\Si{\Sigma}

\def\Om{\Omega}   
%
%
\def\pmb#1{\setbox0=\hbox{#1}       
     \kern-.025em\copy0\kern-\wd0
     \kern.05em\copy0\kern-\wd0
     \kern-.025em\box0}             
\def\0{{\bf 0}}

\def\k{{\bf k}}

\def\cA{{\cal A}}

\def\cF{{\cal F}}

\def\cH{{\cal H}}

\def\cS{{\cal S}}

%
%
\font\tenfrak                 = eufm10
\font\sevenfrak               = eufm7
\font\fivefrak                = eufb5
\newfam\frakfam
     \textfont\frakfam=\tenfrak
     \scriptfont\frakfam=\sevenfrak   
     \scriptscriptfont\frakfam=\fivefrak
\def\frak{\fam\frakfam\tenfrak}
\font \tensans                = cmss10
\font \fivesans               = cmss10 at 5pt
\font \sevensans              = cmss10 at 7pt
\newfam\sansfam
     \textfont\sansfam=\tensans
     \scriptfont\sansfam=\sevensans
     \scriptscriptfont\sansfam=\fivesans
\def\sans{\fam\sansfam\tensans}
%
%
\def\bbbr{{\rm I\!R}}  
\def\bbbn{{\rm I\!N}}

\def\bbbone{{\mathchoice {\rm 1\mskip-4mu l} {\rm 1\mskip-4mu l}    
{\rm 1\mskip-4.5mu l} {\rm 1\mskip-5mu l}}}
\def\bbbc{{\mathchoice {\setbox0=\hbox{$\displaystyle\rm C$}\hbox{\hbox 
to0pt{\kern0.4\wd0\vrule height0.9\ht0\hss}\box0}}
{\setbox0=\hbox{$\textstyle\rm C$}\hbox{\hbox
to0pt{\kern0.4\wd0\vrule height0.9\ht0\hss}\box0}}
{\setbox0=\hbox{$\scriptstyle\rm C$}\hbox{\hbox
to0pt{\kern0.4\wd0\vrule height0.9\ht0\hss}\box0}}
{\setbox0=\hbox{$\scriptscriptstyle\rm C$}\hbox{\hbox
to0pt{\kern0.4\wd0\vrule height0.9\ht0\hss}\box0}}}}
\def\bbbq{{\mathchoice {\setbox0=\hbox{$\displaystyle\rm               
Q$}\hbox{\raise
0.15\ht0\hbox to0pt{\kern0.4\wd0\vrule height0.8\ht0\hss}\box0}}
{\setbox0=\hbox{$\textstyle\rm Q$}\hbox{\raise
0.15\ht0\hbox to0pt{\kern0.4\wd0\vrule height0.8\ht0\hss}\box0}}
{\setbox0=\hbox{$\scriptstyle\rm Q$}\hbox{\raise
0.15\ht0\hbox to0pt{\kern0.4\wd0\vrule height0.7\ht0\hss}\box0}}
{\setbox0=\hbox{$\scriptscriptstyle\rm Q$}\hbox{\raise
0.15\ht0\hbox to0pt{\kern0.4\wd0\vrule height0.7\ht0\hss}\box0}}}}
\def\bbbz{{\mathchoice {\hbox{$\sans\textstyle Z\kern-0.4em Z$}}       
{\hbox{$\sans\textstyle Z\kern-0.4em Z$}}
{\hbox{$\sans\scriptstyle Z\kern-0.3em Z$}}
{\hbox{$\sans\scriptscriptstyle Z\kern-0.2em Z$}}}}
%
%
\def\const{{\rm const}\,}
\def\sgn{{\rm sgn}}
\def\half{\sfrac{1}{2}}

\def\optbar#1{\vbox{\ialign{##\crcr\hfil${\scriptscriptstyle(}\mkern -1mu
         \vrule height 1.2pt width 3pt depth -.8pt
         {\scriptscriptstyle)}$\hfil\crcr
          \noalign{\kern-1pt\nointerlineskip}$\hfil\displaystyle{#1}\hfil$\crcr}}}
\def\<{\left<}
\def\>{\right>}

\def\smprod{\mathop{\textstyle\prod}}
\def\smsum{\mathop{\textstyle\sum}}
\def\set#1#2{\big\{ \ #1\ \big|\ #2\ \big\}}
\def\eval#1{\big|\lower4pt\hbox{$\displaystyle\sst #1$}}
%
%
\font \tafontt                = cmbx10 scaled\magstep2
\font \tbfontt                = cmbx10 scaled\magstep1
\def\titlea#1{\centerline{\tafontt #1 }\vskip.5truein}
\def\titleb#1{\removelastskip\vskip.3truein%
\noindent{\tbfontt #1 }\vskip.25truein}
\def\titlec#1{\removelastskip\vskip.15truein\noindent{\bf #1 }\vskip.1truein}

%
%
\def\newenvironment#1#2#3#4{\long\def#1##1##2{%
\removelastskip\penalty-100\vskip\baselineskip%
\noindent{#3#2\if!##1!.\else\unskip\ \ignorespaces
##1\unskip\fi\ }{#4\ignorespaces##2\vskip\baselineskip}}}
\newenvironment\lemma{Lemma}{\bf}{\it}
\newenvironment\proposition{Proposition}{\bf}{\it}
\newenvironment\theorem{Theorem}{\bf}{\it}
\newenvironment\corollary{Corollary}{\bf}{\it}
\newenvironment\example{Example}{\bf}{\rm}
\newenvironment\problem{Problem}{\bf}{\rm}
\newenvironment\definition{Definition}{\bf}{\rm}
\newenvironment\remark{Remark}{\bf}{\rm}
\newenvironment\hypothesis{Hypothesis}{\bf}{\it}
\newenvironment\convention{Convention}{\bf}{\it}

\def\Item{\vskip.1in\noindent}

%
%
\long\def\proof#1{\removelastskip\penalty-100\vskip\baselineskip\noindent{\bf
            Proof\if!#1!\else\ \ignorespaces#1\fi:\ }\ \ \ignorespaces}
\long\def\prf{\removelastskip\penalty-100\vskip\baselineskip\noindent{\bf
            Proof:\ }\ \ \ignorespaces}
\def\endproof{\hfill\vrule height .6em width .6em depth 0pt\goodbreak\vskip.25in }

\ifundefined{warnForwardRef}  \def\warnForwardRef{n} \fi
\newcount\chapno
\newcount\sectno
\newcount\equano
\newcount\theono
\newcount\probno

\def\IgNoRe#1{}

\chapno=0
\sectno=0
\equano=0
\theono=0
\probno=0
\def\eqhead{}
\def\frefwarning{\if\warnForwardRef y\immediate\write16{   Forward reference on line \the\inputlineno}\fi}
\def\qqqrefwarning{\immediate\write16{   ??? reference on line \the\inputlineno}}

\def\chap#1{\equano=0\sectno=0\theono=0\probno=0\global\advance\chapno by 1%
\def\eqhead{\ifcase\chapno\or I\or II\or III\or IV\or V\or VI\or VII\or
VIII\or IX\or X\or XI\or XII\or XIII\or XIV\or XV\or XVI\or XVII\or XVIII\or
XIX\or XX\or XXI\or XXII\or XXIII\or XXIV\or XXV\or XXVI\or XXVII\or XXVIII\or XXIX\or XXX\or XXXI\or XXXII\or XXXIII\or XXXIV\or XXXV\or XXXVI\or XXXVII\or XXXVIII\or XXXIX\fi.}%
\titlea{\eqhead \hglue 5pt #1}%
}

\def\sect#1{\global\advance\sectno by 1%
\titleb{\eqhead\number\sectno  \hglue 5pt #1}%
}%

\def\appendix#1#2{\equano=0\sectno=0\theono=0\probno=0\def\eqhead{#1.}
\titlea{Appendix #1: #2}%
}

\def\:#1{\def\temp{\expandafter\IgNoRe\string#1}%
\expandafter\ifx\csname\temp\endcsname\relax%
\expandafter\gdef#1{\qqqrefwarning ???}\fi#1}

\def\Eqn{{\hbox{\global\advance\equano by 1}}%
\eqno ({\rm \eqhead\number\equano})}%

\def\Eqno{{\hbox{\global\advance\equano by 1}}%
 ({\rm \eqhead\number\equano})}%

\def\EQN#1{\Eqn\edef\Zwi{\eqhead\number\equano}%
\global\let #1=\Zwi
}

\def\EQNO#1{\Eqno\edef\Zwi{\eqhead\number\equano}%
\global\let #1=\Zwi
}

\def\STM#1{{\global\advance \theono by 1}%
\eqhead\number\theono
\edef\Zwi{\eqhead\number\theono }
\global\let#1=\Zwi
}

\def\PRB#1{{\global\advance \probno by 1}%
\eqhead\number\probno
\edef\Zwi{\eqhead\number\probno }
\global\let#1=\Zwi
}

\def\PG#1{\def\Zwi{\number\pageno }
\global\let#1=\Zwi
}

\def\Stm{{\global\advance \theono by 1}%
\eqhead\number\theono
}

\def\Prb{{\global\advance \probno by 1}%
\eqhead\number\probno
}

\def\EDEF#1#2{
\def\tEmP{#1}\expandafter\gdef\tEmP{#2}
}



\def\suffix{ps}
\newcount\system
\global\system=3   

\def\ifundefined#1{\expandafter\ifx\csname#1\endcsname\relax}
\ifundefined{figdir}\def\figdir{}\fi
%
%
\newcount\firstline
\newdimen\pswidth  \newdimen\xleft
\newdimen\psheight \newdimen\ytop \newdimen\ybot
\newcount\justx \newcount\justy
\global\justx=0 \global\justy=0
\newdimen\vpos \newtoks\labeL 
\newread\labeLfile \newdimen\xcoord \newdimen\ycoord
\newif\ifdoit 
\newbox\labox
\newdimen\xdvikwid 
\newdimen\xdvikht
\newdimen\pspoints
\newdimen\rwi
\pspoints=1bp
\newcount\temp
\def\readdim#1{\global\read\labeLfile to \temp
\global #1=\temp pt}
%
%
%
%
\def\figcrop#1{\par
\openin\labeLfile=\figdir#1.lbl                                              
\global\read\labeLfile to\firstline\message{#1}               
\global\read\labeLfile to\temp
\readdim{\ybot}
\readdim{\xleft}
\readdim{\ytop}
\global\read\labeLfile to\justx
\global\read\labeLfile to\justy
\global\read\labeLfile to\labeL
\readdim{\pswidth}
\global\advance\pswidth by -\xleft
\readdim{\psheight}
\global\advance\ybot by -\psheight
\global\advance\psheight by -\ytop
\global\read\labeLfile to\justx
\global\read\labeLfile to\justy
\global\read\labeLfile to\labeL
\vbox to\psheight{\vfill
\ifnum\system=1
\fi 
\ifnum\system=2
\fi 
\ifnum\system=3
                                                 \fi         
\ifnum\system=4
\fi         
\ifnum\system=1
\hbox to \pswidth{\kern-\xleft\special{postscriptfile \figdir#1.\suffix }\hfil}\fi
\ifnum\system=2
\hbox to \pswidth{\kern-\xleft\special{ps: plotfile \figdir#1.\suffix }\hfil}\fi
\ifnum\system=3
\hbox to \pswidth{\kern-\xleft\includegraphics{\figdir#1.\suffix}\hfil}\fi
\ifnum\system=4
\hbox to \pswidth{\kern-\xleft\includegraphics{\figdir#1.\suffix}\hfil}\fi
\ifnum\system=5
\hbox to \pswidth{\kern-\xleft\includegraphics{\figdir#1.\suffix}\hfil}\fi 
\ifnum\system=6
   \xdvikwid=\pswidth
   \xdvikht=\psheight
   {\global\divide\xdvikwid by \pspoints}
   {\global\divide\xdvikht by \pspoints}
   \rwi=\xdvikwid
    {\global\multiply\rwi by 10}
\hbox to \pswidth{\kern-\xleft\includegraphics{\figdir#1.\suffix\space}\hfil}\fi                   
\vskip -\baselineskip
\vskip -\ybot 
\vskip-\psheight %
\hbox to\pswidth  {\hss}%
\parindent=0pt\offinterlineskip                                       
\vpos=0 pt%
\loop\readdim{\xcoord}                                 
\ifdim \xcoord < -999pt \doitfalse\else\doittrue\fi                        
\ifdoit \advance \xcoord by -\xleft
\readdim{\ycoord}
\advance \ycoord by -\ytop                              
\global\read\labeLfile to\justx                                       
\global\read\labeLfile to\justy                                       
\global\read\labeLfile to\labeL
\global\setbox\labox=\hbox{\labeL\hskip-0.3em}%
\advance\vpos by-\ycoord                                              
\vskip-\vpos \vpos=\ycoord                                         
\hbox to\pswidth{\hskip\xcoord %
\hbox to 0pt{\ifnum\justx>0\hss\fi%
\vbox to0pt{%
\ifnum\justy<2\vss\fi%
\copy\labox\kern0pt%
\ifnum\justy>0\vss\fi}%
\ifnum\justx<2\hss\fi}%
\hss}%
\repeat%
\advance\vpos by-\psheight%
\vskip-\vpos %
}\closein\labeLfile}
%
%
%
\def\figplace#1#2#3{
\openin\labeLfile=\figdir#1.lbl
\ifeof \labeLfile
       \immediate\write16{***Can't find \figdir#1.lbl; Skipping it.***}
\else  \closein\labeLfile
       \null\hskip#2\raise #3 \hbox{\figcrop{#1}}
\fi
}
%
%
%
%
\def\figput#1{
\openin\labeLfile=\figdir#1.lbl
\ifeof \labeLfile
       \immediate\write16{***Can't find \figdir#1.lbl; Skipping it.***}
\else  \closein\labeLfile
       \hbox{\figcrop{#1}}
\fi
}


    \def\squiggle{\raise2pt\hbox{${\scriptstyle\sim}$}}
    \def\stoday{\number\day\space\ifcase\month\or Jan\or Feb\or 
                      Mar\or Apr\or May\or Jun\or Jul\or Aug\or Sep\or 
                      Oct\or Nov\or Dec\fi, \number\year}

    \def\cb{{\frak c}}
    \def\ib{{\rm b}}
    \def\imp{{J}}
    \def\cl{;}
    \def\cont#1#2#3{\mathop{{\rm\ \, Con}_{#3}}\limits_{#1\rightarrow#2}}
    \def\Cont#1#2#3{\mathop{{\rm\ \, {\cal C}on}_{#3}}\limits_{#1\rightarrow#2}}
    \def\smchoose#1#2{{\tst {#1\choose #2}}}
    
    \def\dblint{\int\kern-0.7em\int}
    \def\susywedge{\mathop{\raise2ex\hbox{$\mathchar"030E\mkern-5mu\mathchar"030E\mkern-2.5mu\mathchar"030F$}}}

\def\tailind#1#2#3#4#5#6{\lower2pt\hbox{$({\scriptstyle {#1\,#2\,#3\atop #4\,#5\,#6}})$}}

    \def\cC{{\cal C}}

    \def\cQ{{\cal Q}}
    \def\cR{{\cal R}}
    
    \def\cW{{\cal W}}
    
    \def\cX{{\cal X}}

    \def\fl{{\frak l}}

    \def\fN{{\frak N}}
    \def\fX{{\frak X}}

    \def\tn{|\kern-1pt|\kern-1pt|}
    \def\TN{\big|\kern-1.5pt\big|\kern-1.5pt\big|}
    \def\TTN{\Big|\kern-2pt\Big|\kern-2pt\Big|}

    \def\rw{\mathclose{:}}
    \def\lw{\mathopen{:}}
    \def\lW{\mathopen{{\tst{\hbox{.}\atop\raise 2.5pt\hbox{.}}}}}
    \def\rW{\mathclose{{\tst{{.}\atop\raise 2.5pt\hbox{.}}}}}
    \def\lww{\mathopen{{\tst{\raise 1pt\hbox{.}\atop\raise 1pt\hbox{.}}}}}
    \def\rww{\mathclose{{\tst{\raise 1pt\hbox{.}\atop\raise 1pt\hbox{.}}}}}

     \def\fcirc{\circ}

   \font\sixrm=cmr6   \font\eightrm=cmr8  
   \font\sixi=cmmi6   \font\eighti=cmmi8  
  \font\sixsy=cmsy6  \font\eightsy=cmsy8 
  \font\sixbf=cmbx6  \font\eightbf=cmbx8 
                     \font\eightit=cmti8 
                     \font\eightsl=cmsl8 
                     \font\eighttt=cmtt8 

\font\eightfrak=eufm7 at 8pt

\def\eightpoint{\def\rm{\fam0\eightrm}
 \textfont0=\eightrm \scriptfont0=\sixrm \scriptscriptfont0=\fiverm
 \textfont1=\eighti \scriptfont1=\sixi \scriptscriptfont1=\fivei
 \textfont2=\eightsy \scriptfont2=\sixsy \scriptscriptfont2=\fivesy
 \textfont3=\tenex \scriptfont3=\tenex \scriptscriptfont3=\tenex
 \textfont\itfam=\eightit \def\it{\fam\itfam\eightit}%
 \textfont\slfam=\eightsl \def\sl{\fam\slfam\eightsl}%
 \textfont\ttfam=\eighttt \def\tt{\fam\ttfam\eighttt}%
 \textfont\frakfam=\eightfrak \def\frak{\fam\frakfam\tenfrak}%
 \textfont\bffam=\eightbf \scriptfont\bffam=\sixbf
 \scriptscriptfont\bffam=\fivebf \def\bf{\fam\bffam\eightbf}%
 \normalbaselineskip=9pt
 \setbox\strutbox=\hbox{\vrule height7pt depth2pt width0pt}%
 \let\sc=\sixrm \let\big=\eightbig \normalbaselines\rm}
\catcode`@=11
\def\footnote#1{\edef\@sf{\spacefactor\the\spacefactor}#1\@sf
     \insert\footins\bgroup\eightpoint
     \interlinepenalty100 \let\par=\endgraf
     \leftskip=0pt \rightskip=0pt
     \splittopskip=10pt plus 1pt minus 1pt \floatingpenalty=20000
     \smallskip\item{#1}\bgroup\strut\aftergroup\@foot\let\next}
\skip\footins=12pt plus 2pt minus 4pt
\dimen\footins=30pc
\catcode`@=12


\newcount\CHAPNO
\newcount\APPNO
\CHAPNO=0
\APPNO=1
\def\advCHAPNO{\advance\CHAPNO by 1}
\def\advAPPNO{\advance\APPNO by 1}

\def\caproman#1{\ifcase#1\or I\or II\or III\or IV\or V\or VI\or VII\or
VIII\or IX\or X\or XI\or XII\or XIII\or XIV\or XV\or XVI\or XVII\or XVIII\or
XIX\or XX\or XXI\or XXII\or XXIII\or XXIV\or XXV\or XXVI\or XXVII\or XXVIII\or XXIX\or XXX\or XXXI\or XXXII\or XXXIII\or XXXIV\or XXXV\or XXXVI\or XXXVII\or XXXVIII\or XXXIX\fi}%

\def\capletter#1{\ifcase#1\or A\or B\or C\or D\or E\or F\or G\or
H\or I\or J\or K\or L\or M\or N\or O\or P\or Q\or R\or
S\or T\or U\or V\or W\or X\or Y\or Z\fi}%

\newcount\cHintroI \cHintroI=\CHAPNO \advCHAPNO 
\newcount\cHintroOverview  \cHintroOverview=\CHAPNO \advCHAPNO 
                              \edef\CHintroOverview{\caproman\CHAPNO}  
\newcount\cHrenmap \cHrenmap=\CHAPNO \advCHAPNO 

 \advAPPNO

\newcount\cHintroII \cHintroII=\CHAPNO \advCHAPNO 
                              
\newcount\cHfirstscale \cHfirstscale=\CHAPNO \advCHAPNO
                              
\newcount\cHnewsectors \cHnewsectors=\CHAPNO \advCHAPNO
                              
\newcount\cHphladders \cHphladders=\CHAPNO \advCHAPNO
                              
\newcount\cHfinitescale \cHfinitescale=\CHAPNO \advCHAPNO
                              
\newcount\cHstep \cHstep=\CHAPNO \advCHAPNO
                              
\newcount\cHrecurs \cHrecurs=\CHAPNO \advCHAPNO
                              
 \advAPPNO

\newcount\cHintroIII \cHintroIII=\CHAPNO \advCHAPNO
                              
\newcount\cHtildefinitescale \cHtildefinitescale=\CHAPNO \advCHAPNO
                              
\newcount\cHtildenewsectors \cHtildenewsectors=\CHAPNO \advCHAPNO
                              
\newcount\cHtildephladders \cHtildephladders=\CHAPNO \advCHAPNO
                              
\newcount\cHtildestep  \cHtildestep=\CHAPNO \advCHAPNO

 \advAPPNO
 \advAPPNO


  \IgNoRe{PG}
  \IgNoRe{EQN}
  \IgNoRe{STM Assertion }
  \IgNoRe{PG}
  \IgNoRe{STM Assertion }
  \IgNoRe{STM Assertion }
  \IgNoRe{STM Assertion }
  \IgNoRe{STM Assertion }
  \IgNoRe{STM Assertion }
  \IgNoRe{STM Assertion }
  \IgNoRe{EQN}
  \IgNoRe{STM Assertion }
  \IgNoRe{STM Assertion }
  \IgNoRe{STM Assertion }
  \IgNoRe{STM Assertion }
  \IgNoRe{STM Assertion }
  \IgNoRe{STM Assertion }
  \IgNoRe{STM Assertion }
  \IgNoRe{STM Assertion }
  \IgNoRe{PG}
  \IgNoRe{STM Assertion }
  \IgNoRe{STM Assertion }
  \IgNoRe{STM Assertion }
  \IgNoRe{STM Assertion }
  \IgNoRe{STM Assertion }
  \IgNoRe{STM Assertion }
  \IgNoRe{STM Assertion }
  \IgNoRe{STM Assertion }
  \IgNoRe{EQN}
  \IgNoRe{PG}
  \IgNoRe{PG}
  \IgNoRe{STM Assertion }
  \IgNoRe{STM Assertion }
  \IgNoRe{STM Assertion }
  \IgNoRe{STM Assertion }
  \IgNoRe{STM Assertion }
  \IgNoRe{STM Assertion }
  \IgNoRe{PG}
  \IgNoRe{EQN}
  \IgNoRe{EQN}
  \IgNoRe{EQN}
  \IgNoRe{EQN}
  \IgNoRe{EQN}
  \IgNoRe{EQN}
  \IgNoRe{STM Assertion }
  \IgNoRe{STM Assertion }
  \IgNoRe{STM Assertion }
  \IgNoRe{STM Assertion }
  \IgNoRe{PG}
  \IgNoRe{STM Assertion }
  \IgNoRe{EQN}
  \IgNoRe{STM Assertion }
  \IgNoRe{STM Assertion }
  \IgNoRe{STM Assertion }
  \IgNoRe{STM Assertion }
  \IgNoRe{PG}
  \IgNoRe{STM Assertion }
  \IgNoRe{STM Assertion }
  \IgNoRe{STM Assertion }
  \IgNoRe{STM Assertion }
  \IgNoRe{PG}
  \IgNoRe{EQN}
  \IgNoRe{EQN}
  \IgNoRe{PG}
  \IgNoRe{STM Assertion }
  \IgNoRe{STM Assertion }
  \IgNoRe{STM Assertion }
  \IgNoRe{EQN}
  \IgNoRe{PG}
  \IgNoRe{STM Assertion }
  \IgNoRe{STM Assertion }
  \IgNoRe{EQN}
  \IgNoRe{STM Assertion }
  \IgNoRe{STM Assertion }
  \IgNoRe{STM Assertion }
  \IgNoRe{STM Assertion }
  \IgNoRe{PG}
  \IgNoRe{STM Assertion }
  \IgNoRe{STM Assertion }
  \IgNoRe{STM Assertion }
  \IgNoRe{STM Assertion }
  \IgNoRe{STM Assertion }
  \IgNoRe{STM Assertion }
  \IgNoRe{STM Assertion }
  \IgNoRe{STM Assertion }
  \IgNoRe{PG}
  \IgNoRe{STM Assertion }
  \IgNoRe{STM Assertion }
  \IgNoRe{STM Assertion }
  \IgNoRe{STM Assertion }
  \IgNoRe{STM Assertion }
  \IgNoRe{EQN}
  \IgNoRe{STM Assertion }
  \IgNoRe{STM Assertion }
  \IgNoRe{PG}
  \IgNoRe{STM Assertion }
  \IgNoRe{STM Assertion }
  \IgNoRe{STM Assertion }
  \IgNoRe{STM Assertion }
  \IgNoRe{STM Assertion }
  \IgNoRe{STM Assertion }
  \IgNoRe{STM Assertion }
  \IgNoRe{STM Assertion }
  \IgNoRe{STM Assertion }
  \IgNoRe{EQN}
  \IgNoRe{STM Assertion }
  \IgNoRe{STM Assertion }
  \IgNoRe{PG}
  \IgNoRe{STM Assertion }
  \IgNoRe{STM Assertion }
  \IgNoRe{STM Assertion }
  \IgNoRe{STM Assertion }
  \IgNoRe{STM Assertion }
  \IgNoRe{STM Assertion }
  \IgNoRe{STM Assertion }
  \IgNoRe{PG}
  \IgNoRe{STM Assertion }
  \IgNoRe{PG}
  \IgNoRe{PG}
  \IgNoRe{STM Assertion }
  \IgNoRe{STM Assertion }
  \IgNoRe{PG}
  \IgNoRe{STM Assertion }
  \IgNoRe{STM Assertion }
  \IgNoRe{STM Assertion }
  \IgNoRe{STM Assertion }
  \IgNoRe{STM Assertion }
  \IgNoRe{STM Assertion }
  \IgNoRe{STM Assertion }
  \IgNoRe{STM Assertion }
  \IgNoRe{STM Assertion }
  \IgNoRe{STM Assertion }
  \IgNoRe{STM Assertion }
  \IgNoRe{STM Assertion }
  \IgNoRe{STM Assertion }
  \IgNoRe{STM Assertion }
  \IgNoRe{STM Assertion }
  \IgNoRe{STM Assertion }
  \IgNoRe{EQN}
  \IgNoRe{STM Assertion }
  \IgNoRe{STM Assertion }
  \IgNoRe{PG}
  \IgNoRe{STM Assertion }
  \IgNoRe{EQN}
  \IgNoRe{EQN}
  \IgNoRe{STM Assertion }
  \IgNoRe{EQN}
  \IgNoRe{EQN}
  \IgNoRe{STM Assertion }
  \IgNoRe{EQN}
  \IgNoRe{STM Assertion }
  \IgNoRe{EQN}
  \IgNoRe{STM Assertion }
  \IgNoRe{STM Assertion }
  \IgNoRe{STM Assertion }
  \IgNoRe{STM Assertion }
  \IgNoRe{PG}
  \IgNoRe{STM Assertion }
  \IgNoRe{STM Assertion }
  \IgNoRe{STM Assertion }
  \IgNoRe{STM Assertion }
  \IgNoRe{EQN}
  \IgNoRe{EQN}
  \IgNoRe{STM Assertion }
  \IgNoRe{STM Assertion }
  \IgNoRe{PG}
  \IgNoRe{STM Assertion }
  \IgNoRe{STM Assertion }
  \IgNoRe{STM Assertion }
  \IgNoRe{EQN}
  \IgNoRe{STM Assertion }
  \IgNoRe{EQN}
  \IgNoRe{STM Assertion }
  \IgNoRe{STM Assertion }
  \IgNoRe{EQN}
  \IgNoRe{STM Assertion }
  \IgNoRe{EQN}
  \IgNoRe{EQN}
  \IgNoRe{EQN}
  \IgNoRe{EQN}
  \IgNoRe{STM Assertion }
  \IgNoRe{STM Assertion }
  \IgNoRe{STM Assertion }
  \IgNoRe{STM Assertion }
  \IgNoRe{PG}
  \IgNoRe{STM Assertion }
  \IgNoRe{STM Assertion }
  \IgNoRe{STM Assertion }
  \IgNoRe{STM Assertion }
  \IgNoRe{STM Assertion }
  \IgNoRe{STM Assertion }
  \IgNoRe{STM Assertion }
  \IgNoRe{STM Assertion }
  \IgNoRe{STM Assertion }
  \IgNoRe{STM Assertion }
  \IgNoRe{EQN}
  \IgNoRe{EQN}
  \IgNoRe{STM Assertion }
  \IgNoRe{PG}
  \IgNoRe{STM Assertion }
  \IgNoRe{STM Assertion }
  \IgNoRe{EQN}
  \IgNoRe{STM Assertion }
  \IgNoRe{STM Assertion }
  \IgNoRe{EQN}
  \IgNoRe{EQN}
  \IgNoRe{EQN}
  \IgNoRe{EQN}
  \IgNoRe{EQN}
  \IgNoRe{STM Assertion }
  \IgNoRe{STM Assertion }
  \IgNoRe{STM Assertion }
  \IgNoRe{EQN}
  \IgNoRe{EQN}
  \IgNoRe{EQN}
  \IgNoRe{EQN}
  \IgNoRe{EQN}
  \IgNoRe{EQN}
  \IgNoRe{STM Assertion }
  \IgNoRe{STM Assertion }
  \IgNoRe{STM Assertion }
  \IgNoRe{PG}
  \IgNoRe{STM Assertion }
  \IgNoRe{STM Assertion }
  \IgNoRe{EQN}
  \IgNoRe{EQN}
  \IgNoRe{STM Assertion }
  \IgNoRe{EQN}
  \IgNoRe{EQN}
  \IgNoRe{STM Assertion }
  \IgNoRe{EQN}
  \IgNoRe{EQN}
  \IgNoRe{STM Assertion }
  \IgNoRe{STM Assertion }
  \IgNoRe{PG}
  \IgNoRe{STM Assertion }
  \IgNoRe{STM Assertion }
  \IgNoRe{STM Assertion }
  \IgNoRe{PG}
  \IgNoRe{STM Assertion }
  \IgNoRe{STM Assertion }
  \IgNoRe{STM Assertion }
  \IgNoRe{STM Assertion }
  \IgNoRe{PG}
  \IgNoRe{STM Assertion }
  \IgNoRe{STM Assertion }
  \IgNoRe{STM Assertion }
  \IgNoRe{STM Assertion }
  \IgNoRe{STM Assertion }
  \IgNoRe{STM Assertion }
  \IgNoRe{STM Assertion }
  \IgNoRe{STM Assertion }
  \IgNoRe{STM Assertion }
  \IgNoRe{STM Assertion }
  \IgNoRe{STM Assertion }
  \IgNoRe{STM Assertion }
  \IgNoRe{STM Assertion }
  \IgNoRe{STM Assertion }
  \IgNoRe{PG}
  \IgNoRe{STM Assertion }
  \IgNoRe{PG}
  \IgNoRe{STM Assertion }
  \IgNoRe{STM Assertion }
  \IgNoRe{PG}
  \IgNoRe{STM Assertion }
  \IgNoRe{STM Assertion }
  \IgNoRe{EQN}
  \IgNoRe{PG}
  \IgNoRe{EQN}
  \IgNoRe{STM Assertion }
  \IgNoRe{STM Assertion }
  \IgNoRe{PG}
  \IgNoRe{EQN}
  \IgNoRe{EQN}
  \IgNoRe{EQN}
  \IgNoRe{EQN}
  \IgNoRe{EQN}
  \IgNoRe{STM Assertion }
  \IgNoRe{STM Assertion }
  \IgNoRe{EQN}
  \IgNoRe{STM Assertion }
  \IgNoRe{PG}
  \IgNoRe{PG}
  \IgNoRe{PG}
  \IgNoRe{STM Assertion }
  \IgNoRe{EQN}
  \IgNoRe{STM Assertion }
  \IgNoRe{PG}
  \IgNoRe{STM Assertion }
  \IgNoRe{EQN}
  \IgNoRe{STM Assertion }
  \IgNoRe{STM Assertion }
  \IgNoRe{PG}
  \IgNoRe{EQN}
  \IgNoRe{EQN}
  \IgNoRe{EQN}
  \IgNoRe{STM Assertion }
  \IgNoRe{STM Assertion }
  \IgNoRe{STM Assertion }
  \IgNoRe{EQN}
  \IgNoRe{STM Assertion }
  \IgNoRe{STM Assertion }
  \IgNoRe{STM Assertion }
  \IgNoRe{STM Assertion }
  \IgNoRe{STM Assertion }
  \IgNoRe{STM Assertion }
  \IgNoRe{PG}
  \IgNoRe{STM Assertion }
  \IgNoRe{STM Assertion }
  \IgNoRe{STM Assertion }
  \IgNoRe{STM Assertion }
  \IgNoRe{STM Assertion }
  \IgNoRe{STM Assertion }
  \IgNoRe{STM Assertion }
  \IgNoRe{PG}
  \IgNoRe{PG}


  \IgNoRe{PG}
 \def\defsuperalgebra{\frefwarning II.1} \IgNoRe{STM Assertion }
  \IgNoRe{PG}
  \IgNoRe{PG}
 \def\exsuper{\frefwarning II.2} \IgNoRe{STM Assertion }
  \IgNoRe{PG}
 \def\defrengroup{\frefwarning II.3} \IgNoRe{STM Assertion }
 \def\remnormal{\frefwarning II.4} \IgNoRe{STM Assertion }
  \IgNoRe{EQN}
 \def\defContract{\frefwarning II.5} \IgNoRe{STM Assertion }
  \IgNoRe{STM Assertion }
  \IgNoRe{PG}
  \IgNoRe{STM Assertion }
  \IgNoRe{STM Assertion }
  \IgNoRe{EQN}
 \def\defGrasscontract{\frefwarning II.9} \IgNoRe{STM Assertion }
  \IgNoRe{STM Assertion }
  \IgNoRe{STM Assertion }
 \def\remcontract{\frefwarning II.12} \IgNoRe{STM Assertion }
 \def\lemcontract{\frefwarning II.13} \IgNoRe{STM Assertion }
 \def\defFancynormdomain{\frefwarning II.14} \IgNoRe{STM Assertion }
  \IgNoRe{PG}
  \IgNoRe{STM Assertion }
  \IgNoRe{STM Assertion }
  \IgNoRe{STM Assertion }
 \def\defsymnorm{\frefwarning II.18} \IgNoRe{STM Assertion }
  \IgNoRe{STM Assertion }
  \IgNoRe{STM Assertion }
 \def\defAmnonetonr{\frefwarning II.21} \IgNoRe{STM Assertion }
  \IgNoRe{STM Assertion }
 \def\deffunctnorm{\frefwarning II.23} \IgNoRe{STM Assertion }
 \def\remfunctnorm{\frefwarning II.24} \IgNoRe{STM Assertion }
 \def\defcontractintbound{\frefwarning II.25} \IgNoRe{STM Assertion }
 \def\egIIcompatnorm{\frefwarning II.26} \IgNoRe{STM Assertion }
  \IgNoRe{PG}
  \IgNoRe{EQN}
 \def\defrengroupInfDim{\frefwarning II.27} \IgNoRe{STM Assertion }
 \def\theorII{\frefwarning II.28} \IgNoRe{STM Assertion }
 \def\lemGrasscontractnorm{\frefwarning II.29} \IgNoRe{STM Assertion }
  \IgNoRe{PG}
  \IgNoRe{STM Assertion }
 \def\lemwicknorm{\frefwarning II.31} \IgNoRe{STM Assertion }
 \def\corwicknorm{\frefwarning II.32} \IgNoRe{STM Assertion }
 \def\propestCconnect{\frefwarning II.33} \IgNoRe{STM Assertion }
 \def\eqnPropestCconnect{\frefwarning II.4} \IgNoRe{EQN}
  \IgNoRe{EQN}
 \def\remrenschw{\frefwarning III.1} \IgNoRe{STM Assertion }
  \IgNoRe{PG}
  \IgNoRe{PG}
 \def\theoremIII{\frefwarning III.2} \IgNoRe{STM Assertion }
  \IgNoRe{EQN}
  \IgNoRe{STM Assertion }
 \def\defcalR{\frefwarning III.4} \IgNoRe{STM Assertion }
 \def\propcalR{\frefwarning III.5} \IgNoRe{STM Assertion }
 \def\eqRCdef{\frefwarning III.2} \IgNoRe{EQN}
 \def\remcalR{\frefwarning III.6} \IgNoRe{STM Assertion }
 \def\propnormestR{\frefwarning III.7} \IgNoRe{STM Assertion }
 \def\lemnormestcalR{\frefwarning III.8} \IgNoRe{STM Assertion }
  \IgNoRe{PG}
 \def\cornormestcalR{\frefwarning III.9} \IgNoRe{STM Assertion }
 \def\propnormestcalS{\frefwarning III.10} \IgNoRe{STM Assertion }
  \IgNoRe{PG}
  \IgNoRe{STM Assertion }
 \def\theoremIVa{\frefwarning IV.1} \IgNoRe{STM Assertion }
  \IgNoRe{STM Assertion }
  \IgNoRe{PG}
  \IgNoRe{STM Assertion }
  \IgNoRe{STM Assertion }
  \IgNoRe{STM Assertion }
  \IgNoRe{STM Assertion }
  \IgNoRe{STM Assertion }
  \IgNoRe{STM Assertion }
  \IgNoRe{STM Assertion }
 \def\propBII{\frefwarning A.2} \IgNoRe{STM Assertion }
  \IgNoRe{PG}
 \def\corBIII{\frefwarning A.3} \IgNoRe{STM Assertion }
  \IgNoRe{STM Assertion }
 \def\lemBV{\frefwarning A.5} \IgNoRe{STM Assertion }
  \IgNoRe{STM Assertion }
 \def\lemBVI{\frefwarning A.7} \IgNoRe{STM Assertion }
  \IgNoRe{STM Assertion }
 \def\propGII{\frefwarning B.1} \IgNoRe{STM Assertion }
  \IgNoRe{STM Assertion }
  \IgNoRe{PG}
  \IgNoRe{STM Assertion }
  \IgNoRe{STM Assertion }
  \IgNoRe{PG}
 \def\pgRV{\frefwarning 1} \IgNoRe{PG}
 \def\defimprnorm{\frefwarning VI.1} \IgNoRe{STM Assertion }
 \def\lemGrassimprnorm{\frefwarning VI.2} \IgNoRe{STM Assertion }
 \def\pgRVI{\frefwarning 2} \IgNoRe{PG}
 \def\pgRVIa{\frefwarning 2} \IgNoRe{PG}
 \def\deffunctdegree{\frefwarning VI.3} \IgNoRe{STM Assertion }
 \def\propimprnormest{\frefwarning VI.4} \IgNoRe{STM Assertion }
 \def\eqnPropimprnormest{\frefwarning VI.1} \IgNoRe{EQN}
 \def\defladders{\frefwarning VI.5} \IgNoRe{STM Assertion }
 \def\pgRVIb{\frefwarning 5} \IgNoRe{PG}
 \def\theoremVa{\frefwarning VI.6} \IgNoRe{STM Assertion }
 \def\remtheotheo{\frefwarning VI.7} \IgNoRe{STM Assertion }
 \def\deftails{\frefwarning VI.8} \IgNoRe{STM Assertion }
 \def\pgRVIc{\frefwarning 7} \IgNoRe{PG}
 \def\remformladder{\frefwarning VI.9} \IgNoRe{STM Assertion }
 \def\theoremVb{\frefwarning VI.10} \IgNoRe{STM Assertion }
 \def\deftildeimprconf{\frefwarning VI.11} \IgNoRe{STM Assertion }
 \def\eqnRmodintbnd{\frefwarning VI.2} \IgNoRe{EQN}
 \def\lemtildeimprconfig{\frefwarning VI.12} \IgNoRe{STM Assertion }
 \def\pgRVId{\frefwarning 11} \IgNoRe{PG}
 \def\eqnRmodintbndprf{\frefwarning VI.3} \IgNoRe{EQN}
 \def\defimprconf{\frefwarning VI.13} \IgNoRe{STM Assertion }
 \def\remimprconfig{\frefwarning VI.14} \IgNoRe{STM Assertion }
 \def\lemimprconfig{\frefwarning VI.15} \IgNoRe{STM Assertion }
 \def\remJlimit{\frefwarning VI.16} \IgNoRe{STM Assertion }
 \def\eqJlimit{\frefwarning VI.4} \IgNoRe{EQN}
 \def\eqndefrctilde{\frefwarning VII.1} \IgNoRe{EQN}
 \def\pgRVII{\frefwarning 15} \IgNoRe{PG}
 \def\pgRVIIa{\frefwarning 15} \IgNoRe{PG}
 \def\propoverldecR{\frefwarning VII.1} \IgNoRe{STM Assertion }
 \def\removerldecR{\frefwarning VII.2} \IgNoRe{STM Assertion }
 \def\lemsplittwo{\frefwarning VII.3} \IgNoRe{STM Assertion }
 \def\propsplittwo{\frefwarning VII.4} \IgNoRe{STM Assertion }
 \def\corsplittwo{\frefwarning VII.5} \IgNoRe{STM Assertion }
 \def\lemimprRR{\frefwarning VII.6} \IgNoRe{STM Assertion }
 \def\propimprRR{\frefwarning VII.7} \IgNoRe{STM Assertion }
 \def\defP{\frefwarning VII.8} \IgNoRe{STM Assertion }
 \def\pgRVIIb{\frefwarning 22} \IgNoRe{PG}
 \def\deftailsf{\frefwarning VII.9} \IgNoRe{STM Assertion }
 \def\remtailend{\frefwarning VII.10} \IgNoRe{STM Assertion }
 \def\lemtailI{\frefwarning VII.11} \IgNoRe{STM Assertion }
 \def\lemtailII{\frefwarning VII.12} \IgNoRe{STM Assertion }
 \def\lemtailIII{\frefwarning VII.13} \IgNoRe{STM Assertion }
 \def\defeffecttail{\frefwarning VII.14} \IgNoRe{STM Assertion }
 \def\remeffecttail{\frefwarning VII.15} \IgNoRe{STM Assertion }
 \def\propprodtails{\frefwarning VII.16} \IgNoRe{STM Assertion }
 \def\corprodtails{\frefwarning VII.17} \IgNoRe{STM Assertion }
 \def\defQ{\frefwarning VIII.1} \IgNoRe{STM Assertion }
 \def\pgRVIII{\frefwarning 32} \IgNoRe{PG}
 \def\defsymsupalg{\frefwarning VIII.2} \IgNoRe{STM Assertion }
 \def\exsymsupalg{\frefwarning VIII.3} \IgNoRe{STM Assertion }
 \def\remsupersub{\frefwarning VIII.4} \IgNoRe{STM Assertion }
 \def\pgRVIIIa{\frefwarning 34} \IgNoRe{PG}
 \def\defenlalg{\frefwarning VIII.5} \IgNoRe{STM Assertion }
 \def\remEv{\frefwarning VIII.6} \IgNoRe{STM Assertion }
 \def\defPi{\frefwarning VIII.7} \IgNoRe{STM Assertion }
 \def\lemestEv{\frefwarning VIII.8} \IgNoRe{STM Assertion }
 \def\pgRVIIIb{\frefwarning 36} \IgNoRe{PG}
 \def\propalgtwoWick{\frefwarning VIII.9} \IgNoRe{STM Assertion }
 \def\lemalgtwoWick{\frefwarning VIII.10} \IgNoRe{STM Assertion }
 \def\pgRVIIIc{\frefwarning 37} \IgNoRe{PG}
 \def\propesttwoWick{\frefwarning VIII.11} \IgNoRe{STM Assertion }
 \def\pgRVIIId{\frefwarning 39} \IgNoRe{PG}
 \def\lemQWick{\frefwarning VIII.12} \IgNoRe{STM Assertion }
 \def\pgRVIIIe{\frefwarning 41} \IgNoRe{PG}
 \def\propoverldecQ{\frefwarning VIII.13} \IgNoRe{STM Assertion }
 \def\normestQ{\frefwarning VIII.14} \IgNoRe{STM Assertion }
 \def\propsplittwoQ{\frefwarning IX.1} \IgNoRe{STM Assertion }
 \def\lemsplittwoQ{\frefwarning IX.2} \IgNoRe{STM Assertion }
 \def\pgRIX{\frefwarning 44} \IgNoRe{PG}
 \def\pgRIXa{\frefwarning 44} \IgNoRe{PG}
 \def\propimprQQ{\frefwarning IX.3} \IgNoRe{STM Assertion }
 \def\lemimprRRCD{\frefwarning IX.4} \IgNoRe{STM Assertion }
 \def\eqnVIIIi{\frefwarning IX.1} \IgNoRe{EQN}
 \def\eqnVIIIii{\frefwarning IX.2} \IgNoRe{EQN}
 \def\deftailsec{\frefwarning IX.5} \IgNoRe{STM Assertion }
 \def\remtailendsec{\frefwarning IX.6} \IgNoRe{STM Assertion }
 \def\defPII{\frefwarning IX.7} \IgNoRe{STM Assertion }
 \def\defeffnorm{\frefwarning IX.8} \IgNoRe{STM Assertion }
 \def\pgRIXb{\frefwarning 52} \IgNoRe{PG}
 \def\remeffnorm{\frefwarning IX.9} \IgNoRe{STM Assertion }
 \def\lemtailIsec{\frefwarning IX.10} \IgNoRe{STM Assertion }
 \def\lemtailIIsec{\frefwarning IX.11} \IgNoRe{STM Assertion }
 \def\lemtailIIIsec{\frefwarning IX.12} \IgNoRe{STM Assertion }
 \def\defeffecttailsec{\frefwarning IX.13} \IgNoRe{STM Assertion }
 \def\remeffecttailsec{\frefwarning IX.14} \IgNoRe{STM Assertion }
 \def\propprodtailssec{\frefwarning IX.15} \IgNoRe{STM Assertion }
 \def\pgRIXc{\frefwarning 60} \IgNoRe{PG}
 \def\corprodtailssec{\frefwarning IX.16} \IgNoRe{STM Assertion }
 \def\eqnColourpreserving{\frefwarning X.1} \IgNoRe{EQN}
 \def\eqncptripnorm{\frefwarning X.2} \IgNoRe{EQN}
 \def\pgRX{\frefwarning 71} \IgNoRe{PG}
 \def\propvectorintconst{\frefwarning X.1} \IgNoRe{STM Assertion }
 \def\eqnvectorCsup{\frefwarning X.3} \IgNoRe{EQN}
 \def\thmExample{\frefwarning X.2} \IgNoRe{STM Assertion }
 \def\defAI{\frefwarning C.1} \IgNoRe{STM Assertion }
 \def\lemAII{\frefwarning C.2} \IgNoRe{STM Assertion }
 \def\pgRC{\frefwarning 79} \IgNoRe{PG}
 \def\lemAIII{\frefwarning C.3} \IgNoRe{STM Assertion }
 \def\eqAI{\frefwarning C.1} \IgNoRe{EQN}
 \def\propAIV{\frefwarning C.4} \IgNoRe{STM Assertion }
 \def\pgRIIref{\frefwarning 83} \IgNoRe{PG}
  \IgNoRe{PG}
 \def\pgRIInot{\frefwarning 84} \IgNoRe{PG}

\chapno=4


{\nopagenumbers
\multiply\baselineskip by \spacingDenominator\divide \baselineskip by\spacingNumerator

\null\vskip3truecm

%
%
\centerline{\tafontt Convergence of Perturbation Expansions}
\centerline{\tafontt  in Fermionic Models}
\vskip0.1in
\centerline{\tbfontt Part 2: Overlapping Loops}

\vskip0.75in
\centerline{Joel Feldman{\parindent=.15in\footnote{$^{*}$}{Research supported 
in part by the
 Natural Sciences and Engineering Research Council of Canada and the Forschungsinstitut f\"ur Mathematik, ETH Z\"urich}}}
\centerline{Department of Mathematics}
\centerline{University of British Columbia}
\centerline{Vancouver, B.C. }
\centerline{CANADA\ \   V6T 1Z2}
\centerline{feldman@math.ubc.ca}
\centerline{http:/\hskip-3pt/www.math.ubc.ca/\squiggle
feldman/}
\vskip0.3in
\centerline{Horst Kn\"orrer, Eugene Trubowitz}
\centerline{Mathematik}
\centerline{ETH-Zentrum}
\centerline{CH-8092 Z\"urich}
\centerline{SWITZERLAND}
\centerline{knoerrer@math.ethz.ch, trub@math.ethz.ch}
\centerline{http:/\hskip-3pt/www.math.ethz.ch/\squiggle
knoerrer/}

\vskip0.75in
\noindent
%
{\bf Abstract.\ \ \ } 
We improve on the abstract estimate obtained in Part 1 by assuming that 
there are constraints imposed by `overlapping momentum loops'. These 
constraints are active in a two dimensional,  weakly 
coupled fermion gas with a strictly convex Fermi curve.
The improved estimate is used in another paper to control everything but 
the sum of all ladder contributions to the thermodynamic Green's functions. 

\vfill
\eject

\titleb{Table of Contents}
\halign{\hfill#\ &\hfill#\ &#\hfill&\ p\ \hfil#&\ p\ \hfil#\cr
\noalign{\vskip0.05in}
\S V&\omit Introduction                          \span&\:\pgRV&\omit\cr
\noalign{\vskip0.05in}
\S VI&\omit Overlapping Loops                        \span&\:\pgRVI\cr
&&Norms                                             &\omit&\:\pgRVIa\cr
&&Ladders                                           &\omit&\:\pgRVIb\cr
&&Overlapping loops for the Schwinger functional    &\omit&\:\pgRVIc\cr
&&Configurations of Norms with Improved Power Counting &\omit&\:\pgRVId\cr
\noalign{\vskip0.05in}
\S VII&\omit Finding Overlapping Loops               \span&\:\pgRVII\cr
&&Overlapping loops created by the operator $\cR_{K,C}$ &\omit&\:\pgRVIIa\cr
&&Tails                                                 &\omit&\:\pgRVIIb\cr
\noalign{\vskip0.05in}
\S VIII&\omit The Enlarged Algebra                   \span&\:\pgRVIII\cr
&&Definition of the enlarged algebra                &\omit&\:\pgRVIIIa\cr
&&Norm estimates for the enlarged algebra           &\omit&\:\pgRVIIIb\cr
&&Schwinger Functionals over the Extended Algebra   &\omit&\:\pgRVIIIc\cr
&&A second proof of Theorem \theoremIVa             &\omit&\:\pgRVIIId\cr
&&The Operator $Q$                                  &\omit&\:\pgRVIIIe\cr
\noalign{\vskip0.05in}
\S IX&\omit Overlapping Loops created by the second Covariance\span&\:\pgRIX\cr
&&Implementing Overlapping Loops                     &\omit&\:\pgRIXa\cr
&&Tails                                                 &\omit&\:\pgRIXb\cr
&&Proof of Theorem \:\theoremVb\ in the general case      &\omit&\:\pgRIXc\cr
\noalign{\vskip0.05in}
\S X&\omit Example: A Vector Model                   \span&\:\pgRX\cr
\noalign{\vskip0.05in}
{\bf Appendices}\span\cr
\noalign{\vskip0.05in}
\S C&\omit Ladders expressed in terms of kernels         \span&\:\pgRC\cr
\noalign{\vskip0.05in}
 &\omit References                                    \span&\:\pgRIIref \cr
\noalign{\vskip0.05in}
 &\omit Notation                                      \span&\:\pgRIInot \cr
}
\vfill\eject
\multiply\baselineskip by \spacingNumerator\divide \baselineskip by\spacingDenominator}
\pageno=1


\chap{Introduction to Part 2}\PG\pgRV
In the perturbative analysis of many fermion systems with weak 
short--range interaction in two or more space dimensions, the presence
of an overlapping loop in a Feynman diagram introduces a volume effect in
momentum space that leads to an improvement to ``naive power counting''.
For a detailed discussion of this effect and its consequences, see [FST1-4]. 
For a short description, see subsection 4 of [FKTf1,\S\CHintroOverview].  
In [FKTo3], we use nonperturbative bounds for systems, in two space dimensions, that are based on the cancellation scheme between diagrams 
developed in part 1 of this paper. In this second part, we modify the 
construction so that we can exploit enough overlapping loops to get improved
power counting for the two point function and the non--ladder part of the 
four point function. As in part 1, the treatment is in an abstract setting, 
formulated using systems of seminorms. The postulated volume improvement 
effects are expressed in terms of these seminorms  (Definition \:\defimprnorm).
The main result for the renormalization group map is Theorem \:\theoremVa.
It follows from Theorem \:\theoremVb, which is the main result on the Schwinger functional.

The discussion of the renormalization group map in the first part of the paper 
is based on the representation developed in [FKT1]
(which in turn evolved out of the representation developed in [FMRT]). 
The representation of [FKT1] decomposes Feynman diagrams into annuli.
The first annulus consists of all interaction vertices directly connected
to the external vertices. The second annulus consists of all interaction 
vertices directly connected to the first annulus but not to the 
external vertices. And so on. See the introduction to [FKT1]. Overlapping loops 
that only involve vertices of neighbouring annuli are relatively easy to 
exploit. It turns out, that for the analysis of the two point function and the 
non--ladder part of the four point function, it suffices to use overlapping 
loops that involve only vertices of at most three adjacent annuli. A special 
case of Theorem \:\theoremVb, for which this combinatorial fact is easier to 
see, is proven at
the end of \S VII. After some preparation in \S VIII, the general case 
is proven at the end of \S IX. In \S X, we apply Theorem \:\theoremVa\ 
to a simple vector model. We also describe, by drawing an analogy with 
the vector model, how sectors can be used to 
nonperturbatively implement overlapping loops for many fermion systems.
 A notation table is provided at the end of the paper.

\vfill\eject

\chap{Overlapping Loops}\PG\pgRVI
\sect{ Norms}\PG\pgRVIa

Again, let $A$ be a graded superalgebra and $A'=\bigwedge_AV'$ 
the Grassmann algebra in the variable $\psi$ over $A$.
Also fix two covariances $C$ and $D$ on $V$.

\definition{\STM\defimprnorm}{ 
Let $\|\,\cdot\|$ and $\|\,\cdot\|_{\rm impr}$ be two 
families of symmetric seminorms on the spaces $A_m\otimes V^{\otimes n}$ such that 
$\|\,\cdot\|_{\rm impr} \le \|\,\cdot\|$ and 
$\| f\|_{\rm impr}=0$ if $f\in A_m\otimes V^{\otimes n}$ with $m\ge 1$. 
We say that
$(C,D)$ have improved integration constants $\cb\in\fN_d,\ \ib,\imp\in\bbbr_+$ 
for the families $\|\,\cdot\|$ and $\|\,\cdot\|_{\rm impr}$ of seminorms 
if  $\cb$ is a contraction bound for the covariance $C$  for both seminorms $\|\,\cdot\|$ and $\|\,\cdot\|_{\rm impr}$,
$\ib$ is an integral bound for $C$ and $D$ for
both seminorms and the following triple contraction estimate holds:

{\parindent=.25in
\item{}
Let $n,n'\ge 3$; $1\le i_1,i_2,i_3 \le n$ and $1\le j_1,j_2,j_3 \le n'$ with
$i_1,i_2,i_3$ all different and $j_1,j_2,j_3$ all different. 
Also let the 
covariances $C_1,C_2,C_3$ each be either $C$ or $D$ with 
at least one of these covariances equal to $C$. Then for 
$f\in A_0\otimes V^{\otimes n},\ f' \in A_0\otimes V^{\otimes n'}$
$$
\big\| \Cont{i_1}{j_1}{C_1}\,\Cont{i_2}{j_2}{C_2}\,\Cont{i_3}{j_3}{C_3}\ 
(f \otimes f') \big\|_{\rm impr}
\le \imp\,\ib^4\, \cb \,\|f\|\,\|f'\|
$$
Observe that
$\Cont{i_1}{j_1}{C_1}\,\Cont{i_2}{j_2}{C_2}\,\Cont{i_3}{j_3}{C_3}\ 
(f \otimes f') \in A_0\otimes V^{\otimes (n+n'-6)} $.

}
}

\lemma{\STM\lemGrassimprnorm}{
Assume that $(C,D)$ have improved integration constants $\cb,\ib,\imp$ for the 
families $\|\,\cdot\|$ and $\|\,\cdot\|_{\rm impr}$ of seminorms.
Let $n_1,\cdots,n_r,n_{r+1},\cdots,n_{r+s} \ge 0$, let 
$$\eqalign{
f_1(\xi^{(1)},\cdots,\xi^{(r)}) & \in A_0[n_1,\cdots,n_r] \cr
f_2(\xi^{(r+1)},\cdots,\xi^{(r+s)}) & \in A_0[n_{r+1},\cdots,n_{r+s}] \cr
}$$
and let
$i_1,i_2,i_3 \in \{1,\cdots,r\}$ and $j_1,j_2,j_3 \in \{r+1,\cdots,r+s\}$.
Also let the covariances $C_1,C_2,C_3$ each be either $C$ or $D$ with 
at least one of these covariances is equal to $C$. 
Then 
$$
\Big\|\,\cont{\xi^{(i_1)}}{\xi^{(j_1)}}{C_1}\,
\cont{\xi^{(i_2)}}{\xi^{(j_2)}}{C_2}\, \cont{\xi^{(i_3)}}{\xi^{(j_3)}}{C_3}\, 
(f_1\,f_2)\ 
\Big\|_{\rm impr}
\le \imp\,n_{j_1}n_{j_2}n_{j_3}\,\ib^4\,\cb \,\|f_1\|\,\|f_2\|
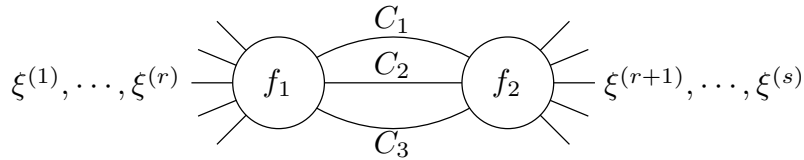
$$

\centerline{\figput{figV1}}
}

\prf The proof is analogous to that of Lemma \lemGrasscontractnorm.i.
\endproof

We define, for a Grassmann function $f$ the improved norm $N_{\rm impr}(f)$ 
as in Definition \deffunctnorm. That is,
$$
N_{\rm impr}(f\cl \al)= 
\sfrac{\cb}{\ib^2}\,\sum_{n_1,\cdots n_r\ge 0}\,
\al^{|n|}\,\ib^{|n|} \,\|f_{0; n_1,\cdots n_r}\|_{\rm impr} 
$$

An abstract example of such norms is described at the end of this Section, 
and this abstract example is made concrete in \S X. 

\definition{\STM\deffunctdegree}{ Let $f(\xi^{(1)},\cdots,\xi^{(r)})$ be a 
Grassmann function and
$I\subset \{1,\cdots,r\}$. We say that $f$ has degree $d$ in the variables
$\xi^{(i)},\ i\in I$ if
$$
f \in \bigoplus_{m;n_1,\cdots,n_r \atop \lower2pt\hbox{${\tst\Si}_{\ssst i\in I}$} n_i=d} A_m[n_1,\cdots,n_r]
$$
where $A_m[n_1,\cdots,n_r]$ was defined in Definition \defAmnonetonr.
We say that $f$ has degree at least $d$ in the variables $\xi^{(i)},\ i\in I$ 
if $f=\smsum_{d'\ge d} f_{d'}$ where each $f_{d'}$  has degree $d'$ in the variables $\xi^{(i)},\ i\in I$. Similarly we say that $f$ has degree at most $d$ in the variables $\xi^{(i)},\ i\in I$ if $f=\smsum_{d'\le d} f_{d'}$ where each $f_{d'}$  has degree $d'$ in the variables $\xi^{(i)},\ i\in I$. 
}

\proposition{\STM\propimprnormest}{
Let $(C,D)$ have improved integration constants $\cb,\ib,\imp$.
Let $r\ge t\ge s\ge 1$, and let $f_1(\xi^{(1)},\cdots,\xi^{(s)},\xi^{(s+1)},\cdots,\xi^{(t)},\xi^{(t+1)}
,\cdots,\xi^{(r)})$ and $f_2(\xi^{(1)},\cdots,\xi^{(r)})$ be Grassmann functions.
Set
$$\eqalign{
g(\xi^{(t+1)},&\cdots,\xi^{(r)}) \cr
&= \int \int 
\lW\lw f_1(\xi^{(1)},\cdots,\xi^{(s)},\cdots,\xi^{(t)},\cdots,\xi^{(r)})
  \rw_{\xi^{(1)},\cdots,\xi^{(s)},C}\, \rW_{\xi^{(s+1)},\cdots,\xi^{(t)},D} \cr
& \hskip 1.2cm \lW\lw f_2(\xi^{(1)},\cdots,\xi^{(r)})
  \rw_{\xi^{(1)},\cdots,\xi^{(s)},C}\, \rW_{\xi^{(s+1)},\cdots,\xi^{(t)},D}
  \smprod_{i=1}^s d\mu_C(\xi^{(i)}) \smprod_{j=s+1}^t d\mu_D(\xi^{(j)}) \cr
}$$
If $f_1$ has degree at least one in the variables $\xi^{(1)},\cdots,\xi^{(s)}$
and degree at least three in the variables $\xi^{(1)},\cdots,\xi^{(t)}$ then
$$
N_{\rm impr}(g\cl \al) \ 
\le  27\sfrac{\imp}{\al^6}  N(f_1\cl \al)\,N(f_2\cl \al)
$$
for $\al \ge 2$.
}

\prf Set 
$\tilde f_i = \lW\lw  f_i\rw_{\xi^{(1)},\cdots,\xi^{(s)},C}\, \rW_{\xi^{(s+1)},\cdots,\xi^{(t)},D}$. 
We first prove the statement in the case that $f_1$ and $f_2$ are both homogeneous, that is
$$\eqalign{
f_1 & \in A_0[n_1,\cdots,n_s,\cdots,n_t,\cdots,n_r] \cr
f_2 & \in A_0[n_1',\cdots,n_r'] \cr
}$$
Then $g=0$ unless $n_i=n_i'$ for $1\le i \le t$, and 
$g \in A_0[n_{t+1}+n_{t+1}',\cdots,n_r+n_r']$. By hypothesis
$n_1+\cdots+n_s \ge 1$ and $n_1+\cdots+n_t \ge 3$. Therefore it is possible to choose $i_1 \in \{1,\cdots,s\}$ with $n_{i_1} \ge 1$, and to choose
$i_2,i_3 \in \{1,\cdots,t\}$ such that
$$\eqalign{
n_{i_2} & \ge \cases{ 1& if $i_2\ne i_1$ \cr
                      2& if $i_2 =  i_1$ \cr} \cr
n_{i_3} & \ge \cases{ 1& if $i_3\ne i_1,i_2$ \cr
                      2& if $i_3 \in  \{i_1,i_2\}$  but $i_1\ne i_2$ \cr
                      3& if $i_3 = i_2 =i_1$ \cr } \cr
}$$
Set 
$$
C'_\nu = \cases{ C & if $ 1 \le i_\nu \le s$ \cr
                D & if $ s+1 \le i_\nu \le t$ \cr}
$$
Clearly, $C'_1=C$. Also set 
${\rm Con}_\nu
=\cont{\xi^{(i_\nu)}}{\ze^{(i_\nu)}}{C'_\nu}$ and
$$\eqalign{
g'(\xi^{(1)},\cdots,\xi^{(r)};\ze^{(1)},\cdots,\ze^{(r)}) 
&= {\rm Con}_1\,{\rm Con}_2\,{\rm Con}_3\,
     f_1(\xi^{(1)},\cdots,\xi^{(r)})\, 
     f_2(\ze^{(1)},\cdots,\ze^{(r)})\cr
g''(\xi^{(1)},\cdots,\xi^{(r)};\ze^{(1)},\cdots,\ze^{(r)}) 
&= {\rm Con}_1\,{\rm Con}_2\,{\rm Con}_3\,
    \tilde f_1(\xi^{(1)},\cdots,\xi^{(r)})\, 
    \tilde f_2(\ze^{(1)},\cdots,\ze^{(r)})\cr
}$$
Observe that 
$$\eqalign{
g'\in A'_0[&n_1 -(\de_{1\,i_1}+\de_{1\,i_2}+\de_{1\,i_3}),
 \cdots, n_r -(\de_{r\,i_1}+\de_{r\,i_2}+\de_{r\,i_3}),\cr
&n'_1 -(\de_{1\,i_1}+\de_{1\,i_2}+\de_{1\,i_3}),
 \cdots, n'_r -(\de_{r\,i_1}+\de_{r\,i_2}+\de_{r\,i_3})]\cr
}$$
and
$$
g''(\xi^{(1)},\cdots,\xi^{(r)};\ze^{(1)},\cdots,\ze^{(r)})
=\lW\lw  g'(\xi^{(1)},\cdots,\xi^{(r)};\ze^{(1)},\cdots,\ze^{(r)})
\rw_{\xi^{(1)},\cdots,\xi^{(s)},C\atop\ze^{(1)},\cdots,\ze^{(s)},C}
\, \rW_{\xi^{(s+1)},\cdots,\xi^{(t)},D\atop \ze^{(s+1)},\cdots,\ze^{(t)},D}
$$
by Remark \remcontract.
By Lemma \lemcontract 
$$
g = \int\!\! \int g''(\xi^{(1)},\cdots,\xi^{(r)};\xi^{(1)},\cdots,\xi^{(r)}) \,
  \smprod_{i=1}^s d\mu_C(\xi^{(i)}) \smprod_{j=s+1}^t d\mu_D(\xi^{(j)})
$$
By Lemma \lemGrasscontractnorm\ and Lemma \lemGrassimprnorm
$$
\|g\|_{\rm impr} 
\ \le\ \ib^{2(n_1+\cdots+n_t-3)}\,\|g'\|_{\rm impr} 
\ \le\ \imp\,\sfrac{n_{i_1}\,n_{i_2}\,n_{i_3}\,\cb}{\ib^2}\, \ib^{2(n_1+\cdots+n_t)}\,\|f_1\|\,\|f_2\|
\EQN\eqnPropimprnormest$$
Therefore
$$\eqalign{
N_{\rm impr}(g) &= \sfrac{\cb}{\ib^2}\,
\al^{n_{t+1}+\cdots+n_r+n'_{t+1}+\cdots+n'_r}\,
\ib^{n_{t+1}+\cdots+n_r+n'_{t+1}+\cdots+n'_r} 
\,\|g\|_{\rm impr} \cr
&\le \sfrac{ n_{i_1}\,n_{i_2}\,n_{i_3}}{\al^{2(n_1+\cdots+n_t)}}\,
\sfrac{\imp\cb^2}{\ib^4}\,
\big(\al\,\ib\big)^{n_1+\cdots+n_r+n'_1+\cdots+n'_r} 
\,\|f_1\|\,\|f_2\| \cr
&\le \imp \sfrac{ n_{i_1}\,n_{i_2}\,n_{i_3}}{\al^{2(n_1+\cdots+n_t)}}\,
  N(f_1)\,N(f_2) \cr
&\le  27\sfrac{\imp}{\al^6}\,N(f_1)\,N(f_2) \cr
}$$
The general case now follows by decomposing $f_1$ and $f_2$ into homogeneous pieces.
\endproof

\sect{ Ladders}\PG\pgRVIb

Theorem \:\theoremVa, below, which is the main result of this paper, shows 
that under appropriate assumptions on an effective interaction $W(\psi)$, 
the two point and non--ladder four point parts of the effective interaction 
$\lw W'(\psi)\rw_{\psi,D}\ =\ \Om_C\big(\lw W\rw_{\psi,C+D}\big)$, 
constructed using the Grassmann Gaussian integral with covariance $C$, 
obeys estimates that are better by a factor $\imp$ 
than those one would expect from Theorem \theoremIVa. To formulate this
precisely, we first give the Definition of ladders.
$$
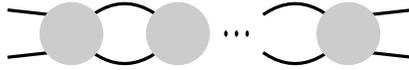
\figput{ladder}
$$
In a ladder, neighbouring four legged vertices are connected by two 
covariances. Since ladders result from integrating with covariance $C$,
at least one of the connecting covariances is equal to $C$. The other 
connecting covariance may be $C$ or $D$.

In the rest of the paper, we will systematically use $\xi,\xi',\xi'',\cdots$ 
for fields associated to the covariance $C$. We will use
$\ze,\ze',\ze'',\cdots$ for fields associated to the covariance $D$ and $\psi$ 
for the external fields.

\definition{\STM\defladders}{
\Item{(i)} A rung is a Grassmann function
$$
\rho(\ze,\xi;\ze',\xi') 
\in A[0,2,0,2]\oplus A[1,1,0,2]\oplus A[0,2,1,1]\oplus A[1,1,1,1]
$$
We think of $\ze,\xi$ as the $D$ resp. $C$ fields on the left side of the rung
and of $\ze',\xi'$ as the $D$ resp. $C$ fields on the right side of the rung.
$$
\figput{figV2apr}
$$
An end is a Grassmann function
$$
E(\psi;\ze,\xi) \in  A[2,0,2]\oplus A[2,1,1]
$$
We think of $\psi$ as the external fields at the end of the ladder and of
$\ze,\xi$ as the $D$ resp. $C$ fields going into the ladder.
$$
\figput{figV2bpr}
$$
\Item{(ii)}
If $E$ is an end and $\rho$ is a rung, we define the end $E\circ \rho$ by
$$
E\circ \rho(\psi;\ze',\xi')
= \int  \hskip -3pt \int
 \lW E(\psi;\ze,\xi)\rW_{\xi,C\atop\ze,D}\ 
 \lW \rho(\ze,\xi;\ze',\xi')\rW_{\xi,C\atop \ze,D}\,
   d\mu_C(\xi)d\mu_D(\ze)
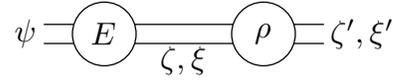
$$
\centerline{\figput{figV2pr}}

\noindent
If $E_1,E_2$ are ends, we define the ladder  $E_1\circ E_2$ 
by
$$
E_1\circ E_2(\psi)
= \int  \hskip -3pt \int
 \lW E_1(\psi;\ze',\xi')\rW_{\xi',C\atop \ze',D}\ 
 \lW E_2(\psi;\ze',\xi')\rW_{\xi',C\atop \ze',D}\,
   d\mu_C(\xi')d\mu_D(\ze')
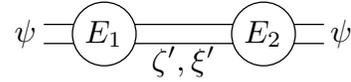
$$
\centerline{\figput{figV3}}

\noindent
\Item{(iii)}
Let $F(\xi) \in A[4]$. Write
$$\eqalign{
F(\xi^{(1)}+\xi^{(2)}+\xi^{(3)})
&= \sum_{n_1+n_2+n_3=4} F_{n_1,n_2,n_3}(\xi^{(1)},\xi^{(2)},\xi^{(3)})\cr
F(\xi^{(1)}+\xi^{(2)}+\xi^{(3)}+\xi^{(4)})
&= \sum_{n_1+n_2+n_3+n_4=4} F_{n_1,n_2,n_3,n_4}(\xi^{(1)},\xi^{(2)},\xi^{(3)},\xi^{(4)})\cr
}$$
with $F_{n_1,n_2,n_3} \in A[n_1,n_2,n_3]$ and 
$F_{n_1,n_2,n_3,n_4} \in A[n_1,n_2,n_3,n_4]$.
The rung associated to $F$ is
$$
\rho(F)(\ze,\xi;\ze',\xi') =
F_{0,2,0,2}+ F_{1,1,0,2}+ F_{0,2,1,1}+ F_{1,1,1,1}
$$
The end associated to $F$ is
$$
E(F)(\psi;\ze',\xi') =
F_{2,0,2}+ F_{2,1,1}
$$
The ladder of length $r\ge 1$ with vertex $F$ is defined as
$$
L_r(F)(\psi) = E(F) \circ \rho(F) \circ \rho(F) \circ \cdots \rho(F) \circ E(F)
$$
with $(r-1)$ copies of $\rho(F)$.

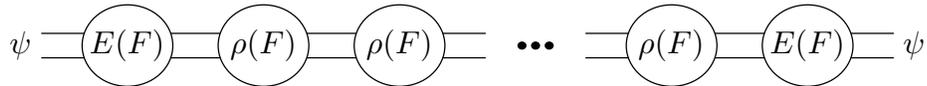
\centerline{\figput{figV4}}

\noindent
In Appendix C, we describe ladders in terms of kernels. 
}

The main result of this paper is

\theorem{\STM\theoremVa}{
Let $W(\psi)$ be an even Grassmann function with coefficients in $A$. 
Assume that  $\,N\big(W\cl 64\al\big)_\0 < \sfrac{1}{8} \al$, and that 
$\al\ge 8$. Set
$$\eqalign{
:W'(\psi):_{\psi,D}\ &=\ \Om_C\big(:W:_{\psi,C+D}\big)\cr
}$$
If $(C,D)$ have improved integration constants $\cb,\ib,\imp$,
then
\Item{(i)}
$$\eqalign{
N\big( W'-W\cl \al\big) & \le \sfrac{1}{2\al^2}\,
     \sfrac{N( W\cl 32\al)^2}{1-{1\over\al^2}N( W\cl 32\al)}  \cr
N_{\rm impr}\big( W'-W\cl \al\big) &\le \sfrac{1}{2\al^2}\,
     \sfrac{N( W\cl 32\al)^2}{1-{1\over\al^2}N( W\cl 32\al)} \cr
}$$
\Item{(ii)} Write $W(\psi) =\smsum_{m;n} W_{m;n}(\psi)$, 
$W'(\psi) =\smsum_{m;n} W'_{m;n}(\psi)$ with $W_{m;n},W'_{m;n}\in A'_m[n]$.
If $W_{0;2} = 0$, then
$$\eqalign{
N_{\rm impr}\big( W'_{0,2}\cl \al\big) 
&\le  \sfrac{2^{10}\,\imp}{\al^6}\,\sfrac{N(W\cl 64\al)^2}
{1-{8\over\al}N(W\cl 64\al)} \cr
N_{\rm impr}\big( W'_{0,4}-W_{0,4}-\sfrac{1}{2} \sum_{r=1}^\infty L_r(W_{0,4})\cl \al\big) 
&\le  \sfrac{2^{10}\,\imp}{\al^6}\,\sfrac{N(W\cl 64\al)^2}
{1-{8\over\al}N(W\cl 64\al)}\cr
}$$
}

\remark{\STM\remtheotheo}{ 
\Item i)
Part (i) of the Theorem follows directly from Theorem \theoremIVa.
For the proof of part (ii) one can replace the algebra $A$ by $A_0$, since
$W'_{0,2},\, W'_{0,2}$ and $L_r(W_{0,4})$ depend only on 
$\smsum_{n=0}^\infty W_{0;n}$.
\Item ii)
The hypothesis that $W_{0;2}=0$ in part (ii) of Theorem \theoremVa\ prevents strings 
of two--legged vertices from appearing in diagrammatic expansions. The 
expansion used in the proof of part (ii) cannot detect certain overlapping loops containing such strings. In practice a nonzero $W_{0;2}$ can be absorbed in the propagator.

}

The proof of part (ii) of Theorem \theoremVa\ is based on an analysis of 

\sect{ Overlapping loops for the Schwinger functional}\PG\pgRVIc

We first generalise the concept of a ladder.

\noindent
If $U(\psi;\xi)$ is a Grassmann function we write
$$
U(\psi+\ze+\ze';\xi+\xi') = \sum_{p_1,p_2\atop n_0,n_1,n_2} 
U_{n_0;p_1,p_2;n_1,n_2}(\psi;\ze,\ze';\xi,\xi')
$$
with $U_{n_0;p_1,p_2;n_1,n_2}\in A[n_0;p_1,p_2,n_1,n_2]$.

\definition{\STM\deftails}{ Let $U$ be as above.
\Item{(i)} The rung associated to $U$ is
$$
{\rm Rung}(U)(\ze,\xi;\ze',\xi') =
U_{0;0,2;0,2} + U_{0;1,1;0,2} + U_{0;0,2;1,1} + U_{0;1,1;1,1}
$$
\Item{(ii)} The tail $T_\ell(U)$ associated to $U$ is recursively defined as
$$\eqalign{
T_1(U)(\psi;\ze',\xi') &= U_{2;0,0;0,2} + U_{2;0,1;0,1} \cr
T_{\ell+1}(U)(\psi;\ze',\xi') &= T_\ell(U) \circ {\rm Rung}(U) \cr
}$$
Observe that $T_\ell(U)(\psi;\ze',\xi')$ lies in $A[2,0,2]\oplus A[2,1,1]$.
}

Later we need

\remark{\STM\remformladder}{
Let $E_1,E_2$ be ends whose coefficients are even elements of $A$ and let $g(\psi;\xi)$ a Grassmann function. Set
$$\eqalign{
h(\psi) &= \int \hskip -6pt \int 
\lW E_1(\psi;\ze',\xi')\,E_2(\psi;\ze',\xi')\rW_{\ze',D\atop \xi',C}\ \
\lW g(\psi+\ze';\xi')\rW_{\ze',D\atop \xi',C}\,
d\mu_D(\ze')\,d\mu_C(\xi') \cr
h(\psi) &= \smsum_{n=4}^\infty h_n(\psi) \qquad {\rm with}\ h_n\in A[n]
}$$
Then
$$
h_4 = E_1\circ {\rm Rung}(g) \circ E_2
$$
}
\prf
By Lemma \:\lemBV
$$\eqalign{
h(\psi) = \int \hskip -6pt \int 
\lW E_1(\psi;\ze,\xi)\rW_{\ze,D\atop \xi,C} \ \ 
\lW g(\psi+\ze+&\ze';\xi+\xi')\rW_{\ze,\ze',D\atop \xi,\xi',C}\cr
&\lW E_2(\psi;\ze',\xi')\rW_{\ze',D\atop \xi',C}\,
d\mu_D(\ze,\ze')\,d\mu_C(\xi,\xi') \cr
}$$
As $E_1$ is of degree  at most one in $\ze$ and $E_2$ is of degree at most
one in $\ze'$,
$$\eqalign{
h_4(\psi) &= \int \hskip -6pt \int \hskip -2pt
\lw E_1(\psi;\ze,\xi)\rw_{\xi,C} \  
\lw {\rm Rung}(g)(\ze,\xi;\ze',\xi')\rw_{\xi,\xi',C}\  
\lw E_2(\psi;\ze',\xi')\rw_{\xi',C}\,
d\mu_D(\ze,\ze')\,d\mu_C(\xi,\xi') \cr
&= \big(E_1\circ {\rm Rung}(g) \circ E_2\big)(\psi)
}$$
\endproof

The main estimate on the Schwinger functional is:

\theorem{\STM\theoremVb}{
Let $A$ be a superalgebra, with all elements having degree zero (that 
is $A=A_0$), $\|\,\cdot\|$ and $\|\,\cdot\|_{\rm impr}$ be two families 
of symmetric seminorms on the spaces $A_m\otimes V^{\otimes n}$ and 
let $(C,D)$ have improved integration constants $\cb,\ib,\imp$.
Let $\hat U(\psi,\xi), \hat f(\psi,\xi)$ be Grassmann functions with coefficients in $A$ of degree at least four and $\hat U$ even. Set 
$$\eqalign{
U(\psi,\xi) &= \ \lW \hat U(\psi,\xi)\rW_{\psi,D\atop\xi,C}  \cr
f(\psi,\xi) &= \ \lW \hat f(\psi,\xi)\rW_{\psi,D\atop\xi,C}  \cr
}$$
Assume that $\al \ge 8$ and $N({\hat U}\cl 32\al)_\0 < \sfrac{1}{8}\al$.
By Proposition \propnormestcalS
$$
\lw f'(\psi)\rw_{\psi,D}= \cS_{U,C}(f)  
$$
exists. Write
$$
\hat f(\psi,\xi) = \smsum_{n_0,n_1}\hat f_{n_0,n_1}(\psi,\xi) \qquad,\qquad
 f'(\psi) = \smsum_{n} f'_n(\psi)
$$
with $\hat f_{n_0,n_1} \in A[n_0,n_1],\ f'_n\in A[n]$.
Then
$$
N_{\rm impr}(f'_2\cl \al) \le
\sfrac{2^{10}\,\imp}{\al^6}\, 
\sfrac{N(\hat U\cl 32\al) }{1-{8\over\al}N(\hat U\cl 32\al)}
N\big(\hat f\cl 32\al\big)
$$
and there exists a Grassmann function $g(\psi)$ such that
$$
f'_4 = \hat f_{4,0} +\smsum_{\ell=1}^\infty T_\ell(\hat U) \circ T_1(\hat f)\ 
   +\sfrac{1}{2}\, \smsum_{\ell,\ell'\ge 1}  
    T_\ell(\hat U)\circ {\rm Rung}(\hat f) \circ T_{\ell'}(\hat U)\ 
     + \ g 
$$
and
$$
N_{\rm impr}(g\cl \al) \le
\sfrac{2^{10}\,\imp}{\al^6}\, 
\sfrac{N(\hat U\cl 32\al) }{1-{8\over\al}N(\hat U\cl 32\al)}
N\big(\hat f\cl 32\al\big)
$$
}

In the case $D=0$ this Theorem is proven in Section VI; in the general case it is proven in Section VIII.

\proof{ that Theorem \theoremVb\ implies Theorem \theoremVa}
By part (i) of Remark \remtheotheo\ we may assume that $A=A_0$. We write $W_n$ for $W_{0;n}$ and $W'_n$ for $W'_{0;n}$. Set
$$\eqalign{
\hat U(\psi,\xi) &= W(\psi+\xi) \in \bigwedge\nolimits_{A'}V \cr
U(\psi,\xi) &= \ \lW \hat U(\psi,\xi)\rW_{\psi,D\atop\xi,C}  \cr
:U_t'(\psi):_{\psi,D}\ &=\ \cS_{t U, C}(U) 
}$$
By Remark \remfunctnorm, 
$N(\hat U\cl \al) \le N(W\cl 2\al)$.
As in the proof of Theorem \theorII\
$$\eqalign{
\lw W'(\psi)\rw_{\psi,D} - \lw W(\psi)\rw_{\psi,D} = 
&\int_0^1 \big(\lw U_t'(\psi)\rw_{\psi,D} -\lw \hat U(\psi,0)\rw_{\psi,D} \big)\, dt
\qquad{\rm mod}\, A_0
}$$
so
$$
W'(\psi) -  W(\psi) = 
\int_0^1 \big( U_t'(\psi) - \hat U(\psi,0) \big)\, dt
\qquad{\rm mod}\, A_0  
$$
In particular, for $n=2,4$
$$
W'_n - W_n = 
\int_0^1 \big(U_{t,n}' -\hat U_{n,0} \big)\, dt
$$
Therefore, by Theorem \theoremVb
$$\eqalign{
N_{\rm impr}(W'_2 )
&\le \max_{0\le t \le 1}\,N_{\rm impr}\big(U_{t,2}'\big)\cr
& \le \sfrac{2^{10}\,\imp}{\al^6}\,
\sfrac{N(\hat U\cl 32\al)}{1-{8\over\al}N(\hat U\cl 32\al)}\,
N(\hat U\cl 32\al) \cr
& \le \sfrac{2^{10}\,\imp}{\al^6}\,\sfrac{N(W\cl 64\al)^2}
{1-{8\over\al}N(W\cl 64\al)} \cr
}$$

Observe that 
$$\eqalign{
{\rm Rung}(\hat U) &= \rho(W_4) \cr
T_\ell(\hat U) &= E(W_4) \circ \rho(W_4) \circ\cdots\circ \rho(W_4)
}$$
with $\ell-1$ copies of $\rho(W_4)$. Therefore
$$\eqalign{
T_\ell(t \hat U) \circ T_1(\hat U) &= t^\ell L_\ell(W_4)  \cr
T_\ell(t \hat U) \circ {\rm Rung}(\hat U) \circ T_{\ell'}(t \hat U) 
&= t^{\ell+\ell'} L_{\ell+\ell'}(W_4)  \cr
}$$
Hence, by Theorem \theoremVb
$$
W_4'=W_4 +\smsum_{\ell=1}^\infty \int_0^1 t^\ell L_\ell(W_4)\, dt\ 
   +\sfrac{1}{2} \,\smsum_{\ell,\ell'\ge 1} \int_0^1
       t^{\ell+\ell'} L_{\ell+\ell'}(W_4) \,dt\  
     + g
$$
with
$$
N_{\rm impr}(g)
\ \le\ \sfrac{2^{10}\,\imp}{\al^6}\,
\sfrac{N(\hat U\cl 32\al)}{1-{8\over\al}N(\hat U\cl 32\al)}\,
N(\hat U\cl 32\al)
\le \sfrac{2^{10}\,\imp}{\al^6}\,\sfrac{N(W\cl 64\al)^2}
{1-{8\over\al}N(W\cl 64\al)}
$$
Now
$$\eqalign{
\smsum_{\ell=1}^\infty \int_0^1 t^\ell L_\ell(W_4)\, dt\ 
   +\sfrac{1}{2} \smsum_{\ell,\ell'\ge 1} \int_0^1
       t^{\ell+\ell'} L_{\ell+\ell'}(W_4) \,dt\  
& =\sfrac{1}{2} \smsum_{r=1}^\infty \int_0^1 (r+1)\,t^r L_r(W_4)\, dt\  \cr
& = \sfrac{1}{2} \smsum_{r=1}^\infty L_r(W_4) \cr
}$$
\endproof

\goodbreak
\sect{ Configurations of Norms with Improved Power Counting}\PG\pgRVId

\definition{\STM\deftildeimprconf}{
Let $q$ be an even natural number.
For $p=1,2,3,\cdots,q$, let $\|\,\cdot\,\|_p$ be a system of symmetric seminorms on the spaces $A_m\otimes V^{\otimes n}$.
We say that $(C,D)$ have integration constants $\cb,\ib$ for the configuration 
$\|\,\cdot\,\|_1$, $\|\,\cdot\,\|_2$, $\cdots$, $\|\,\cdot\,\|_q$ of 
seminorms if the  following estimates hold:
\medskip
{\parindent=.25in
\item{}
Let $m,m'\ge 0$ and $1\le i\le n,\ 1\le j\le n'$. Also let 
$f\in A_m\otimes V^{\otimes n}$, $f'\in A_{m'}\otimes V^{\otimes n'}$
Then for all natural numbers $p\le q$ the simple contraction estimate
$$
\big\|\Cont{i}{n+j}{C} (f\otimes f')\big\|_p
\le \cb\,\smsum_{p_1+p_2=p+1\atop{\rm at\ least\atop one\ odd}}
\|f\|_{p_1}\,\|f'\|_{p_2}
$$
holds. 

\item{}
Furthermore, if $C_2,C_3 \in\{C,D\}$, $m=m'=0$,  
$1\le i_1,i_2,i_3 \le n$ with $i_1,i_2,i_3$ all different and
 $1\le j_1,j_2,j_3\le n'$  with  $j_1,j_2,j_3$ all different,
 the improved contraction estimate
$$
\big\|\Cont{i_1}{n+j_1}{C}\,\Cont{i_2}{n+j_2}{C_2}\,\Cont{i_3}{n+j_3}{C_3}\,
 (f\otimes f')\big\|_p
\le \ib^4\, \cb \,\smsum_{p_1+p_2=p+3\atop{\rm at\ least\atop one\ odd}}
\|f\|_{p_1}\,\|f'\|_{p_2}
$$
holds for $p\le q-2$.

\item{}
For every $f\in A_m\otimes V^{\otimes n}$ and every $n'\le n$ the
modified integral bound
$$
\Big\| \int Ant_{n'}(f) d\mu_C \Big\|_p,\ 
\Big\| \int Ant_{n'}(f) d\mu_D \Big\|_p
      \le \half\ (\ib/2)^{n'} \Big[\|f\|_p+\|f\|_{p-{(-1)}^p}\Big]
\EQN\eqnRmodintbnd$$
holds. The partial antisymmetrization $Ant_{n'}$ was defined in  
Definition \defcontractintbound.ii. 

}
}

\lemma{\STM\lemtildeimprconfig}{
Let $q$ be an even natural number.
Assume that $(C,D)$ have integration constants $\cb,\ib$ 
for the configuration $\|\,\cdot\,\|_1$, $\|\,\cdot\,\|_2$, $\cdots$,
$\|\,\cdot\,\|_q$ of seminorms and let $\imp>0$. For 
$f\in A_m\otimes V^{\otimes n}$, set
$$\eqalign{
\|f\| &= \smsum_{p=1}^q \imp^{-[(p-1)/2]} \|f\|_p
= \|f\|_1+\|f\|_2+\sfrac{1}{J}\|f\|_3+\sfrac{1}{J}\|f\|_4+\cdots
+\sfrac{1}{J^{(q-2)/2}}\|f\|_q\cr
\|f\|_{\rm impr} &= \cases{
\smsum\limits_{p=1}^{q-2} \imp^{-[(p-1)/2]} \|f\|_p& if $m=0$\cr
\noalign{\vskip.1in}
0  & if $m\ne 0$\cr}
 \cr
}$$ 
Here $[(p-1)/2]$ is the integer part of $\sfrac{p-2}{2}$.
Then $(C,D)$ have improved integration constants $\cb,\ib,\imp$ for the 
families $\|\,\cdot\|$ and $\|\,\cdot\|_{\rm impr}$ of seminorms.
}
\prf
Clearly $\|\cdot\|_{\rm impr} \le \|\cdot\|$. 
 To verify that $\cb$ is a contraction bound for $C$
let $f\in A_m\otimes V^{\otimes n}$, $f'\in A_{m'}\otimes V^{\otimes n'}$
and $1\le i \le n,\ 1\le j \le n'$. 
Observe that if $p_1+p_2=p+1$ with at least one of $p_1$ and $p_2$ odd, then
$$\eqalign{
\frac{1}{J^{\raise1pt\hbox{$\sst[(p_1-1)/2]$}}}
\frac{1}{J^{\raise1pt\hbox{$\sst[(p_2-1)/2]$}}}
=\frac{1}{J^{\raise1pt\hbox{$\sst[(p_1+p_2-2)/2]$}}}
=\frac{1}{J^{\raise1pt\hbox{$\sst[(p-1)/2]$}}}
}
$$
Consequently,
$$\eqalign{
\big\|\Cont{i}{n+j}{C} (f\otimes f')\big\|
& = \smsum_{p=1}^q \imp^{-[(p-1)/2]} \|\Cont{i}{n+j}{C} f\otimes f'\|_p \cr  
& \le \sum_{p=1}^q\ 
\cb\smsum_{p_1+p_2=p+1\atop{\rm at\ least\atop one\ odd}} \imp^{-[(p-1)/2]} 
\|f\|_{p_1}\,\|f'\|_{p_2} \cr
& \le \sum_{p=1}^q\ 
\cb\smsum_{p_1+p_2=p+1} \imp^{-[(p_1-1)/2]} 
\|f\|_{p_1}\,\imp^{-[(p_2-1)/2]}\|f'\|_{p_2} \cr
& \le \cb \,\|f\|\,\|f'\|\cr  
}$$
Replacing $q$ by $q-2$ gives the corresponding bound for 
$\|\ \cdot\ \|_{\rm impr}$.
To verify the triple contraction estimate of Definition \defimprnorm,
let $C_2,C_3 \in\{C,D\}$, $m=m'=0$, $1\le i_1,i_2,i_3 \le n$ with $i_1,i_2,i_3$ all different and
 $1\le j_1,j_2,j_3\le n'$  with  $j_1,j_2,j_3$ all different. Then
$$\eqalign{
\big\|\Cont{i_1}{n+j_1}{C}\,\Cont{i_2}{n+j_2}{C_2}\,\Cont{i_3}{n+j_3}{C_3}\,
 (f\otimes f')\big\|_{\rm impr}
& = \smsum_{p=1}^{q-2} \imp^{-[{p-1\over2}]} 
\big\|\Cont{i_1}{n+j_1}{C}\,\Cont{i_2}{n+j_2}{C_2}\,\Cont{i_3}{n+j_3}{C_3}\,
 (f\otimes f')\big\|_p \cr
&\le \imp\,\ib^4\, \cb \,\sum_{p=1}^{q-2}
\smsum_{p_1+p_2=p+3\atop{\rm at\ least\atop one\ odd}}^q  \imp^{-[(p+1)/2]}\, 
\|f\|_{p_1}\,\|f'\|_{p_2} \cr
&\le \imp\,\ib^4\, \cb \,\smsum_{p_1,p_2=1}^q  \imp^{-[(p_1-1)/2]} 
\|f\|_{p_1}\,\imp^{-[(p_2-1)/2]}\|f'\|_{p_2} \cr
&= \imp\,\ib^4\,\cb \,\|f\|\,\|f'\| \cr
}$$
We verify that $\ib$ is an integral bound for $C$ for the norm $\|\cdot\|$.
The other cases are similar.
Let $f\in A_m\otimes V^{\otimes n}$ and $n'\le n$.  Then
$$\eqalign{
\Big\| \int Ant_{n'}(f) d\mu_C \Big\|
&=\sum_{p=1}^q \imp^{-[(p-1)/2]}\Big\| \int Ant_{n'}(f) d\mu_C \Big\|_p   \cr
&=\half\ (\ib/2)^{n'} \sum_{p=1}^q \imp^{-[(p-1)/2]}
 \Big[\|f\|_p+\|f\|_{p-{(-1)}^p}\Big]   \cr
&\le (\ib/2)^{n'} \|f\|\cr
}\EQN\eqnRmodintbndprf$$

\endproof

In our main application, we use a special case of Definition 
\deftildeimprconf\ in which only norms $\|\,\cdot\,\|_p$, with $p$ odd,
appear.

\definition{\STM\defimprconf}{
Let $q$ be an odd natural number.
For $p=1,3,5,\cdots,q$, let $\|\,\cdot\,\|_p$ be a system of symmetric 
seminorms on the spaces $A_m\otimes V^{\otimes n}$. 
We say that $(C,D)$ have integration constants $\cb,\ib$ for the 
configuration $\|\,\cdot\,\|_1$, $\|\,\cdot\,\|_3$, $\cdots$,
$\|\,\cdot\,\|_q$ of 
seminorms if $\ib$ is an integral bound for both $C$ and $D$ and all
of the seminorms $\|\,\cdot\,\|_p$ (see Definition \defcontractintbound.ii) and
the  following contraction estimates hold:
\medskip
{\parindent=.25in
\item{}
Let $m,m'\ge 0$ and $1\le i\le n,\ 1\le j\le n'$. Also let 
$f\in A_m\otimes V^{\otimes n}$, $f'\in A_{m'}\otimes V^{\otimes n'}$
Then for all odd natural numbers $p\le q$ 
$$
\big\|\Cont{i}{n+j}{C} (f\otimes f')\big\|_p
\le \cb\,\smsum_{p_1+p_2=p+1\atop{p_1,p_2{\rm\ odd}}}\|f\|_{p_1}\,\|f'\|_{p_2}
$$

\item{}Furthermore, if $C_2,C_3 \in\{C,D\}$, $m=m'=0$,  
$1\le i_1,i_2,i_3 \le n$ with $i_1,i_2,i_3$ all different and
 $1\le j_1,j_2,j_3\le n'$  with  $j_1,j_2,j_3$ all different,
 then, for all odd $p\le q-2$,
$$
\big\|\Cont{i_1}{n+j_1}{C}\,\Cont{i_2}{n+j_2}{C_2}\,\Cont{i_3}{n+j_3}{C_3}\,
 (f\otimes f')\big\|_p
\le \ib^4\, \cb \,\smsum_{p_1+p_2=p+3\atop{p_1,p_2{\rm\ odd}}}
\|f\|_{p_1}\,\|f'\|_{p_2}
$$

}
}
\remark{\STM\remimprconfig}{
If, in the setting of Definition \defimprconf,
the norm $\|\,\cdot\,\|_p$ is defined to be zero for all even $p$, then
the conditions of Definition \deftildeimprconf\ are fulfilled, except 
that the factor of $\half$ in (\eqnRmodintbnd) does not appear in the
Definition \defcontractintbound.ii of integral bound.

}

\lemma{\STM\lemimprconfig}{
Let $q$ be an odd natural number.
Assume that $(C,D)$ have integration constants $\cb,\ib$ 
for the configuration $\|\,\cdot\,\|_1$, $\|\,\cdot\,\|_3$, $\cdots$,
$\|\,\cdot\,\|_q$ of seminorms and let $\imp>0$. For 
$f\in A_m\otimes V^{\otimes n}$, set
$$\eqalign{
\|f\| &= \smsum_{p=1\atop p{\rm odd}}^q \imp^{(1-p)/2} \|f\|_p \cr
\|f\|_{\rm impr} &= \cases{
\smsum\limits_{p=1\atop p{\rm odd}}^{q-2} \imp^{(1-p)/2} \|f\|_p& if $m=0$\cr
\noalign{\vskip.1in}
0  & if $m\ne 0$\cr}
 \cr
}$$ 
Then $(C,D)$ have improved integration constants $\cb,\ib,\imp$ for the 
families $\|\,\cdot\|$ and $\|\,\cdot\|_{\rm impr}$ of seminorms.
}
\prf 
By Remark \remimprconfig, Lemma \lemtildeimprconfig\ implies all of the conditions of Definition \defimprnorm, except that $\ib$ be an integral bound 
for $C$ and $D$ for both seminorms. However, the proof of this condition
is virtually identical to (\eqnRmodintbndprf).
\endproof
\remark{\STM\remJlimit}{
Lemma \lemimprconfig\ holds for all $\imp>0$. In applications,
$\imp$ is chosen so that
$$
\| f\|_p\le \const \imp^{(p-1)/2}\| f\|_1
\EQN\eqJlimit$$
for all $f$ of interest. If $\imp$ satisfying (\eqJlimit) can be chosen sufficiently small, Lemma \lemimprconfig\ can be used in conjunction with 
Proposition \propimprnormest\ to obtain improved bounds, as the following example illustrates.

For simplicity, we assume that $q=3$. 
Let $f\big(\xi^{(1)},\xi^{(2)}\big)\in A_0[n_1,n_2]$ with $n_2\ge 3$ and set
$$
g\big(\xi^{(1)}\big)=\int \Big(\lW f\big(\xi^{(1)},\xi^{(2)}\big)\rW_{\xi^{(2)},C}\Big)^2
\ d\mu_C\big(\xi^{(2)}\big)
$$
The standard bound, without improvement, follows from (\eqnPropestCconnect) in the proof of Proposition \propestCconnect:
$$
\|g\|_1
\ \le\ n_2\,\cb\,\ib^{2(n_2-1)}\,\|f\|_1^2
$$
On the other hand, by (\eqnPropimprnormest), in the proof of 
Proposition \propimprnormest,
$$
\|g\|_1= \|g\|_{\rm impr} 
\ \le\ n_2^3\,\imp\cb\, \ib^{2(n_2-1)}\,\|f\|^2
\ =\ n_2^3\,\imp\cb\, \ib^{2(n_2-1)}\,\big(\|f\|_1+\sfrac{1}{\imp}\|f\|_3\big)^2
$$
If $\|f\|_3\le\const\imp \|f\|_1$,
$$
\|g\|_1
\ \le\ \const n_2^3\,\imp\cb\, \ib^{2(n_2-1)}\,\|f\|_1^2
$$

}

\vfill\eject

\chap{ Finding Overlapping Loops}\PG\pgRVII

In this chapter, we give the proof of Theorem \theoremVb\ in the case $D=0$, 
using the representation 
$\ \cS_{U,C} = \int  \sfrac{1}{\bbbone - R_{U,C}} \, d\mu_C \ $  
of Theorem \theoremIII. We assume that the coefficient algebra $A$ contains 
only elements of degree zero, that is $A=A_0$. Let $\|\,\cdot\|$ and 
$\|\,\cdot\|_{\rm impr}$ be two families of symmetric seminorms on the spaces 
$A_m\otimes V^{\otimes n}$ such that
$(C,0)$ has improved integration constants $\cb,\ib,\imp$ for these families of 
seminorms.

Recall from Remark \remcalR\ that the operator $\cR_{K,C}$ is written as a sum of operators $R_C(K_1,\cdots,K_\ell)$ with even Grassmann functions
$K_1(\xi,\xi',\eta),\cdots, K_\ell(\xi,\xi',\eta)$. If one of these Grassmann functions, say $K_1$ has degree at least three in the variables $\xi',\eta$
then, for any Grassmann function $f(\xi)$, there is a pair of overlapping
loops in each Feynman diagram contributing to $R_C(K_1,\cdots,K_\ell)(:f:)$.
The way these overlapping loops can occur is indicated in the figures below.

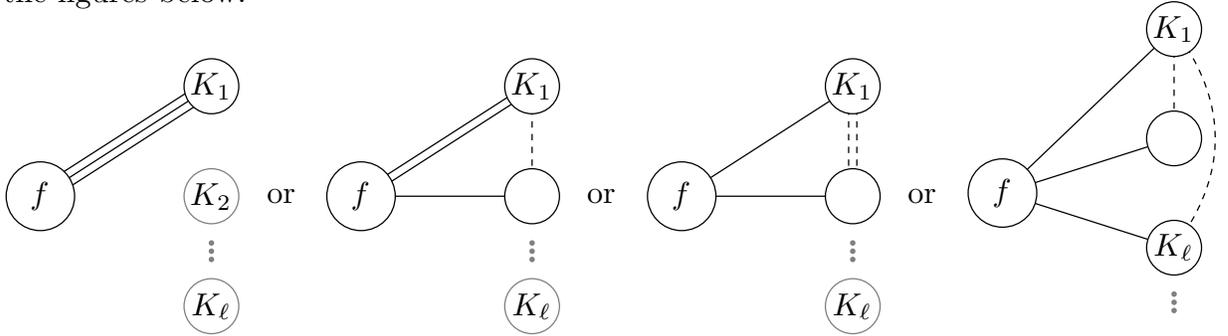
\centerline{\figplace{figVI7}{0 in}{-.47in} or\ 
            \figplace{figVI8}{0 in}{-.47in} or\ 
            \figplace{figVI9}{0 in}{-.47in} or\ 
            \figplace{figVI10}{0 in}{-.75in}}

\sect{ Overlapping loops created by the operator $\cR_{K,C}$}\PG\pgRVIIa

In this subsection, we suppress the external fields $\psi$ by working in the 
Grassmann algebra $\bigwedge_{A'}V$ with coefficients in the algebra 
$A'=\bigwedge_A V$ generated by the fields $\psi$. This algebra was defined in 
subsection III.2. Recall that $\|\,\cdot\|$ and $\|\,\cdot\|_{\rm impr}$ induce 
a family of symmetric seminorms on the spaces $A'_m\otimes V^{\otimes n}$, 
which we here denote by the same symbols.

We split up the operators
$R_C(K_1,\cdots,K_\ell)$ of (\eqRCdef) in order to exhibit possible overlapping loops. For Grassmann functions 
$K_2(\xi,\xi',\eta),\cdots,K_\ell(\xi,\xi',\eta)$ and $f(\xi)$
we define
$$
\tilde R_C(K_2,\cdots,K_\ell)(f)\ =\ 
\int \hskip-6pt\int \lW \big( \smprod_{i=2}^\ell \lw K_i(\xi,\xi'+\xi^{\prime\prime},\eta')\rw_{\xi^{\prime\prime}} \big) \rW_{\eta'} \,f(\eta+\eta')\,d\mu_{C}(\xi^{\prime\prime})\,d\mu_{C}(\eta') 
\EQN\eqndefrctilde$$
This is a Grassmann function of $\xi,\xi',\eta$ that is schematically 
represented in the figure below.

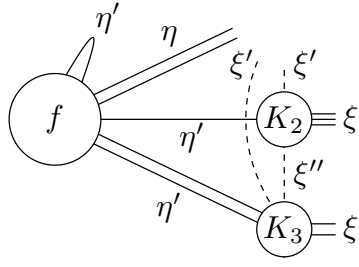
\centerline{\figput{figVI5}}

\goodbreak
\proposition{\STM\propoverldecR}{ For even Grassmann functions $K_1(\xi,\xi',\eta),\cdots,K_\ell(\xi,\xi',\eta)$ and $f(\xi)$
$$
R_C(K_1,\cdots,K_\ell)(f)=\lW
\int \hskip-6pt\int \lw K_1(\xi,\xi',\eta)\rw_{\xi',\eta}\ 
\lw \tilde R_C(K_2,\cdots,K_\ell)(f)(\xi,\xi',\eta)\rw_{\xi',\eta}
\,d\mu_{C}(\xi')\,d\mu_{C}(\eta) \rW_\xi 
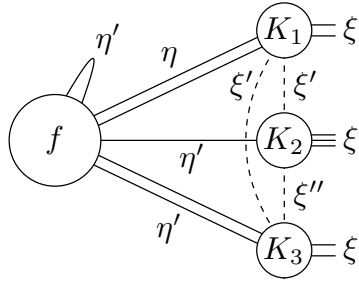
$$
}
\centerline{\figput{figVI6}}
\prf If
$$
\lw f'\rw_\xi\ =\ R_C(K_1,\cdots,K_\ell)(f)
$$
then by part (iii) of Proposition \:\propBII\ (applied to the variable $\xi'$) and Lemma \:\lemBV\ (applied to the variable $\eta$) 
$$\eqalign{
f'(\xi) 
&=
\int \hskip -6pt\int \lW\big[\lw K_1(\xi,\xi',\eta)\rw_{\xi'} 
\smprod_{i=2}^\ell \lw K_i(\xi,\xi',\eta)\rw_{\xi'} \big] \rW_{\eta} \,f(\eta)\,d\mu_C(\xi')\,d\mu_C(\eta) \cr
&=
\int\hskip -6pt \int\! \lww \big[ \lw K_1(\xi,\xi',\eta)\rw_{\xi'} 
\lW\! \Big( \int \hskip -2pt \smprod_{i=2}^\ell 
\lw K_i(\xi,\xi'+\xi^{\prime\prime},\eta)
\rw_{\xi^{\prime\prime}}d\mu_C(\xi^{\prime\prime}) \Big)\rW_{\xi'}
 \big] \rww_{\eta}\ f(\eta)\,d\mu_C(\xi')\,d\mu_C(\eta) \cr
&=
\int\hskip -6pt \int\lw K_1(\xi,\xi',\eta)\rw_{\xi',\eta} 
\lW \Big( \int\hskip -6pt \int \smprod_{i=2}^\ell \lw K_i(\xi,\xi'+\xi^{\prime\prime},\eta')
\rw_{\xi^{\prime\prime}}d\mu_C(\xi^{\prime\prime}) \Big)\rW_{\xi',\eta'} \cr
& \hskip 6cm
\lw f(\eta+\eta')\rw_\eta\,d\mu_C(\eta')\,d\mu_C(\xi')\,d\mu_C(\eta) \cr
& =
\int \hskip-6pt\int  \lw K_1(\xi,\xi',\eta)\rw_{\xi',\eta}\,
\lw \tilde R_C(K_2,\cdots,K_\ell)(f)(\xi,\xi',\eta)\rw_{\xi',\eta}
\,d\mu_{C}(\xi')\,d\mu_{C}(\eta)  \cr
}$$
\endproof

\remark{\STM\removerldecR}{ 
Set
$$\eqalign{
\hat K^{(i)}(\xi,\xi';\xi^{\prime\prime},\eta') &=  
K^{(i)}(\xi,\xi'+\xi^{\prime\prime},\eta') \cr
\hat f(\xi,\xi')&= f(\xi+\xi') \cr
}$$
Then the map $f \mapsto\ \lw \tilde R_C(K_2,\cdots,K_\ell)(f)\rw_\xi$ 
over the algebra $A'$ agrees with the map 
$f \mapsto  R_C(\hat K_2,\cdots,\hat K_\ell)(\hat f)$ of (\eqRCdef)
over the algebra $\tilde A$ of Grassmann functions in the variables $\xi',\eta$ 
with coefficients in $A'$. Therefore we can use the results of \S III
to obtain estimates on $\tilde R_C$.

}

\lemma{\STM\lemsplittwo}{ Let $K(\xi,\xi',\eta)$ be an even Grassmann function with $K(\xi,\xi',0)=0$. Decompose
$$
K(\xi,\xi',\eta) = K'(\xi,\xi',\eta) + K^{\prime\prime} (\xi,\xi',\eta)
$$
where $K'$ has degree at most two in the variables $\xi',\eta$ and
$K^{\prime\prime}$ has degree at least three in the variables $\xi',\eta$.
Let each of the functions $K^{(1)},\cdots,K^{(\ell)}$ be one of 
$K'$, $K''$ or $K$, and assume that at least one of them is equal to 
$K^{\prime\prime}$. Let $f(\xi)\in \bigwedge_{A'} V$, and set
$$
\sfrac{1}{\ell !}\, R_C(K^{(1)},\cdots,K^{(\ell)})(\lw f\rw)(\xi) = \ \lw f'(\xi)\rw
$$
Then, if $\al\ge 2$ 
$$
N_{\rm impr}(f'\cl \al) \le \sfrac{2^5\,\imp}{\ell\,\al^{\ell+5}}\,
N(f\cl 2\al)\,N(K\cl 2\al)^\ell
$$
}

\prf
We may assume that $K^{(1)}=K^{\prime\prime}$. Set
$$
g(\xi,\xi',\eta) = \tilde R_C(K^{(2)},\cdots, K^{(\ell)})(\lw f\rw)(\xi,\xi',\eta)
$$
By Remark \removerldecR, in the algebra $\tilde A$
$$
\lw g\rw_{\xi'}\ =\ R_C(\hat K^{(2)},\cdots,\hat K^{(\ell)})(\lw \hat f\rw)
$$
Therefore, by part (ii) of Proposition \propnormestR\ and Remark \remfunctnorm
$$\eqalign{
\sfrac{1}{(\ell-1)!} \,N(g) 
&\le \sfrac{1}{\al^{\ell-1}}\, N(\hat f)\,\smprod_{i=2}^\ell N(\hat K^{(i)}) \cr
&\le \sfrac{1}{\al^{\ell-1}}\, N(f\cl 2\al)\,
\smprod_{i=2}^\ell  N(K^{(i)}\cl 2\al) \cr
&\le \sfrac{1}{\al^{\ell-1}}\, N(f\cl 2\al)\,N(K\cl 2\al)^{\ell-1} \cr
}$$
By Proposition \propoverldecR
$$
f'(\xi) = \sfrac{1}{\ell !}
\int \hskip-6pt\int \lw K^{\prime\prime}(\xi,\xi',\eta)\rw_{\xi',\eta}
 \lw g(\xi,\xi',\eta)\rw_{\xi',\eta} \,d\mu_{C}(\xi')\,d\mu_{C}(\eta)  
$$
Proposition \propimprnormest\ implies that
$$\eqalign{
N_{\rm impr} (f'\cl \al) 
&\le  \sfrac{ 27\imp}{\ell!\,\al^6}\,N(K''\cl \al)\,N(g\cl \al)  \cr
&\le  \sfrac{2^5\, \imp}{\ell\,\al^{\ell+5}}\,
N(f\cl 2\al)\,N(K\cl 2\al)^\ell  \cr
}$$
\endproof

\proposition{\STM\propsplittwo}{ Let $K(\xi,\xi',\eta)$ be an even Grassmann function. Decompose
$$
K(\xi,\xi',\eta) = K'(\xi,\xi',\eta) + K^{\prime\prime} (\xi,\xi',\eta)
$$
where $K'$ has degree at most two in the variables $\xi',\eta$ and
$K^{\prime\prime}$ has degree at least three in the variables $\xi',\eta$.
Let $f(\xi)\in \bigwedge_{A'} V$, and set
$$
\lw g(\xi)\rw \ =\ \cR_{K,C}(\lw f\rw)  \qquad\qquad
\lw g'(\xi)\rw \ =\ \cR_{K',C}(\lw f\rw)
$$
Then, if $\al \ge 2$ and $N(K\cl 2\al)_\0 \le \al$
$$
N_{\rm impr}(g-g'\cl \al) 
 \le \sfrac{2^{5}\,\imp}{\al^6}\,
N(f\cl 2\al)\,\sfrac{N(K\cl 2\al)}{1-{1\over\al}N(K\cl 2\al)}
$$
}

\prf
By Remark \remcalR
$$
g = \smsum_{\ell=1}^\infty g_\ell \qquad  \qquad 
g' = \smsum_{\ell=1}^\infty g'_\ell
$$
where
$$\eqalign{
\lw g_\ell\rw\  &= \ \sfrac{1}{\ell !}\,R_C(K,\cdots,K)(\lw f\rw) \cr
\lw g'_\ell\rw\ &= \ \sfrac{1}{\ell !}\,R_C(K',\cdots,K')(\lw f\rw)  \cr
}$$
Since
$$\eqalign{
R_C&(K,\cdots,K) - R_C(K',\cdots,K') \cr
&=  R_C(K-K',K,\cdots,K)
 + R_C(K',K-K',K,\cdots,K) + \cdots +R_C(K',\cdots,K-K') \cr
&=  R_C(K^{\prime\prime},K,\cdots,K)
 + R_C(K',K^{\prime\prime},K,\cdots,K) + \cdots    +R_C(K',\cdots,K',K^{\prime\prime}) \cr
}$$ 
it follows from Lemma \lemsplittwo\  that
$$
N_{\rm impr} (g_\ell-g'_\ell) \le \sfrac{2^5\,\imp}{\al^{\ell+5}}\,
N(f\cl 2\al)\,N(K\cl 2\al)^\ell
$$
Therefore
$$
N_{\rm impr}(g-g') \le \smsum_{\ell=1}^\infty N_{\rm impr}(g_\ell-g'_\ell)
 \le \sfrac{2^{5}\,\imp}{\al^6}\,
N(f\cl 2\al)\,\sfrac{N(K\cl 2\al)}{1-{1\over\al}N(K\cl 2\al)}
$$
\endproof

\corollary{\STM\corsplittwo}{ 
Under the hypotheses of Proposition \propsplittwo, set
$$
\lw h\rw\ =\ \frac{1}{\bbbone - \cR_{K,C}}(\lw f\rw) - \frac{1}{\bbbone - \cR_{K',C}}(\lw f\rw)
$$
If $N(K\cl 2\al)_\0 < \sfrac{\al}{6}$, then
$$
N_{\rm impr}(h\cl \al) 
 \le \sfrac{2^{5}\,\imp}{\al^6}\,
\sfrac{N(K\cl 2\al)}{1-{6\over\al}N(K\cl 2\al)}\,
  N(f\cl 2\al)
$$
}

\prf Since
$$
\lw h\rw\ =\ \big(\bbbone - \cR_{K,C}\big)^{-1} \big( \cR_{K,C} -\cR_{K',C} \big)
    \big(\bbbone - \cR_{K',C}\big)^{-1}(\lw f\rw)
$$
it follows from Corollary \cornormestcalR, with $N_{\rm impr}$ in place of $N$,
and Proposition \propsplittwo\ that
$$\eqalign{
N_{\rm impr}(h) 
&\le 
\Big( 1+\sfrac{2}{\al^2}\sfrac{N_{\rm impr}(K)}{1-{4\over\al^2}N_{\rm impr}(K)} \Big)\,
N_{\rm impr}\Big(\big( \cR_{K,C} -\cR_{K',C} \big)
    \big(\bbbone - \cR_{K',C}\big)^{-1}(\lw f\rw)\Big)\cr 
&\le 
\Big( 1+\sfrac{2}{\al^2}\sfrac{N_{\rm impr}(K)}{1-{4\over\al^2}N_{\rm impr}(K)} \Big)\,
 \sfrac{2^{5}\,\imp}{\al^6}\,
\sfrac{N(K\cl 2\al)}{1-{1\over\al}N(K\cl 2\al)}\,
 N\Big( \big(\bbbone - \cR_{K',C}\big)^{-1}(\lw f\rw)\cl 2\al\Big)
}$$
By Definition \defimprnorm\ and Remark \remfunctnorm, 
$N_{\rm impr}(K) \le N(K)\le N(K\cl 2\al) $ 
so that  Corollary \cornormestcalR\ implies
$$\eqalign{
N_{\rm impr}(h) 
&\le\Big( 1+\sfrac{2}{\al^2}\sfrac{N(K\cl 2\al)}{1-{4\over\al^2}N(K\cl 2\al)} \Big)\,
 \sfrac{2^{5}\,\imp}{\al^6}\,
\sfrac{N(K\cl 2\al)}{1-{1\over\al}N(K\cl 2\al)}\,
\Big( 1+\sfrac{1}{2\al^2}\sfrac{N(K'\cl 2\al)}{1-{1\over\al^2}N(K'\cl 2\al)} \Big)\,
  N(f\cl 2\al)  \cr 
&\le\sfrac{2^{5}\,\imp}{\al^6}\,
\Big(\sfrac{1-{2\over\al^2}N(K\cl 2\al)}{1-{4\over\al^2}N(K\cl 2\al)} \Big)\,
\sfrac{N(K\cl 2\al)}{1-{1\over\al}N(K\cl 2\al)}\,
\Big(\sfrac{1-{1\over{2 \al^2}}N(K\cl 2\al)}{1-{1\over\al^2}N(K\cl 2\al)} 
\Big)\,
  N(f\cl 2\al)  \cr 
&\le\sfrac{2^{5}\,\imp}{\al^6}\,
\sfrac{1}{1-{4\over\al^2}N(K\cl 2\al)}\,
\sfrac{N(K\cl 2\al)}{1-{1\over\al}N(K\cl 2\al)}\,
\sfrac{1}{1-{1\over\al^2}N(K\cl 2\al)}\,
  N(f\cl 2\al)  \cr 
&\le\sfrac{2^{5}\,\imp}{\al^6}\,
\sfrac{N(K\cl 2\al)}{[1-{2\over\al}N(K\cl 2\al)]^3}\,
  N(f\cl 2\al)  \cr 
&\le\sfrac{2^{5}\,\imp}{\al^6}\,
\sfrac{N(K\cl 2\al)}{1-{6\over\al}N(K\cl 2\al)}\,
  N(f\cl 2\al)  \cr 
}$$
since, for $X\in\fN_d$,
$$
\sfrac{1}{[1-X]^3}=\sum_{r=0}^\infty\smchoose{-3}{r}(-X)^r
=\sum_{r=0}^\infty\sfrac{3\cdot 4\cdot 5\cdots(3+r-1)}{1\cdot 2\cdot 3\cdots r}X^r
\le\sum_{r=0}^\infty(3X)^r
=\sfrac{1}{1-3X}
$$
\endproof

Proposition \propsplittwo \  exploits overlapping loops that are created by one application of the operator $\cR_{K,C}$. There are additional overlapping loops created by the composite operator  $\cR_{K,C}\circ \cR_{K,C}$ which we shall exploit now.

\lemma{\STM\lemimprRR}{ Let $B(\eta',\eta^{\prime\prime}) \in A'[1,1]$ and let 
$H(\xi,\xi',\eta),\,K(\xi,\xi',\eta)$ be even Grassmann functions that vanish for $\eta =0$. Assume that $H$ or $K$ has degree at least two in the variables $\xi',\eta$.
Let $f(\xi) \in \bigwedge_{A'} V$ and set
$$\eqalign{
g(\xi) = \int \lW 
\Big[ \int\hskip -6pt\int\hskip -6pt\int  B(\eta',\eta^{\prime\prime})
\,\lw H(\xi+\eta',\xi',\eta)\rw_{\eta',\xi'}
\,\lw K(\xi+\eta^{\prime\prime},\xi',\eta)\rw_{\eta^{\prime\prime},\xi'}\,
&d\mu_C(\eta',\eta^{\prime\prime},\xi') \Big]\rW_\eta    \cr
& \lw f(\eta)\rw_\eta \,d\mu_C(\eta) \cr
}
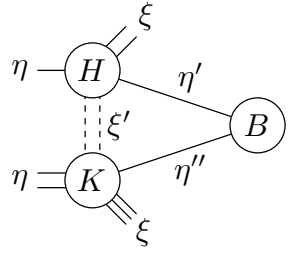
$$

\centerline{\figput{figVI1}}

\noindent
Then, if $\al \ge 2$
$$
N_{\rm impr}(g\cl \al) \le \sfrac{2^5\,\imp}{\al^{10}}\,
N(f\cl 2\al)\,N(H\cl 2\al)\,N(K\cl 2\al)\,N(B\cl \al)
$$
}

\prf We may assume that $K$ has degree at least two in the variables $\xi',\eta$. Set 
$$
h(\xi,\eta,\xi',\eta^{\prime\prime})
= \int\hskip -6pt\int B(\eta',\eta^{\prime\prime})
\,\lw H(\xi+\eta',\xi',\ze)\rw_{\eta',\xi,\ze}\,\lw f(\eta+\ze)\rw_\ze \,d\mu_C(\eta')d\mu_C(\ze)
$$
By Lemma \:\lemBV, applied to the variable $\eta$
$$\eqalign{
g(\xi) &= \int  
 B(\eta',\eta^{\prime\prime})
\,\lw H(\xi+\eta',\xi',\ze)\rw_{\eta',\xi',\ze}
\,\lw K(\xi+\eta^{\prime\prime},\xi',\eta)\rw_{\eta^{\prime\prime},\xi',\eta}\,
 \lw f(\eta+\ze)\rw_{\eta,\ze} \cr
& \hskip 8cm d\mu_C(\eta',\eta^{\prime\prime},\xi',\eta,\ze)    \cr
& = \int \lw h(\xi,\eta,\xi',\eta^{\prime\prime})\rw_{\xi',\eta}\,
\lw K(\xi+\eta^{\prime\prime},\xi',\eta)\rw_{\eta^{\prime\prime},\xi',\eta}\,
d\mu_C(\eta,\eta^{\prime\prime},\xi')    \cr
}$$
Since $B$ is of degree one in $\et'$ and $H$ is of degree at least one
in $\ze$, iterated application of Proposition \propestCconnect\ and Remark \remfunctnorm\
yields
$$
N(h) \le \sfrac{1}{\al^4} N(B)\,N(H\cl 2\al)\,N(f\cl 2\al)
$$
Since $h$ is of degree one in $\et''$, only the part of $K$ that is of
 degree at least three in the variables $\eta^{\prime\prime}, \xi',\eta$
can contribute. Hence  Proposition \propimprnormest\ implies that
$$\eqalign{
N_{\rm impr}(g)
&\le \sfrac{27\,\imp}{\al^6}\, N(h)\,N(K\cl 2\al) \cr
&\le \sfrac{2^5\,\imp}{\al^{10}}\, N(B)\,N(H\cl 2\al)\,N(f\cl 2\al)\,N(K\cl 2\al) \cr
}$$
\endproof

\proposition{\STM\propimprRR}{
Let $K(\xi,\xi',\eta)$ be a Grassmann function of degree at most two in the variables $\xi',\eta$ and of degree at least one in $\eta$. Write
$$
K(\xi,\xi',\eta) = \sum_{n_1,n_2,n_3} K_{n_1,n_2,n_3}(\xi,\xi',\eta)
\qquad {\rm with\ }\qquad  K_{n_1,n_2,n_3} \in A'[n_1,n_2,n_3]
$$
Also set $K_{\cdot,n_2,n_3} = \smsum_{n_1}  K_{n_1,n_2,n_3}$. Furthermore let $T(\eta)\in A'[2]$ and write
$$
T(\eta+\eta') = T(\eta) +T(\eta') +T_{\rm mix}(\eta,\eta')
$$
with $T_{\rm mix} \in A'[1,1]$. Set
$$
\tilde K(\xi,\xi',\eta) 
= \int \lw T(\eta')\rw_{\eta'}\,\lw K(\xi+\eta',\xi',\eta)\rw_{\eta'}\,d\mu_C(\eta')
\hskip.3in \figplace{figVI2}{0in}{-.2in}
$$
$$
 \hat K(\xi,\xi',\eta) 
= \int T_{\rm mix}(\eta',\eta^{\prime\prime})
\,\lw K_{\cdot,0,1}(\xi+\eta',\xi',\eta)\rw_{\eta'}
\,\lw K_{\cdot,0,1}(\xi+\eta^{\prime\prime},\xi',\eta)\rw_{\eta''}\,
d\mu_C(\eta',\eta^{\prime\prime})
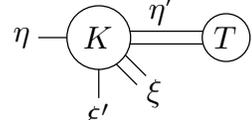
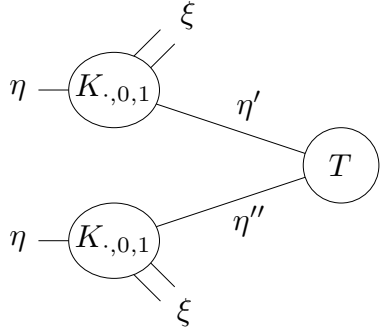
$$
\centerline{\figput{figVI3}}
Observe that $\hat K$ is independent of $\xi'$.
Finally let $f(\xi)\in\bigwedge_{A'} V$ and set
$$\eqalign{
f'(\eta) &= \tilde R_C(T) R_C(K,K)(\lw f\rw) \cr
\lw \tilde f(\xi)\rw_\xi &= 2\,R_C(\tilde K,K)(\lw f\rw) \cr
\lw \hat f(\xi)\rw_\xi &= R_C(\hat K)(\lw f\rw) \cr
}$$
Then
$$
N_{\rm impr}(f'-(\tilde f+\hat f)\cl \al) \le \sfrac{2^{6}\,\imp}{\al^{10}}\,
N(f\cl 2\al)\,N(K\cl 2\al)^2\,\,N(T\cl 2\al)
$$
}

\prf  By definition  
$$
f'(\eta) 
= \int \lw T(\eta')\rw_{\eta'} \,R_C(K,K)(\lw f\rw)(\eta+\eta')\,d\mu_C(\eta') 
$$
Consequently, by part (ii) of Proposition \:\propBII\ and Lemma \:\lemBVI, in the variable
$\et'$,
$$\eqalignno{
f'(\xi) 
&= \int \lw T(\eta')\rw_{\eta'}\,R_C(K,K)(\lw f\rw)(\xi+\eta')\ d\mu_C(\eta')\cr 
&= \int \lw T(\eta')\rw_{\eta'} 
   \lww\big[\lW \big(\lw
   K(\xi+\eta',\xi',\eta)\rw_{\xi'}\,\lw K(\xi+\eta',\xi',\eta)\rw_{\xi'}
    \big)\rW_\eta\,\lw f(\eta)\rw_\eta            
    \big]\rww_{\eta'}\,d\mu_C(\xi',\eta,\eta') \cr
&= 2\int \lww\Big[\lW
   \Big( \int \lw T(\eta')\rw_{\eta'}
    \lw K(\xi+\eta',\xi',\eta)\rw_{\eta'}\,d\mu_C(\eta')\Big)\rW_{\xi'}
    \lw K(\xi,\xi',\eta)\rw_{\xi'}\Big]\rww_\eta\cr
 & \hskip 9cm \lw f(\eta)\rw_\eta  \, d\mu_C(\xi',\eta) \cr
&\ \ \ + \int \lww
    \Big[ \int T_{\rm mix}(\eta',\eta^{\prime\prime})
    \lw K(\xi+\eta',\xi',\eta)\rw_{\xi',\eta'}\,
    \lw K(\xi+\eta^{\prime\prime},\xi',\eta)\rw_{\xi',\eta^{\prime\prime}}\,     d\mu_C(\xi',\eta',\eta^{\prime\prime}) \Big]\rww_\eta\cr
 & \hskip 9cm \lw f(\eta)\rw_\eta  \, d\mu_C(\eta) \cr
}$$
$$\eqalign{
\phantom{f'(\xi)}
&= 2\int \lW \big[
    \lw\tilde K(\xi,\xi',\eta)\rw_{\xi'}\,
    \lw K(\xi,\xi',\eta)\rw_{\xi'}\big]\rW_\eta\,
    \lw f(\eta)\rw_\eta\,d\mu_C(\xi',\eta) \cr
&\ \ \ + \int  \lw\hat K(\xi,\xi',\eta)\rw_{\xi',\eta}\, \lw f(\eta)\rw_\eta\,d\mu_C(\xi',\eta)  \cr
&\ \ \ + \int \lW
     \Big[ \int T_{\rm mix}(\eta',\eta^{\prime\prime}) \Big(
    \lw K(\xi+\eta',\xi',\eta)\rw_{\xi',\eta'}\,
    \lw K(\xi+\eta^{\prime\prime},\xi',\eta)\rw_{\xi',\eta^{\prime\prime}} \cr
& \hskip 2cm     -\lw K_{\cdot,0,1}(\xi+\eta',\xi',\eta)\rw_{\xi',\eta'}\,
  \lw K_{\cdot,0,1}(\xi+\eta^{\prime\prime},\xi',\eta)\rw_{\xi',\eta^{\prime\prime}}                         \Big)d\mu_C(\xi',\eta',\eta^{\prime\prime}) \Big]\rW_\eta\cr
 & \hskip 9cm   \lw f(\eta)\rw_\eta  \  d\mu_C(\eta) \cr
&= \tilde f(\xi) + \hat f(\xi) +g'(\xi) +g^{\prime\prime}(\xi) \cr
}$$
with
$$\eqalign{
g'(\xi) &=  \int \lW
     \Big[ \int T_{\rm mix}(\eta',\eta^{\prime\prime})\,
    \lw H(\xi+\eta',\xi',\eta)\rw_{\xi',\eta'}\,
    \lw K(\xi+\eta^{\prime\prime},\xi',\eta)\rw_{\xi',\eta^{\prime\prime}}\,     d\mu_C(\xi',\eta',\eta^{\prime\prime}) \Big]\rW_\eta\cr
 & \hskip 9cm \lw f(\eta)\rw_\eta  \, d\mu_C(\eta) \cr
g^{\prime\prime}(\xi) &=   \int \lW
     \Big[  \int T_{\rm mix}(\eta',\eta^{\prime\prime})
    \lw K_{\cdot,0,1}(\xi+\eta',\xi',\eta)\rw_{\xi',\eta'}\,
    \lw H(\xi+\eta^{\prime\prime},\xi',\eta)\rw_{\xi',\eta^{\prime\prime}}          d\mu_C(\xi',\eta',\eta^{\prime\prime}) \Big]\rW_\eta\cr
 & \hskip 9cm \lw f(\eta)\rw_\eta  \, d\mu_C(\eta) \cr
}$$
where $H=K-K_{\cdot,0,1}$.
In the last equality, we used the fact that 
$K_{\cdot,0,1}(\xi+\eta',\xi',\eta)$ and hence $\hat K$ is independent of 
$\xi'$, so that we are free to drop the Wick ordering with respect to $\xi'$
in the expression yielding $\hat f(\xi)$.
By Lemma \lemimprRR\ and the observations that
$$\eqalign{
&N(K_{\cdot,0,1}\cl 2\al),\ 
N(K-K_{\cdot,0,1}\cl 2\al)
\le N(K\cl 2\al)\cr
& N(T_{\rm mix}\cl \al)
\le N(T(\et+\et')\cl \al)
\le N(T\cl 2\al)\cr
}$$
we have
$$
N_{\rm impr}(f'-\tilde f -\hat f) = N_{\rm impr}(g' +g^{\prime\prime}) 
 \le \sfrac{2^6\,\imp}{\al^{10}}\,N(f\cl 2\al)\,
N(K\cl 2\al)^2\,\,N(T\cl 2\al) 
$$
\endproof

\sect{ Tails}\PG\pgRVIIb

In this subsection let $K(\psi;\xi,\xi',\eta)$ be an even Grassmann function 
with coefficients in $A$ that has  degree at least four in the variables
$\psi,\xi,\xi',\eta$ and degree at least one in the variable $\eta$. We always write
$$
K = \sum_{n_0,\cdots,n_3} K_{n_0,n_1,n_2,n_3}  \qquad {\rm with} \qquad
 K_{n_0,n_1,n_2,n_3} \in A[n_0,n_1,n_2,n_3]
$$
For $f(\psi;\xi): \in \bigwedge_A(V'\oplus V)$ we are interested in the two and four legged contributions to 
$\sfrac{1}{\bbbone - \cR_{K,C}}(\lw f\rw_\xi)$. Therefore we make the following
\definition{\STM\defP}{
The  projection 
$$\eqalign{
P:\bigwedge\nolimits_A(V'\oplus V) &\longrightarrow \bigwedge\nolimits_A V' \cr
   \lw f(\psi;\xi)\rw_\xi & \longmapsto f_{4,0}(\psi,0) + f_{2,0}(\psi,0) 
}$$
where $f(\psi;\xi) = \smsum_{n_0,n_1} f_{n_0,n_1}(\psi;\xi)$ with
$f_{n_0,n_1} \in A[n_0,n_1]$.
}

\definition{\STM\deftailsf}{
\Item{(i)} An $n$--legged tail is a Grassmann function 
 $T(\psi;\eta) \in \bigoplus_{d\ge 2} A[d,n]$. We say that an $n$--legged tail
$T$ has at least $d$ external legs if $T \in \bigoplus_{d'\ge d} A[d',n]$.
\Item{(ii)} If $T$ is a two--legged tail we define the two--legged tail 
$T\circ K$ by
$$
(T\circ K)(\psi;\eta) = 
\int\lw T(\psi;\eta')\rw_{\eta'}\, \lw K_{0,2,0,2}(\psi;\eta',\xi',\eta)\rw_{\eta'}\,
d\mu_C(\eta')\hskip.3in\figplace{figVI4}{0in}{-.25in}
$$


\noindent
Observe that $T\circ K$ depends only on the part of $K$ that has degree at most two in the variables $\xi',\eta$.
}

\remark{\STM\remtailend}{
A two--legged tail with two external legs is an end in the sense of Definition \defladders.i. If $K(\psi;\xi,\xi',\eta) = U(\psi;\xi+\xi'+\eta) - U(\psi;\xi+\xi')$
for some even Grassmann function $U(\psi;\xi)$ then $T\circ K$ agrees with 
$T\circ {\rm Rung}(U)$ of Definition \defladders.iii.
}

\lemma{\STM\lemtailI}{ 
Assume that $K$ has degree at most two in the variables $\xi',\eta$. Let $T$ be a two--legged tail. Then there exists a one--legged tail $t_1$ with at least three external legs and a two--legged tail $t_2$ with at least three external legs such that for any
$f(\psi;\xi) \in \bigwedge_A(V'\oplus V)$ the following holds:

\noindent
Set
$$
f'(\psi) = P\Big[ R_C(T)\cR_{K,C}(\lw f\rw) - R_C(T\circ K)(\lw f\rw)
 -  R_C(T\circ K,K_{2,0,0,2})(\lw f\rw) - R_C(t_1+t_2)(\lw f\rw) \Big]
$$
where $\lw f\rw$ is shorthand for $\lw f\rw_\xi$. Then
$$
N_{\rm impr}(f'\cl \al) \le \sfrac{2^{6}\,\imp}{\al^{10}}\,
N(f\cl 2\al)\,N(K\cl 2\al)^2\,\,N(T\cl 2\al)
$$
If $T$ has at least three external legs then
$$
 P R_C(T)\cR_{K,C}(\lw f\rw) = P R_C(T\circ K)(\lw f\rw) + P R_C(t_1+t_2)(\lw f\rw)
$$
}

\prf
By assumption, $K(\psi;\xi,\xi',\eta)$ is of degree at least two in 
$\psi,\xi$, so $R_C(K,\cdots,K)(\lw f\rw)$ (with $\ell$ $K$'s) is of degree at 
least $2\ell$ in $\psi,\xi$. Since $T$ is two--legged, 
$R_C(T)R_C(K,\cdots,K)(\lw f\rw)$ is of degree at least $2\ell-2+2=2\ell$
(with the last $+2$ coming from the $d\ge 2$ external legs of $T$) in $\psi$
and is independent of $\xi$.
So, by Remark \remcalR
$$
 P\big[ R_C(T)\cR_{K,C}(\lw f\rw)\big] 
= P \big[ R_C(T)R_C(K)(\lw f\rw)\big]+ \sfrac{1}{2} P \big[ R_C(T)R_C(K,K)(\lw f\rw)\big]
$$
Set 
$$\eqalign{
t_{11}(\psi;\eta) &= \int \lw T(\psi;\eta')\rw_{\eta'}\,\lw K_{\cdot,2,0,1}(\psi;\eta',\xi',\eta)\rw_{\eta'}\,
      d\mu_C(\eta') \cr
t_{21}(\psi;\eta) &= \int \lw T(\psi;\eta')\rw_{\eta'}\lW\big( K_{\cdot,2,0,2}(\psi;\eta',\xi',\eta)-K_{0,2,0,2}(\psi;\eta',\xi',\eta)
     \big)\rW_{\eta'}\, d\mu_C(\eta')  \cr
}$$
Since $K$ has degree at least four overall and $T$ has degree
at least two in $\psi$, $t_{11}$ is a one--legged tail and $t_{21}$ is a 
two--legged tail, both having at least three external legs. 
As $K$ has degree at most two in the variables $\xi',\eta$ and degree at
least one in $\et$ and $T$ has degree two in $\et$ and the definition of
$R_C(K)$ Wick orders $K$ with respect to $\xi'$,
$$\eqalign{
P \big[ R_C(T)R_C(K)(\lw f\rw)\big] 
&= P \big[ R_C(T)R_C(K_{\cdot,2,0,1})(\lw f\rw)\big] 
+ P \big[ R_C(T)R_C(K_{\cdot,2,0,2})(\lw f\rw)\big] \cr
&= P \big[ R_C(t_{11}+t_{21})(\lw f\rw)\big] 
+ P \big[ R_C(T\circ K)(\lw f\rw)\big]\cr
}$$

Define the projection
$$\eqalign{
P':\bigwedge\nolimits_A(V'\oplus V) &\longrightarrow \bigwedge\nolimits_A V' \cr
   f(\psi;\xi) & \longmapsto f_{4,0}(\psi,0) + f_{2,0}(\psi,0) 
}$$
As $P$ and $P'$ only differ by Wick ordering in the $\xi$--argument,
$$\eqalign{
P \big[ R_C(T)R_C(K,K)(\lw f\rw)\big] 
&=P' \big[ R_C(T)R_C(K,K)(\lw f\rw)\big] \cr
&= P' \big[ \tilde R_C(T)R_C(K,K)(\lw f\rw)(\psi;0,0,\xi)\big]\cr
}$$
so that we can apply Proposition \propimprRR. Modulo a term whose improved norm 
$N_{\rm impr}$ is bounded by 
$\sfrac{2^{6}\,\imp}{\al^{10}}\,
N(f\cl 2\al)\,N(K\cl 2\al)^2\,\,N(T\cl 2\al)$
$$
P' \big[ \tilde R_C(T)R_C(K,K)(\lw f\rw)(\psi;0,0,\xi)\big]
= 2P \big[ R_C(\tilde K,K)(\lw f\rw)\big] + 
  P \big[ R_C(\hat K)(\lw f\rw)\big]
$$
with
$$\eqalign{
\tilde K(\psi;\xi,\xi',\eta) 
&= \int \lw T(\psi;\eta')\rw_{\eta'}\,\lw K(\psi;\eta',\xi',\eta)\rw_{\eta'}\,d\mu_C(\eta') \cr
\hat K(\psi;\xi,\xi',\eta) 
&= \int T_{\rm mix}(\psi;\eta',\eta^{\prime\prime})
\,\lw K_{\cdot,1,0,1}(\psi;\eta',\xi',\eta)\rw_{\eta'}
\,\lw K_{\cdot,1,0,1}(\psi;\eta^{\prime\prime},\xi',\eta)\rw_{\eta^{\prime\prime}}\,
d\mu_C(\eta',\eta^{\prime\prime}) \cr
}$$
Here we have used the projection $P$ to set $\xi=0$ and we used that 
$ T_{\rm mix}(\psi;\eta',\eta^{\prime\prime})$ is of degree one in $\et'$
and in $\et''$.
Again, since $K$ has degree at least four overall and $T$ is of degree
at least two in $\psi$, the tail 
$\hat K$ has at least six external legs, so that
$P \big[ R_C(\hat K)(\lw f\rw)\big]=0$. Finally, since
\item{$\bullet$} $P$ sets the $\xi$'s in the argument $K$ of 
$P \big[ R_C(\tilde K,K)(\lw f\rw)\big]$ to zero.
\item{$\bullet$} $\tilde K$ is of degree at least two in $\psi$
\item{$\bullet$} $K_{\cdot,0,0,1}$ is of degree at least three in $\psi$
\item{$\bullet$} $K$ is of degree at most two in 
$\xi',\et$ and degree at least one in $\et$.

\noindent
we have
$$\eqalign{
P \big[ R_C(\tilde K,K)(\lw f\rw)\big] 
&= P \big[ R_C(\tilde K_{\cdot,0,\cdot,\cdot},K_{\cdot,0,0,2})(\lw f\rw)\big]
   +P \big[ R_C(\tilde K_{\cdot,0,\cdot,\cdot},K_{\cdot,0,1,1})(\lw f\rw)\big] \cr
&= P \big[ R_C(\tilde K_{\cdot,0,0,\cdot},K_{\cdot,0,0,2})(\lw f\rw)\big]
   +P \big[ R_C(\tilde K_{\cdot,0,1,\cdot},K_{\cdot,0,1,1})(\lw f\rw)\big]
}$$
and since
\item{$\bullet$} $K$ is of degree at least four overall and
$\tilde K$ is of degree at least two in $\psi$

\noindent
we have
$$\eqalign{
P \big[ R_C(\tilde K,K)(\lw f\rw)\big] 
&= P \big[ R_C(\tilde K_{2,0,0,\cdot},K_{2,0,0,2})(\lw f\rw)\big]
   +P \big[ R_C(\tilde K_{2,0,1,\cdot},K_{2,0,1,1})(\lw f\rw)\big] \cr
&= P \big[ R_C(\tilde K_{2,0,0,2},K_{2,0,0,2})(\lw f\rw)\big]
   +P \big[ R_C(\tilde K_{2,0,1,1},K_{2,0,1,1})(\lw f\rw)\big] \cr
&= P \big[ R_C(T\circ K,K_{2,0,0,2})(\lw f\rw)\big]
   +P \big[ R_C(t_{22})(\lw f\rw)\big] \cr
}$$

\noindent
where 
$$
t_{22}(\psi;\eta)= \int \tilde K_{2,0,1,1}(\psi;\xi,\xi',\eta) \,
K_{2,0,1,1}(\psi;\xi,\xi',\eta)\,d\mu_C(\xi')
$$
is a two--legged tail with at least four external legs. 

Setting $t_1=t_{11}$ and $t_2=t_{21}+2t_{22}$ yields the main result.
If $T$ has at least three external legs, then 
$R_C(T)R_C(K,\cdots,K)(\lw f\rw)$ (with $\ell$ $K$'s) is of degree at least 
$2\ell-2+3=2\ell+1$, so
$$
 P\big[ R_C(T)\cR_{K,C}(\lw f\rw)\big] 
= P \big[ R_C(T)R_C(K)(\lw f\rw)\big]
= P \big[ R_C(t_{11}+t_{21})(\lw f\rw)\big] 
+ P \big[ R_C(T\circ K)(\lw f\rw)\big]
$$

\endproof

\lemma{\STM\lemtailII}{ Assume that $K$ has degree at most two in the variables $\xi',\eta$.
Let $T_1,T_2$ be two--legged tails. Then there exists a one--legged tail $t_1$ and a two--legged tail $t_2$, each with at least four external legs, such that for any
$f(\psi;\xi) \in \bigwedge_A(V'\oplus V)$ the following holds:

\noindent
Set
$$
f'(\psi) = P\Big[ R_C(T_1,T_2)\cR_{K,C}(\lw f\rw) - R_C(T_1\circ K,T_2\circ K)(\lw f\rw)
 - R_C(t_1+t_2)(\lw f\rw) \Big]
$$
Then, if $\al \ge 2$
$$
N_{\rm impr}(f'\cl \al) \le \sfrac{2^{6}\,\imp}{\al^{12}}\,
N(f\cl 2\al)\,N(K\cl 2\al)^2\,\,N(T_1\cl 2\al)\,N(T_2\cl 2\al)
$$
}

\prf{} 
Again, since $K$ has degree at most two in the variables $\xi',\eta$ and degree at least four in all variables
$$
P\big[ R_C(T_1,T_2)\cR_{K,C}(\lw f\rw)\big] 
= P\big[ R_C(T_1,T_2)R_C(K)(\lw f\rw)\big]  + 
\sfrac{1}{2}\, P\big[ R_C(T_1,T_2)R_C(K,K)(\lw f\rw)\big] 
$$
and
$$\eqalign{
P\big[ R_C(T_1,T_2)R_C(K)(\lw f\rw)\big] 
&= P\big[ R_C(T_1,T_2)R_C(K_{0,4,0,\cdot})(\lw f\rw)\big] \cr
&= P\big[ R_C(t_{11}+t_{21})(\lw f\rw)\big] \cr
}$$
where
$$\eqalign{
t_{11}(\psi;\eta) &= \int \lw T_1(\psi;\eta')T_2(\psi;\eta')\rw_{\eta'}\,
K_{0,4,0,1}(\psi;\eta',\xi',\eta)\,d\mu_C(\eta')\cr
t_{21}(\psi;\eta) &= \int \lw T_1(\psi;\eta')T_2(\psi;\eta')\rw_{\eta'}\,
K_{0,4,0,2}(\psi;\eta',\xi',\eta)\,d\mu_C(\eta')\cr
}$$
are one-- resp. two--legged tails with at least four external legs. 

\noindent
Similarly
$$\eqalign{
P&\big[ R_C(T_1,T_2)R_C(K,K)(\lw f\rw)\big] \cr
&= P\big[ R_C(T_1,T_2)R_C(K_{0,2,0,2},K_{0,2,0,2})(\lw f\rw)\big]
   +P\big[ R_C(T_1,T_2)R_C(K_{0,2,1,1},K_{0,2,1,1})(\lw f\rw)\big] \cr
}$$
The second term is $PR_C(t_{22})(\lw f\rw)$ where
$$
t_{22}(\psi;\eta) = \int \lw T_1(\psi;\eta')T_2(\psi;\eta')\rw_{\eta'}\,
\lw K_{0,2,1,1}(\psi;\eta',\xi',\eta)K_{0,2,1,1}(\psi;\eta',\xi',\eta)\rw_{\eta'}\,
d\mu_C(\eta')
$$
is a two--legged tail with at least four external legs. To deal with the first term we use Proposition \propoverldecR\ to see that
$$
R_C(T_1,T_2)R_C(K_{0,2,0,2},K_{0,2,0,2})(\lw f\rw)
=\int \lw T_1(\psi,\eta')\rw_{\eta'}\,\lw g(\psi;\eta')\rw_{\eta'}\,d\mu_C(\eta')
$$
where
$$
g(\psi;\et) = \tilde R_C(T_2)R_C(K_{0,2,0,2},K_{0,2,0,2})(\lw f\rw)
$$
By Proposition \propimprRR, $\lw g(\psi;\xi)\rw_\xi $ is -- modulo terms whose improved norm can be bounded by 
$\sfrac{2^{6}\,\imp}{\al^{10}}\,
N(f\cl 2\al)\,N(K\cl 2\al)^2\,\,N(T_2\cl 2\al)\ $ -- equal to
$2\, R_C(K_{0,2,0,2},T_2\circ K)(\lw f\rw)$. 
Therefore by Proposition  \propestCconnect, modulo terms whose improved norm can be bounded by 
$\sfrac{2^{6}\,\imp}{\al^{12}}\,
N(f\cl 2\al)\,N(K\cl 2\al)^2\,\,
N_{\rm impr}(T_1\cl \al)\,N(T_2\cl 2\al)$
$$\eqalign{
\sfrac{1}{2}\, P\big[& R_C(T_1,T_2)R_C(K_{0,2,0,2},K_{0,2,0,2})(\lw f\rw)\big] \cr
&= P\,\int \lw T_1(\psi,\eta')\rw_{\eta'}\,
 R_C(K_{0,2,0,2},T_2\circ K)(\lw f\rw)(\psi;\eta')\,d\mu_C(\eta') \cr
&= P \big[ R_C(T_1\circ K,T_2\circ K)(\lw f\rw) \big] \cr
}$$
Since $N_{\rm impr}(T_1\cl \al) \le N(T_1\cl \al)
\le N(T_1\cl 2\al)$, the Lemma follows.
\endproof

\lemma{\STM\lemtailIII}{Assume that $K$ has degree at most two in the variables $\xi',\eta$.
Let $T$ be a one--legged tail with at least three external legs. Then there is a
two--legged tail $t_2$ with at least four external legs such that for all $f(\psi;\xi) \in \bigwedge_A(V'\oplus V)$
$$
P\big[ R_C(T)\cR_{K,C}(\lw f\rw)\big] = P\big[ R_C(t_2)(\lw f\rw)\big]
$$
}

\prf 
Since $T$ is one--legged and $K$ has degree at least four
$$\eqalign{
P\,R_C(T)\,\cR_{K,C} & = P\,R_C(T)\,R_C(K) \cr
& = P\,R_C(T)\,
   R_C\big(\smsum_{d\ge 2} K_{d,1,0,1} +\smsum_{d\ge 1} K_{d,1,0,2}\big) \cr
& = P\,R_C(T)\,R_C(K_{1,1,0,2}) \ =\ P\,R_C(t_2)
}$$
with 
$$
t_2(\psi;\eta) = \int T(\psi;\eta')\,K_{1,1,0,2}(\psi;\eta',\xi',\eta)\,d\mu_C(\eta')
$$
\endproof

\definition{\STM\defeffecttail}{ 
The two--legged tails $T_\ell(K)$ are recursively defined by
$$\eqalign{
T_1(K)(\psi;\eta) &= K_{2,0,0,2}(\psi;\xi,\xi',\eta) \cr
T_{\ell+1}(K) &= T_\ell \circ K \qquad \qquad {\rm for\ } \ell \ge 1  \cr
}$$
}

\remark{\STM\remeffecttail}{
\Item{(i)}
Clearly $T_\ell(K)$ has two external legs. Using Proposition \propestCconnect\ one proves by induction that for $\al \ge 2$
$$
N\big( T_\ell(K) \big) \le \sfrac{1}{\al^{2\ell-2}} N(K)^\ell
$$
\Item{(ii)} 
If $K(\psi;\xi,\xi',\eta) = U(\psi;\xi+\xi'+\eta) - U(\psi;\xi+\xi')$
for some even Grassmann function $U(\psi;\xi)$ then, by Remark \remtailend,
$T_\ell(K) = T_\ell(U)$, where $T_\ell(U)$ was defined in 
Definition \deftails.ii.
\Item {(iii)} $T_\ell(K)$ depends only on the part of $K$ that has degree at most two in the variables $\xi',\eta$.
}

\proposition{\STM\propprodtails}{
Assume that $K$ has degree at most two in the variables $\xi',\eta$.
For each $\ell \ge 1$ there exists a one--legged tail $t_1$ and a two--legged tail $t_2$, each with at least three external legs such that for any $f(\psi;\xi) \in \bigwedge_A(V'\oplus V)$ the following holds:
Set
$$
f'_\ell(\psi) = P \big[ \cR_{K,C}^\ell(\lw f\rw) - R_C\big(T_\ell(K)\big)(\lw f\rw) 
 -\sfrac{1}{2}\hskip-9pt\smsum_{\ell',\ell''\ge 1\atop\max\{\ell',\ell''\}=\ell}
\hskip-9pt R_C\big(T_{\ell'}(K),T_{\ell''}(K)\big)(\lw f\rw)
 -R_C(t_1+t_2)(\lw f\rw) \Big]
$$ 
Then, if $\al \ge 2$ and $N(K\cl 2\al)_\0 < \al^2$
$$
N_{\rm impr}(f'_\ell\cl \al) 
\le \,\imp\,\sfrac{2^{7}}{\al^{2\ell+6}}\,N\big(f\cl 2\al\big)\,
\sfrac{N(K\cl 2\al)^{\ell+1}}
{\big[1-{1\over \al^2}N(K\cl 2\al)\big]^{\ell-1}}
$$
}

\proof{ by induction on $\ell$} Set $N_K= N(K\cl 2\al)$.
Since $K$  has degree at most two in the variables $\xi',\eta$ and degree at least four overall, for $\ell =1$ 
$$\eqalign{
P\, \cR_{K,C}
& = P\,R_C(K_{\cdot,0,0,2} + K_{\cdot,0,0,1}) 
  +\sfrac{1}{2}\, P\,R_C(K_{2,0,1,1} , K_{2,0,1,1})
  +\sfrac{1}{2}\, P\,R_C(K_{2,0,0,2} , K_{2,0,0,2}) \cr
& = P\,R_C\big(T_1(K)\big) 
  + P\,R_C(t_1+t_2)
  +\sfrac{1}{2}\, P\,R_C\big(T_1(K),T_1(K)\big) \cr
}$$
with
$$\eqalign{
t_1 &= K_{3,0,0,1} + K_{4,0,0,1} \cr
t_2 &= K_{3,0,0,2} + K_{4,0,0,2} 
 + \sfrac{1}{2}\,\int K_{2,0,1,1}(\psi;\xi,\xi',\eta)\, K_{2,0,1,1}(\psi;\xi,\xi',\eta)\,d\mu_C(\xi') \cr
}$$

Now assume that the statement of the Lemma is true for $\ell$. By the induction hypothesis there exist a one--legged tail $t_1$ and a two--legged tail $t_2$, each with at least three external legs, such that
$$
P \big[ \cR_{K,C}^{\ell+1}(\lw f\rw)\big] = 
P \big[ \cR_{K,C}^\ell\big(\cR_{K,C}(\lw f\rw)\big)\big]
$$
differs from
$$\eqalign{
&P\Big[ R_C\big(T_\ell(K)\big)
\big(\cR_{K,C}(\lw f\rw)\big)
+ \sfrac{1}{2}\hskip-9pt\smsum_{\ell',\ell''\ge 1\atop\max\{\ell',\ell''\}=\ell}
\hskip-9pt R_C\big(T_{\ell'}(K),T_{\ell''}(K)\big)\big(\cR_{K,C}(\lw f\rw)\big)
\cr&\hskip3in +R_C(t_1+t_2)\big(\cR_{K,C}(\lw f\rw)\big) \Big]
}$$
by a function $g_0(\psi)$ with
$$\eqalign{
N_{\rm impr}(g_0) 
&\le \sfrac{2^{7}}{\al^{2\ell+6}}\,\imp\,
\,\sfrac{N_K^{\ell+1}}
{\big[1-{1\over \al^2}N_K\big]^{\ell-1}}N\big(\tilde f\cl 2\al\big) \cr
&\le \sfrac{2^{7}}{\al^{2\ell+6}}\,\imp\,
\,\sfrac{N_K^{\ell+1}}
{\big[1-{1\over \al^2}N_K\big]^{\ell-1}}
\ \sfrac{1}{2\al^2}\,\sfrac{N_K}
{1-{1\over 2\al^2}N_K}
N\big(f\cl 2\al\big) \cr
&\le \sfrac{2^{6}}{\al^{2\ell+8}}\,\imp\,
\,\sfrac{N_K^{\ell+2}}
{\big[1-{1\over \al^2}N_K\big]^{\ell}}N\big(f\cl 2\al\big) 
}$$
where $\lw \tilde f\rw =\cR_{K,C}(\lw f\rw)$.
Here, we used Lemma \lemnormestcalR, with $\al$ replaced by $2\al$,
to estimate $N\big(\tilde f\cl 2\al\big)$.

\noindent
By Lemma \lemtailI, there exists a one--legged tail $t_{11}$  and a two--legged 
tail $t_{21}$, with at least three external legs each, such that
$$\eqalign{
P\Big[ R_C\big(T_\ell(K)\big)
&\big(\cR_{K,C}(\lw f\rw)\big)\Big]
= P\big[ R_C\big(T_{\ell+1}(K)\big)(\lw f\rw) \big]
 +  P\big[ R_C\big(T_{\ell+1}(K),T_1(K)\big)(\lw f\rw) \big]\cr
&\hskip2.5in+ P\big[ R_C\big(t_{11}+t_{21})(\lw f\rw)+g_1(\psi)\big]\cr
&= P\big[ R_C\big(T_{\ell+1}(K)\big)(\lw f\rw) \big]
 + \half P\big[ R_C\big(T_{\ell+1}(K),T_1(K)\big)(\lw f\rw) \big]\cr
&\hskip.5in+ \half P\big[ R_C\big(T_1(K),T_{\ell+1}(K)\big)(\lw f\rw) \big]+ P\big[ R_C\big(t_{11}+t_{21})(\lw f\rw)+g_1(\psi)\big]\cr
}$$
with
$$\eqalign{
N_{\rm impr}(g_1) 
&\le \sfrac{2^{6}\,\imp}{\al^{10}}\,N(f\cl 2\al)\,N_K^2\ 
N(T_\ell(K)\cl 2\al)  \cr
&\le \sfrac{2^{6}\,\imp}{\al^{2\ell+8}}\,N(f\cl 2\al)\,N_K^{\ell+2} \cr
}$$
Here, we used Remark \remeffecttail\ to bound $N(T_\ell(K)\cl 2\al)$ by $\sfrac{1}{\al^{2\ell-2}}N_K^\ell$.

\noindent
Similarly, for $\ell', \ell''\ge 1$ with $\max\{\ell',\ell''\}=\ell$, by Lemma \lemtailII, there exists a one--legged tail $t_{12}^{(\ell',\ell'')}$ and a two--legged tail $t_{22}^{(\ell',\ell'')}$ with at least three external legs such that
$$\eqalign{
P\Big[ R_C\big(T_{\ell'}(K),T_{\ell''}(K)\big)
\big(\cR_{K,C}(\lw f\rw)\big) \Big]
&= P\Big[ R_C\big(T_{\ell'+1}(K),T_{\ell''+1}(K)\big)
(\lw f\rw) \Big] \cr
&\ \ \ + P\big[ R_C\big(t_{12}^{(\ell',\ell'')}+t_{22}^{(\ell',\ell'')})
(\lw f\rw)
+g_{2,\ell',\ell''}(\psi) \cr
}$$
with
$$\eqalign{
N_{\rm impr}(g_{2,\ell',\ell''}) 
&\le \sfrac{2^{6}\,\imp}{\al^{12}}\,
N(f\cl 2\al)\,N_K^2\ 
N(T_{\ell'}(K)\cl 2\al) \,N(T_{\ell''}(K)\cl 2\al) \cr
&\le\,\imp\, \sfrac{2^{6}}{\al^{2(\ell'+\ell'')+8}}\,N(f\cl 2\al)\,N_K^{\ell'+\ell''+2}
 \cr 
}$$
By Lemma \lemtailIII, there exists a two--legged tail $t_{23}$ with at least four external legs such that
$$
P\Big[R_C(t_1)\big(\cR_{K,C}(\lw f\rw)\big) \Big]
= P\Big[R_C(t_{23})(\lw f\rw) \Big]
$$
Finally, by Lemma \lemtailI
$$
P\Big[R_C(t_2)\big(\cR_{K,C}(\lw f\rw)\big) \Big]
= P\Big[R_C(t_2\circ K+t_{14}+t_{24})(\lw f\rw) \Big]
$$
where $t_{14},t_{24}$ are one-- resp. two--legged tails with at least three external legs.

Combining the results above, we see that
$$\eqalign{
P \big[ \cR_{K,C}^{\ell+1}&(\lw f\rw) \big] 
= f'_{\ell+1}(\psi)\cr
&+P\Big[ R_C\big(T_{\ell+1}(K)\big)(\lw f\rw) 
 +\sfrac{1}{2}\hskip-9pt
\smsum_{\ell',\ell''\ge 1\atop\max\{\ell',\ell''\}=\ell+1}
\hskip-9pt
 R_C\big(T_{\ell'}(K),T_{\ell''}(K)\big)
(\lw f\rw)
 +R_C(t_1'+t_2')(\lw f\rw) \Big] 
}$$ 
with one-- resp. two--legged tails $t_1',t_2'$ with at least three external legs, and
$$
f'_{\ell+1} = g_0 +g _1 +\half
\hskip-9pt\smsum_{\ell',\ell''\ge 1\atop\max\{\ell',\ell''\}=\ell}
\hskip-9pt g_{2,\ell',\ell''}
$$
By the triangle inequality
$$\eqalign{
N_{\rm impr}(f'_{\ell+1}) 
&\le N_{\rm impr}(g_0) + N_{\rm impr}(g_1) +\half
    \hskip-9pt\smsum_{\ell',\ell''\ge 1\atop\max\{\ell',\ell''\}=\ell}
\hskip-9pt
 N_{\rm impr}(g_{2,\ell',\ell''}) \cr
&\le \sfrac{2^{6}}{\al^{2\ell+8}}\,\imp\,\,N(f\cl 2\al)\,N_K^{\ell+2}
\Big[ \sfrac{1}{\big[1-{1\over \al^2}N_K\big]^{\ell}} + 1 
 +  \smsum_{\ell'=1}^\ell \sfrac{N_K^{\ell'}}{\al^{2\ell'}} \Big]\cr
&\le \sfrac{2^{6}}{\al^{2\ell+8}}\,\imp\,\,N(f\cl 2\al)\,N_K^{\ell+2}
\Big[ \sfrac{1}{\big[1-{1\over \al^2}N_K\big]^{\ell}}  
 +  \smsum_{\ell'=0}^\infty \big(\sfrac{N_K}{\al^2}\big)^{\ell'} \Big]\cr
&\le \sfrac{2^{7}}{\al^{2\ell+8}}\,\imp\,\,N(f\cl 2\al)\,
\sfrac{N_K^{\ell+2}}{\big[1-{1\over \al^2}N_K\big]^{\ell}}  \cr
}$$
\endproof

\corollary{\STM\corprodtails}{
Let $K(\psi;\xi,\xi',\eta)$ be an even Grassmann function that has degree at most two in the variables $\xi',\eta$,
degree at least one in $\et$ and degree at least four in the variables
$\psi,\xi,\xi',\eta$. Furthermore let $f(\psi;\xi)$ be a Grassmann function of degree at least four in the variables $\psi,\xi$. Set
$$
h(\psi) = P\Big[ \frac{1}{\bbbone - \cR_{K,C}}(\lw f\rw) 
     - \smsum_{\ell=0}^\infty R_C\big(T_\ell(K)\big)(\lw f\rw) 
      - \sfrac{1}{2} \smsum_{ \ell' , \ell''\ge 1}
              R_C\big(T_{\ell'}(K),T_{\ell''}(K)\big)(\lw f\rw) \Big]
$$
If $\al \ge 2$ and $N(K\cl 2\al)_\0 < \sfrac{\al^2}{2}$, then
$$
N_{\rm impr}(h)  
\le \sfrac{2^{7}\,\imp}{\al^8}\,N(f\cl 2\al)\,
\sfrac{N(K\cl 2\al)^2}{1-{2\over\al^2}N(K\cl 2\al)}
$$
}
\prf
By Proposition \propprodtails, 
$$\eqalign{
N_{\rm impr}(h) &\le \smsum_{\ell=1}^\infty N_{\rm impr}(f'_\ell) \cr
&\le \sfrac{2^{7}\,\imp}{\al^8}\,N(f\cl 2\al)\,N(K\cl 2\al)^2\,
\smsum_{\ell = 1}^\infty \Big[\sfrac{1}{\al^2}\sfrac{N(K\cl 2\al)}
{1-{1\over\al^2}N(K\cl 2\al)}\Big]^{\ell-1} \cr
&= \sfrac{2^{7}\,\imp}{\al^8}\,N(f\cl 2\al)\,N(K\cl 2\al)^2\,
\Big[1-\sfrac{1}{\al^2}\sfrac{N(K\cl 2\al)}
{1-{1\over\al^2}N(K\cl 2\al)}\Big]^{-1} \cr
&= \sfrac{2^{7}\,\imp}{\al^8}\,N(f\cl 2\al)\,N(K\cl 2\al)^2\,
\sfrac{{1-{1\over\al^2}N(K\cl 2\al)}}{1-{2\over\al^2}N(K\cl 2\al)} \cr
&\le \sfrac{2^{7}\,\imp}{\al^8}\,N(f\cl 2\al)\,
\sfrac{N(K\cl 2\al)^2}{1-{2\over\al^2}N(K\cl 2\al)} \cr
}$$ 
\endproof

\titlec{ Proof of Theorem \theoremVb\ in the case $D=0$}

Set
$K(\psi;\xi,\xi',\eta) = \hat U(\psi;\xi+\xi'+\eta) - \hat U(\psi;\xi+\xi')$.
By Theorem \theoremIII\ and Proposition \propcalR
$$
Pf' \ =\ P\cS_{U,C}(\lw \hat f\rw)\ 
=\ P\big[ \sfrac{1}{\bbbone - \cR_{K,C}}(\lw \hat f\rw) \Big]
$$
Observe that $K$ has degree at least one in $\et$ and four overall and
that, by part i of Remark \remfunctnorm\ (twice) 
$$
N(K\cl 2\al)_\0\le N(\hat U(\psi;\xi+\xi'+\eta)\cl 2\al)_\0
\le N(\hat U(\psi;\xi)\cl 8\al)_\0
< \sfrac{\al}{8}
$$
Decompose
$$
K(\psi;\xi,\xi',\eta) = K'(\psi;\xi,\xi',\eta) 
+ K^{\prime\prime} (\psi;\xi,\xi',\eta)
$$
where $K'$ has degree at most two in the variables $\xi',\eta$ and
$K^{\prime\prime}$ has degree at least three in the variables $\xi',\eta$.
By Corollary \corsplittwo\ and Corollary \corprodtails\ there exists a Grassmann function $g(\psi)$
with
$$\eqalign{
N_{\rm impr}(g) 
&\le \sfrac{2^{5}\,\imp}{\al^6}\,N\big(\hat f\cl 2\al\big)
\,N(K\cl 2\al)
\Big[\sfrac{1}{1-{6\over \al}N(K\cl 2\al)}
+\sfrac{{4\over \al^2}N(K\cl 2\al)}{1-{2\over \al^2}N(K\cl 2\al)}\Big]\cr
&\le \sfrac{2^{5}\,\imp}{\al^6}\,N\big(\hat f\cl 2\al\big)
\,N(K\cl 2\al)
\sfrac{1+{2\over \al}N(K\cl 2\al)}{1-{6\over \al}N(K\cl 2\al)}\cr
&\le \sfrac{2^{5}\,\imp}{\al^6}\,N\big(\hat f\cl 2\al\big)
\,N(K\cl 2\al)
\sfrac{1}{1-{6\over \al}N(K\cl 2\al)}\sfrac{1}{1-{2\over \al}N(K\cl 2\al)}\cr
&\le \sfrac{2^{5}\,\imp}{\al^6}\,N\big(\hat f\cl 2\al\big)
\,N(K\cl 2\al)
\sfrac{1}{1-{8\over \al}N(K\cl 2\al)}\cr
&\le \sfrac{2^{5}\,\imp}{\al^6}\,N\big(\hat f\cl 2\al\big)
\,\sfrac{N(\hat U\cl 8\al)}{1-{8\over \al}N(\hat U\cl 8\al)}\cr
}$$
and
$$
Pf'=P \Big[ \hat f(\psi,0) + 
\smsum_{\ell=1}^\infty R_C\big(T_\ell(K)\big)(\lw \hat f\rw) 
      + \sfrac{1}{2} \smsum_{\ell, \ell'}
              R_C\big(T_\ell(K),T_{\ell'}(K)\big)(\lw \hat f\rw) \Big] \ +\ g
$$
We have used that $P\lw \hat f(\psi,\xi)\rw=P\hat f(\psi,0)$ and
$T_\ell(K')=T_\ell(K)$.
Since $\hat f$ and $K$ have degree at least four overall,
$$
P\,R_C\big(T_\ell(K)\big)(\lw \hat f\rw) = T_\ell(K) \circ T_1(\hat f)
$$
(where the $\circ$ composition was defined in Definition \defladders.ii)
and by Remark \remformladder
$$
P\big[ R_C\big(T_\ell(K),T_{\ell'}(K)\big)(\lw \hat f\rw) \big] = 
T_\ell(K) \circ {\rm Rung}(\hat f) \circ T_{\ell'}(K)
$$
Furthermore, by part (ii) of Remark \remeffecttail, $T_\ell(K)=T_\ell(\hat U)$. Thus
$$
Pf'=P \hat f(\psi,0) + 
\smsum_{\ell=1}^\infty T_\ell(\hat U) \circ T_1(\hat f) 
      + \sfrac{1}{2} \smsum_{\ell, \ell'\ge 1}
             T_\ell(\hat U) \circ {\rm Rung}(\hat f) \circ T_{\ell'}(\hat U)
 \ +\ g
$$
\endproof

\vfill\eject

\chap{ The Enlarged Algebra}\PG\pgRVIII
 
The estimate of Theorem \theoremIVa\ on $W'(\psi)$, defined by
$$
\lw W'(\psi)\rw_{\psi,D}\ =\ \Om_C(\lw W\rw_{\psi,C+D})
$$
was proven in the following way. We applied the results of 
Theorem \theorII, combined with the estimates on Wick ordering (Corollary \corwicknorm) to get estimates on 
$$
W^{\prime\prime}(\psi)=\Om_C(\lw W\rw_{\psi,C+D}) = \Om_C(\lW\lw W\rw_{\psi,D}\rW_{\psi,C})
$$ 
in terms of the norm of $W$. Then we used Corollary \corwicknorm\ again to estimate the norm of $W'$ in terms of the norm of $W^{\prime\prime}$. The 
transition from $W^{\prime\prime}$ to $W'=\lw W^{\prime\prime}\rw_{\psi,-D}$ 
creates new two and four legged
vertices, whose improved norm cannot be estimated by the technique of Section VI. This transition also creates new ladder diagrams.

The same difficulty would occur if we tried to reduce the general case of Theorem \theoremVb\ to the special case $D=0$ by Wick ordering $\cS_{U,C}(f)$
at the end of the construction. Therefore we monitor the Wick ordering with respect to $D$ throughout the construction. To do this we introduce fields $\ze,\ze',\varphi$ for these Wick contractions and
 use an analogue of the operator $\cR_{K,C}$ of Definition \defcalR.

\definition{\STM\defQ}{
\Item{(i)}
Let 
$\bar K_1(\psi;\ze,\ze',\varphi;\xi,\xi',\eta),\cdots,
 \bar K_\ell(\psi;\ze,\ze',\varphi;\xi,\xi',\eta)$ be even Grassmann functions with
$\bar K_i(\psi;\ze,\ze',\varphi;\xi,\xi',0)=0$.
For a Grassmann function $g(\psi;\ze;\xi)$ we define
$$\eqalign{
Q(\bar K_1,\cdots,\bar K_\ell)(g)(\psi;\ze;\xi) 
=  \ \lww  \int \hskip -6pt\int  \lW\big( \smprod_{i=1}^\ell &
\lW\bar K_i(\psi;\ze,\ze',\varphi;\xi,\xi',\eta)\rW_{\ze',D\atop \xi',C} 
\big)\rW_{\varphi,D\atop \eta,C} \cr
&\lw g(\psi;\ze+\varphi;\eta)\rw_{\varphi,D}\,
\,d\mu_{D}(\ze',\varphi)\,d\mu_{C}(\xi',\eta)\ \rww_{\xi,C}  \cr
}$$

\noindent
The structure of $Q$ is illustrated in the figure

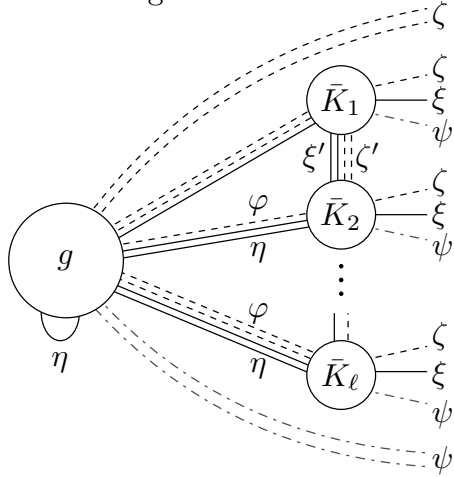
\centerline{\figput{figVII1}}

\Item{(ii)}
For a Grassmann function $\bar K(\psi;\ze,\ze',\varphi;\xi,\xi',\eta)$  with
$\bar K(\psi;\ze,\ze',\varphi;\xi,\xi',0)=0$ we
define the operator $\cQ_{\bar K}$ by
$$
\cQ_{\bar K}(g) =\ \smsum_{\ell=1}^\infty \sfrac{1}{\ell !}\, 
Q(\bar K,\cdots,\bar K)(g)
$$
}

In this definition, the fields $\xi,\xi',\eta$ involving the covariance $C$ are treated in the same way as in Section III. The fields $\varphi$ are analogous to the fields $\eta$ and describe Wick contractions between $g$ and the $K_i$.
The fields $\ze'$ are analogous to the fields $\xi'$ and describe Wick contractions among $K_i$. Similar to the $\xi$ fields, the fields $\ze$ are not integrated out and are later used for Wick contractions in further applications of $\cQ_{\bar K}$. In contrast to the $\xi$--fields, however, also Wick contractions between fields of $g$ and kernels appearing in future applications of $\cQ_{\bar K}$ have to be allowed. This is the reason for the field $\ze$ 
that appears in the term $g(\psi;\ze+\varphi;\eta)\rw_{\varphi,D}$ 
in Definition \defQ.i.

The Definition \defQ.i\ of $Q$ was chosen so that
$$
\lw Q(\bar K_1,\cdots,\bar K_\ell)(g)\rw_{\ze,D}
=R(\lw K_1\rw_{\ze,D},\cdots,\lw K_\ell\rw_{\ze,D})(\lw g\rw_{\ze,D})
$$
when 
$
\bar K_i(\psi;\ze,\ze',\varphi;\xi,\xi',\eta)
= K_i(\psi;\ze+\ze'+\varphi;\xi,\xi',\eta)
$. 
This formula is proven in Lemma \:\lemQWick.i.\footnote{$^{(1)}$}{To see this, choose  
$f= \lw g\rw_{\ze,D}$. } Consequently
$$
\lw \cQ_{\bar K}^n(h)\rw_{\ze,D}=\cR_{\lw K\rw_{\ze,D}}^n(\lw h\rw_{\ze,D})
$$
for $n=0,\ 1,\ 2, \cdots$. From this one can deduce, as in Proposition 
\:\propalgtwoWick, that
$$
\cS_{U,C}(\lw f\rw_{\psi,D}) 
= \lww\ \int \sum_{n=0}^\infty \cQ^n_{\bar K}(h)(\psi;0;\xi)\ d\mu_{C}(\xi)\ \rww_{\psi,D}
$$
where, given an even Grassmann function $\hat U(\psi;\xi)$ and a Grassmann
function $f(\psi;\xi)$,
$$\eqalign{
U(\psi;\xi) &= \lW \hat U(\psi;\xi)\rW_{\xi,C\atop\psi,D}\cr
h(\psi;\ze;\xi)&= f(\psi+\ze;\xi)\cr
\bar K(\psi;\ze,\ze',\varphi;\xi,\xi',\eta) 
&=\hat U(\psi+\ze+\ze'+\varphi;\xi+\xi'+\eta) - 
\hat U(\psi+\ze+\ze'+\varphi;\xi+\xi') \cr
}$$
However, this formula would not be good enough to get an estimate on
$\cS_{U,C}(\lw f\rw_{C+D})$. At each application of $\cQ_{\bar K}$, the passage from 
$g(\psi;\ze;\xi)$ to $g(\psi;\ze+\varphi;\eta)$, that is the separation between the $D$- fields that are to be Wick contracted at the present step and those that are to be Wick contracted at a future step, leads to a deterioration in the norm:
$$
N\big(g(\psi;\ze+\varphi;\eta)\cl \al\big) \le 
N\big(g(\psi;\ze;\xi)\cl 2\al\big)
$$
by Remark \remfunctnorm. In iterated applications of $\cQ_{\bar K}$ this deterioration in the norm would build up excessively. For this reason, we avoid Wick contractions at most steps of the construction and perform them only before and
during steps in which overlapping loops are exploited. For bookkeeping of
the partially Wick ordered fields in intermediate steps we introduce an enlarged  algebra.

\sect{Definition of the enlarged algebra}\PG\pgRVIIIa

\definition{\STM\defsymsupalg}{
\Item{(i)} A $\bbbz_2$--graded vector space is a complex vector space $E$, together with a decomposition $E=E_+\oplus E_-$. The elements of $E_+$ are called even, the elements of $E_-$ odd. A graded vector space is a complex vector space $E$, together with a decomposition $E=\bigoplus_{m=0}^\infty E_m$.
Every graded vector space is considered as a $\bbbz_2$--graded vector space with
$$
E_+ = \bigoplus\limits_{r\ {\rm even}} E_m
\qquad\qquad\qquad
E_- = \bigoplus\limits_{r\ {\rm odd}}  E_m
$$
\Item{(ii)} If $E$ is a ($\bbbz_2$--) graded vector space, the tensor algebra
$T(E)$ has a natural ($\bbbz_2$--) grading. The symmetric superalgebra over $E$
is denoted $S(E)$ and is defined as the quotient of $T(E)$ by the two sided ideal $I(E)$ generated by
$$\deqalign{
& a \otimes b - b \otimes a    \qquad
                     &{\rm with}\ a\in E_+\ {\rm or}\ b \in E_+ \cr
& a \otimes b + b \otimes a   &{\rm with}\ a,b \in E_- \cr
}$$ 
It is a superalgebra, and it is a graded superalgebra if $E$ is a graded vector space (see [BS]).
}

\example{\STM\exsymsupalg}{ 
Let $E$ be a complex vector space. Setting $E_+=\{0\},\ E_-=E$, we give $E$ the structure of a $\bbbz_2$--graded vector space in which all elements are odd. Then the symmetric superalgebra over $E$ is the Grassmann algebra 
$\bigwedge E$ over $E$. Setting $E_+=E,\ E_-=\{0\}$, $S(E)$ is the classical
symmetric algebra $\cS(E)$ over $E$.
} 

\remark{\STM\remsupersub}{ There is a natural isomorphism $\iota$ between
$S(E)$ and the algebra $\cS(E_+)\otimes\bigwedge E_-$, constructed in the following way:
{\parindent=.25in
\item{} Observe that
$$
T(E) \cong 
\bigoplus_{m_1,n_1,\cdots,m_r,n_r\ge 0\atop n_1,m_2,\cdots,n_{r-1},m_r\ge 1}
E_+^{\otimes m_1}\otimes E_-^{\otimes n_1}\otimes\cdots\otimes
E_+^{\otimes m_r}\otimes E_-^{\otimes n_r}
$$
Define the algebra homomorphism $\iota':T(E)\rightarrow \cS(E_+)\otimes\bigwedge E_-$ by
$$
\iota'\Big(v_1^+\otimes v_1^-\otimes\cdots\otimes v_r^+\otimes v_r^-\Big)
=\big(v_1^+\cdot\ldots\cdot v_r^+\big)\otimes\big(v_1^-\cdot\ldots\cdot v_r^-\big)
$$
for $v_i^+\in E_+^{\otimes m_i},\ v_i^-\in E_-^{\otimes m_i}$. Clearly $I(E)$ lies in the kernel of $\iota'$, so $\iota'$ induces an algebra homomorphism $\iota:S(E)\rightarrow \cS(E_+)\otimes\bigwedge E_-$. One can check that $\iota$ is an isomorphism.

}
\noindent
As $\cS(E)$ and $\bigwedge E$ are subalgebras of the tensor algebra $T(E)$, $S(E)$ can also be viewed as a subalgebra of $T(E)\otimes T(E)\cong T(E)$.

}

\definition{\STM\defenlalg}{
\Item{(i)} Let $E$ be a complex vector space. The symmetric superalgebra $S(\bigwedge E)$ over $\bigwedge E$ (considered as a graded vector space) is called the enlarged algebra $\susywedge  E$ over $E$. It is a graded superalgebra.
\Item{(ii)} Multiplication 
$$
f_1 \otimes \cdots \otimes f_r \longmapsto f_1\cdots f_r
$$
defines an algebra homomorphism from the tensor algebra $T(\bigwedge E)$ to $\bigwedge E$. The ideal $I(\bigwedge E)$ of part (ii) of Definition \defsymsupalg\
 lies in the kernel of this homomorphism. Therefore it induces an algebra homomorphism
$$
{\rm Ev}:\ \susywedge E \longrightarrow \bigwedge E
$$
called the evaluation map. It is graded, that is
$$
{\rm Ev}(x) \in \bigwedge\nolimits^m E \ \ {\rm when} \ 
x\in \big(\susywedge E\big)_m
$$
\Item{(ii)} If $A$ is a superalgebra, we define the enlarged algebra over $E$
with coefficients in $A$ as the tensor product
$$
\susywedge\nolimits_A E = A \otimes \susywedge E
$$
in the sense of Definition \defsuperalgebra.iv. The evaluation map extends by $A$-linearity to an algebra homomorphism
$$
{\rm Ev}:\ \susywedge\nolimits_AE \longrightarrow \bigwedge\nolimits_A E
$$
} 

\remark{\STM\remEv}{
\Item{(i)} $\bigwedge E$ is identified with the subspace $(\susywedge V)_1$ of $\susywedge V$.
\Item{(ii)} ${\rm Ev}\big(x\cdot y\big) = {\rm Ev}\big(x\cdot {\rm Ev}(y)\big) =
{\rm Ev}\big({\rm Ev}(x)\cdot y\big)$ for all $x,y\in \susywedge V$.
\Item{(iii)} By Remark \remsupersub, there is a natural inclusion of
$\susywedge E=S\big(\bigwedge E\big)$ as a graded subalgebra of
$T\big(\bigwedge E\big)\subset T\big(T(E)\big)\cong T(E)$.
 
}

\sect{Norm estimates for the enlarged algebra}\PG\pgRVIIIb

 As in subsection III.3, let $A'=\bigwedge_A V'$ be the Grassmann algebra in the variables $\psi_i$ with coefficients in $A$. Furthermore let $E$ be a copy of $V$ with generators $\ze_i$ corresponding to the fields $\psi_i$. We will use the enlarged algebra 
$$
\cA = \susywedge\nolimits_{A'}E
$$
over $E$ with coefficients in $A'$. Elements of $\cA$ will be written as Grassmann functions $f(\psi;\vec \ze)$. 
They are linear combinations of monomials of the form
 $$
 \psi_{i_1}\cdots \psi_{i_n} \ (\ze_{j^{(1)}_1}\cdots\ze_{j^{(1)}_{p_1}})
 \otimes (\ze_{j^{(2)}_1}\cdots\ze_{j^{(2)}_{p_2}}) \otimes\cdots\otimes
 (\ze_{j^{(r)}_1}\cdots\ze_{j^{(r)}_{p_r}})
 $$
The evaluation map ${\rm Ev}:\cA\rightarrow \bigwedge_{A'} E \cong 
\bigwedge_A(V\oplus E)$ maps such a monomial to 
$\ \psi_{i_1}\cdots \psi_{i_n} \ \ze_{j^{(1)}_1}\cdots\ze_{j^{(1)}_{p_1}}\,
 \ze_{j^{(2)}_1}\cdots\ze_{j^{(r)}_{p_r}}\ $.
As in Remark \remEv, the Grassmann algebra
$\bigwedge_{A'}E$ of Grassmann functions $g(\psi;\ze)$ is viewed as a subspace of $\cA$.

The evaluation map ${\rm Ev}:\cA\rightarrow \bigwedge_{A'} E \cong \bigwedge_A(V\oplus E)$ extends to a map
$$
{\rm Ev}:\bigwedge\nolimits_\cA (V^{(1)}\oplus \cdots\oplus V^{(r)})
\longrightarrow 
\bigwedge\nolimits_{A'} (E\oplus V^{(1)}\oplus \cdots\oplus V^{(r)})
\cong \bigwedge_A(V\oplus E\oplus V^{(1)}\oplus \cdots\oplus V^{(r)})
$$
where $V^{(j)},\,j=1,\cdots, r$ are copies of $V$ with generators
$\xi^{(j)}$.

\definition{\STM\defPi}{
By Remark \remEv.iii, $\cA$ can be viewed as a graded subalgebra of $A'\otimes
T(E)\cong A'\otimes T(V)$. Therefore the family $\|\ \cdot\ \|'$ of seminorms on the spaces $A'_m\otimes V^{\otimes n}$, introduced in subsection III.3, 
induces a 
family of symmetric seminorms on the spaces $\cA_{m'}\otimes V^{\otimes n'}$,
which we again denote by $\|\ \cdot\ \|'$.
This Definition extends the Definition of $\|\,\cdot\,\|'$ given in subsection III.3. Also, $\cb$ is a
contraction bound and $\ib$ an integral bound for the covariance $C$
 with respect to these norms.

}

\lemma{\STM\lemestEv}{
Let $f\in \bigwedge_\cA(V_1\oplus\cdots\oplus V_r)$. Then
$$
N'\big( {\rm Ev}(f)\cl \al\big) \le N'(f\cl \al) 
$$
}
\prf ${\rm Ev}(f)$ is obtained from $f$ by antisymmetrization.
\endproof

\sect{ Schwinger Functionals over the Extended Algebra}\PG\pgRVIIIc

Recall that for any even $U(\psi;\xi)\in \bigwedge_{A'}V$, the Schwinger functional
with respect to $U$ and $C$ is the map from $\bigwedge_{A'}V$ to $A'$ given by
$$
\cS_{U,C}(f)(\psi) = \sfrac{1}{Z} \int e^{U(\psi;\xi)} f(\psi;\xi)\  d\mu_C(\xi)
\qquad\qquad
Z= \int e^{U(\psi;\xi)}\  d\mu_C(\xi)
$$
Also recall that for an even Grassmann function $K(\psi,\vec\ze;\xi,\xi',\eta)$
in the variables $\xi,\xi',\eta$ over the extended algebra $\cA$ and any Grassmann function $f(\psi;\vec\ze;\xi)\in \bigwedge_\cA V$
$$
\cR_{K,C}(f) = \lww \int \!\!\int \lW 
e^{\lw K(\psi,\vec\ze;\xi,\xi',\eta)\rw_{\xi',C}} -1 \rW_{\eta,C} \,f(\psi;\vec\ze;\eta)\,d\mu_C(\xi')\,d\mu_C(\eta) \rww_{\xi,C}
$$

\proposition{\STM\propalgtwoWick}{
Let 
$$
U(\psi;\xi) = \ \lW \hat U(\psi;\xi)\rW_{\psi,D\atop\xi,C}
$$
be even and set 
$$
K(\psi,\ze;\xi,\xi',\eta) = 
\lw \hat U(\psi+\ze;\xi+\xi'+\eta)\rw_{\ze,D}- \lw \hat U(\psi+\ze;\xi+\xi')\rw_{\ze,D}
$$
Let $f(\psi;\xi)\in \bigwedge_{A'}V$ and set 
$\tilde f(\psi;\ze;\xi)=\lw f(\psi+\ze;\xi)\rw_{\ze,D}\,\in \bigwedge_\cA V$. Then
$$\eqalign{
\cS_{U,C}\big(\lw f\rw_{\psi,D}\big)(\psi) 
&= \ \lww\int\!\!\int {\rm Ev} \Big(
 \sfrac{1}{\bbbone - \cR_{K,C}}\big(\tilde f\big)\Big)\, d\mu_C(\xi)\, d\mu_D(\ze)\rww_{\psi,D}\cr
}$$
Observe that
$ \sfrac{1}{\bbbone - \cR_{K,C}}(\tilde f) \in \bigwedge_\cA V$,
so that $ \int \sfrac{1}{\bbbone - \cR_{K,C}}(\tilde f)\, d\mu_C(\xi) \in \cA$.
}

For the proof of this Proposition we use

\lemma{\STM\lemalgtwoWick}{
 Let $f(\psi;\xi)$ be a Grassmann function and set 
$$
\tilde f(\psi;\ze;\xi) = \lw f(\psi+\ze;\xi)\rw_{\ze,D}
$$
Furthermore, let $n\ge 1$.

\Item (i)
Let  $\ell_1,\cdots,\ell_n \ge 1$ and let
$$
\hat K_{ji}(\psi;\xi,\xi',\eta), \qquad\qquad j=1,\cdots n,\ \ i=1,\cdots,\ell_j
$$
be even Grassmann functions with coefficients in $A$. Let
$$
K_{ji}(\psi;\ze ;\xi,\xi',\eta) = \lw \hat K_{ji}(\psi+\ze;\xi,\xi',\eta)\rw_{\ze,D}
$$
considered as Grassmann functions in the variables $\xi,\xi',\eta$ with coefficients in the enlarged algebra $\cA$.  Then
$$\eqalign{
\Big( \smprod_{j=1}^n R_C
\big(\lw \hat K_{j1}\rw_{\psi,D},\cdots,\lw \hat K_{j\ell_j}\rw_{\psi,D}&\big)
\Big) \big(\lw f\rw_{\psi,D}\big)  \cr 
&= \ \lww \int {\rm Ev} \Big(
\smprod_{j=1}^n R_C \big( K_{j1},\cdots,K_{j\ell_j} \big)(\tilde f) \Big) \,d\mu_D(\ze)\rww_{\psi,D} \cr
}$$
On the left hand side of the equation above, the operator $R_C$ (defined in (\eqRCdef) ) is considered over the algebra $A'$, while on the right hand side it is considered over the extended algebra $\cA$.

\Item (ii) 
Let
$\ 
\hat K(\psi;\xi,\xi',\eta)
\ $
be an even Grassmann function and
$$ 
K(\psi;\ze ;\xi,\xi',\eta) = \lw \hat K(\psi+\ze;\xi,\xi',\eta)\rw_{\ze,D}
$$
 Then
$$ 
\cR^n_{\lw \hat K\rw_{\psi,D},C} \big(\lw f\rw_{\psi,D}\big) 
= \ \lww \int {\rm Ev}\  \cR^n_{ K,C}(\tilde f) \,d\mu_D(\ze)\rww_{\psi,D} 
$$
}

\prf (i) Let
$$
h(\psi;\xi)=\lw g(\psi;\xi)\rw_{\psi,D}\ =\ \Big( \smprod_{j=1}^n 
R_C\big(\lw \hat K_{j1}\rw_{\psi,D},\cdots,\lw \hat K_{j\ell_j}\rw_{\psi,D}\big)
\Big) \big(\lw f\rw_{\psi,D}\big)
$$
By part (ii) of Proposition \:\propBII, in the Grassmann algebra over $A$ with variables
$\psi,\ze,\xi,\xi',\eta$
$$
h(\psi+\ze;\xi)=\lw g(\psi+\ze;\xi)\rw_{\ze,D}\ =\ \Big( \smprod_{j=1}^n 
R_C\big(K_{j1},\cdots,K_{j\ell_j}\big) \Big) \big(\lw \tilde f\rw_{\ze,D}\big)
$$
Hence, by the construction of $\cA$ and ${\rm Ev}$,
$$
h(\psi+\ze;\xi)=\lw g(\psi+\ze;\xi)\rw_{\ze,D}\ =\ {\rm Ev}\Big[\Big( \smprod_{j=1}^n 
R_C\big(K_{j1},\cdots,K_{j\ell_j}\big) \Big) \big(\lw \tilde f\rw_{\ze,D}\big)\big]
$$
Therefore
$$\eqalign{
g(\psi;\xi) &= \int \lw g(\psi+\ze;\xi)\rw_{\ze,D}\,d\mu_D(\ze) \cr
&= \int {\rm Ev} \Big(
\smprod_{j=1}^n R_C \big( K_{j1},\cdots,K_{j\ell_j} \big)(\tilde f) \Big) \,d\mu_D(\ze)
}$$
\Item (ii) follows from part (i) and Remark \remcalR.
\endproof

\proof{ of Proposition \propalgtwoWick}
Set
$$
\lw\hat K(\psi;\xi,\xi',\eta)\rw_{\psi,D} = 
\lw \hat U(\psi;\xi+\xi'+\eta)\rw_{\psi,D} - \lw \hat U(\psi;\xi+\xi')\rw_{\psi,D}
$$
By Proposition \propcalR, with $\hat U(\xi)$ replaced by 
$\lw\hat U(\psi;\xi)\rw_{\psi,D}$ and $K(\xi,\xi',\et)$ replaced by 
$\lw\hat K(\psi;\xi,\xi',\et)\rw_{\psi,D}$,
 and part ii of Lemma \lemalgtwoWick
$$\eqalign{
R_{U,C}^n\big(\lw f\rw_{\psi,D}\big)
&= \ \cR^n_{\lw \hat K\rw_{\psi,D},C} \big(\lw f\rw_{\psi,D}\big) \cr
&= \ \lww \int {\rm Ev}\,\big(\cR_{K,C}^n\big)(\tilde f)  
\,d\mu_D(\ze)\rww_{\psi,D} \cr
}$$
By Theorem \theoremIII
$$\eqalign{
\cS_{U,C}\big(\lw f\rw_{\psi,D}\big)(\psi) 
&= \int \smsum_{n\ge 0}\, R_{U,C}^n\big(\lw f\rw_{\psi,D}\big)\ d\mu_C(\xi) \cr
&=\ \lww\int {\rm Ev} \Big( \int \smsum_{n\ge 0}\,
  \cR_{K,C}^n(\tilde f)\, d\mu_C(\xi) \Big)\,d\mu_D(\ze) \rww_{\psi,D} \cr
&= \ \lww\int {\rm Ev} \Big(
\int  \sfrac{1}{\bbbone - \cR_{K,C}}\big(\tilde f\big)\, d\mu_C(\xi) \Big)\,d\mu_D(\ze) \rww_{\psi,D}
}$$
\endproof

We use the extended algebra to give

\sect{ A second proof of Theorem \theoremIVa}\PG\pgRVIIId

\proposition{\STM\propesttwoWick}{
Let $\hat U(\psi;\xi), \hat f(\psi;\xi) \in \bigwedge_{A'} V$ with $\hat U$ even. Set
$$\eqalign{
U(\psi;\xi) &= \ \lW \hat U(\psi;\xi)\rW_{\psi,D\atop\xi,C}  \cr
f(\psi;\xi) &= \ \lW \hat f(\psi;\xi)\rW_{\psi,D\atop\xi,C}  \cr
}$$
Assume that $\cb$ is a contraction bound for the covariance $C$ and 
$\ib$ is an integral bound for $C$ and for $D$ and that
$N({\hat U}\cl 8\al)_\0 < \sfrac{\al^2}{4}$. Then $\cS_{U,C}(f)$ exists. 
If
$$
\cS_{U,C}(f) = \ \lw f'(\psi)\rw_{\psi,D}
$$
then
$$
N\big(f'(\psi)-\hat f(\psi;0)\cl \al \big) \le
\sfrac{2}{\al^2}\,N(\hat f\cl 4\al) \,
\sfrac{N({\hat U}\cl 8\al)}{1-{4\over\al^2}N({\hat U}\cl 8\al)} 
$$
}

\prf As above, set
$$\eqalign{
K(\psi;\ze;\xi,\xi',\eta) &= 
\lw \hat U(\psi+\ze;\xi+\xi'+\eta)\rw_{\ze,D} - \lw \hat U(\psi+\ze;\xi+\xi')\rw_{\ze,D} \cr
\tilde f(\psi;\ze;\xi) &= \lw \hat f(\psi+\ze;\xi)\rw_{\ze,D} \cr
}$$
By Remark \remfunctnorm\ and Corollary \corwicknorm
$$\eqalign{
N(K) &\ \le \ N({\hat U}\cl 8\al) \cr
N(\tilde f) &\ \le\ \,N(\hat f\cl 4\al) \cr
}$$
By  Corollary \cornormestcalR, applied with $A$ replaced by $\cA$
$$
\sfrac{1}{\bbbone -\cR_{K,C}}(\lw \tilde f\rw_{\xi,C} ) - \lw \tilde f\rw_{\xi,C}
\ =\ \lw g(\psi,\vec\ze;\xi)\rw_{\xi,C}
$$
with 
$$
N'(g) \le \sfrac{2}{\al^2}\,N'(\tilde f)\,\sfrac{N'( K)}{1-{4\over\al^2}N'(K)} 
=\sfrac{2}{\al^2}\,N(\tilde f)\,\sfrac{N( K)}{1-{4\over\al^2}N(K)} 
\le \sfrac{2}{\al^2}\,N(\hat f\cl 4\al) \,
\sfrac{N({\hat U}\cl 8\al)}{1-{4\over\al^2}N({\hat U}\cl 8\al)}
$$
Consequently, by Lemma \lemwicknorm,
$$\eqalign{
N' \bigg[ \int \Big( \frac{1}{\bbbone -\cR_{K,C}}(\lw \tilde f\rw_{\xi,C} ) - 
\lw \tilde f\rw_{\xi,C} \Big) d\mu_C(\xi) \bigg]
&= N'\Big(\int \lw g(\psi;\vec\ze;\xi)\rw_{\xi,C} \, d\mu_C(\xi) \Big) \cr
&\le N'\big(g(\psi;\vec\ze;\xi)\big) \cr
&\le \sfrac{2}{\al^2}\,N(\hat f\cl 4\al) \,
\sfrac{N({\hat U}\cl 8\al)}{1-{4\over\al^2}N({\hat U}\cl 8\al)}
}$$
Observe that
$$
\int \lw \tilde f\rw_{\xi,C}\,d\mu_C(\xi) = \tilde f(\psi;\ze;0) = 
\lw \hat f(\psi+\ze;0)\rw_{\ze,D}
$$
Hence, by Proposition \propalgtwoWick, with $f(\psi;\xi)$ replaced by 
$\lw\hat f(\psi;\xi)\rw_{\xi,C}$, Lemma \lemwicknorm\ and Lemma \lemestEv
$$\eqalign{
N\big(f'-\hat f(\psi,0)\big) 
&=N'\big(f'-\hat f(\psi,0)\big)\cr 
&= N'\bigg[ \int {\rm Ev}
\Big( \int \sfrac{1}{\bbbone -\cR_{K,C}}(\lw \tilde f\rw_{\xi,C} )\, d\mu_C(\xi) -\lw \hat f(\psi+\ze,0)\rw_{\ze,D} \Big)\,d\mu_D(\ze) \bigg] \cr 
&\le N'\bigg[ {\rm Ev}
\Big( \int \sfrac{1}{\bbbone -\cR_{K,C}}(\lw \tilde f\rw_{\xi,C} )\, d\mu_C(\xi) -\lw \hat f(\psi+\ze,0)\rw_{\ze,D} \Big) \bigg] \cr 
&\le N'\bigg[ \int \Big( \sfrac{1}{\bbbone -\cR_{K,C}}(\lw \tilde f\rw_{\xi,C} ) 
- \lw \tilde f\rw_{\xi,C} \Big)\,d\mu_C(\xi) \bigg] \cr
&\le \sfrac{2}{\al^2}\,N(\hat f\cl 4\al) \,
\sfrac{N({\hat U}\cl 8\al)}{1-{4\over\al^2}N({\hat U}\cl 8\al)}\cr
}$$
\endproof

\proof{of Theorem \theoremIVa, again}
By Remark \remrenschw.i, with $W$
replaced by $\lw W\rw_{C+D}$
$$\eqalign{
\lw W'(\psi)-W(\psi)\rw_{\psi,D}&=
\Om_C(\lw W\rw_{C+D})(\psi) -\lw W(\psi)\rw_{\psi,D} \cr 
&=\int_0^1 \big( \cS_{t U, C}( U ) 
  - \lw W(\psi)\rw_{\psi,D} \big)\,dt\qquad{\rm mod}\,A_0\cr
}$$
where
$$\
U(\psi;\xi) = \lW W(\psi+\xi)\rW_{\xi,C\atop\psi,D}
$$
Define $f'_t$ and $\hat f$ by
$$\eqalign{
\hat f(\psi;\xi) &= W(\psi+\xi)\cr
 \lw f'_t(\psi)\rw_{\psi,D} &= \cS_{t U, C}( U ) \cr
}$$
Then
$$
W'(\psi)-W(\psi)
=\int_0^1 \big(f'_t(\psi)-\hat f(\psi,0) \big)\,dt
\qquad{\rm mod}\,A_0
$$
By Proposition \propesttwoWick, with $\hat U_t(\psi;\xi)=tW(\psi+\xi)$,
$$\eqalign{
N(W'-W\cl \al)
&\le \max_{0\le t\le 1}N(f_t'(\psi)-\hat f(\psi,0)\cl \al)\cr
&\le \sfrac{2}{\al^2}\max_{0\le t\le 1}
\,N(\hat f\cl 4\al) \,
\sfrac{N({\hat U_t}\cl 8\al)}{1-{4\over\al^2}N({\hat U_t}\cl 8\al)}\cr
&\le \sfrac{2}{\al^2}\,
\sfrac{N(W\cl 16\al)^2}{1-{4\over\al^2}N(W\cl 16\al)}\cr
}$$
for all $W$ with $N(W\cl 16\al)_\0<\sfrac{\al^2}{4}$. 
Replacing $\al$ by $2\al$ gives
$$
N(W'-W\cl 2\al)
\le \sfrac{1}{2\al^2}\,\sfrac{N(W\cl 32\al)^2}{1-{1\over\al^2}N(W\cl 32\al)}
$$
for all  $W$ with $N(W\cl 32\al)_\0< \al^2$. 
\endproof

\goodbreak
\sect{ The Operator $Q$}\PG\pgRVIIIe


\lemma{\STM\lemQWick}{
Let $f(\psi,\vec \ze;\xi)$ be a Grassmann function over the extended algebra $\cA$ and set 
$\lw f'(\psi,\ze ; \xi)\rw_{\ze,D} = {\rm Ev}(f)$.
\Item{(i)} Let  
$ K_i(\psi,\ze;\xi,\xi',\eta),\ i=1,\cdots,\ell$ be Grassmann functions with
$ K_i(\psi,\ze;\xi,\xi',0)=0$. Set 
$$
\bar K_i(\psi;\ze,\ze',\varphi;\xi,\xi',\eta)
= K_i(\psi;\ze+\ze'+\varphi;\xi,\xi',\eta)
$$
 Then
$$
{\rm Ev}\ R_C(\lw K_1\rw_{\ze,D},\cdots,\lw K_\ell\rw_{\ze,D})(f)
= \lW Q(\bar K_1,\cdots,\bar K_\ell)(f')\rW_{\ze,D}
$$
\Item{(ii)} Let  
$ K(\psi,\ze;\xi,\xi',\eta)$ be a Grassmann functions with
$ K(\psi,\ze;\xi,\xi',0)=0$. Set 
$$
\bar K(\psi;\ze,\ze',\varphi;\xi,\xi',\eta)
= K(\psi;\ze+\ze'+\varphi;\xi,\xi',\eta)
$$
Then, when $\cR_{\lw K\rw_{\ze,D},C}$ is considered as an operator over $\cA$,
$$
{\rm Ev}\Big( \cR_{\lw K\rw_{\ze,D},C}(f)\Big)
 = \lW\cQ_{\bar K}(f')\rW_{\ze,D}
$$
}
\prf
Applying Remark \remEv\  and Corollary \:\corBIII\ to the variable $\ze$ we see that
$$\eqalign{
{\rm Ev}& \big( \lw K_1(\psi,\ze;\xi,\xi',\eta)\rw_{\ze,D} \cdot\cdots\cdot 
  \lw K_\ell(\psi,\ze;\xi,\xi',\eta)\rw_{\ze,D} \cdot f(\psi,\vec\ze;\eta)\big)\cr 
&= {\rm Ev} \big( \lw K_1\rw_{\ze,D} \cdot\cdots\cdot 
  \lw K_\ell\rw_{\ze,D} \cdot {\rm Ev} f(\psi,\vec\ze;\eta)\big) \cr
&= {\rm Ev} \big( \lw K_1\rw_{\ze,D} \cdot\cdots\cdot 
  \lw K_\ell\rw_{\ze,D} \cdot \lw f'\rw_{\ze,D}\big) \cr
&= \lww \int\lW\Big( \smprod_{i=1}^\ell \lw K_i(\psi,\ze+\ze'+\varphi;\xi,\xi',\eta)\rw_{\ze',D}\Big)\rW_{\varphi,D}
\ \lw f'(\psi,\ze+\varphi;\eta)\rw_{\varphi,D}
  \,d\mu_D(\ze',\varphi)\rww_{\ze,D} \cr
}$$
Applying each of the following operations to both sides of this
equation
\item{$\bullet$} Wick order each $\lw K_i\rw_{\ze,D}$ with respect to $C$
in the $\xi'$ variable.
\item{$\bullet$} Wick order the product $\smprod_i\lW K_i\rW_{\xi',C\atop
\ze,D}$ with respect to $C$ in the $\et$ variable. 
\item{$\bullet$} Integrate using $\int\ \cdot\ d\mu_C(\xi',\et)$.
\item{$\bullet$} Wick order the result with respect to $C$
in the $\xi$ variable.

\noindent
yields ${\rm Ev}\, R_C(\lw K_1\rw_{\ze,D},\cdots,\lw K_\ell\rw_{\ze,D})(f) $ on 
the left and $\lW Q(\bar K_1,\cdots,\bar K_\ell)(f')\rW_{\ze,D}$ on the right. 
 \Item{(ii)} follows from part (i), the definition of $\cQ_{\bar K}$ and Remark \remcalR.
\endproof

As in subsection VII.1, we define for Grassmann functions
$\bar K_2(\psi;\ze,\ze',\varphi;\xi,\xi',\eta)$,$\cdots$,
$ \bar K_\ell(\psi;\ze,\ze',\varphi;\xi,\xi',\eta)$ and $f(\psi,\ze;\xi)$
the Grassmann function 
$\tilde Q(\bar K_2,\cdots,\bar K_\ell)$
by
$$\eqalign{
\tilde Q(\bar K_2,\cdots,\bar K_\ell)&(f)(\psi;\ze,\ze',\varphi;\xi,\xi',\eta) \cr
& = \int \hskip-6pt \int \lww \big( \smprod_{i=2}^\ell 
\lW \bar K_i(\psi;\ze,\ze'+\ze'',\varphi';\xi,\xi'+\xi'',\eta')
\rW_{\ze'',D\atop\xi'',C} 
\big)\rww_{\varphi',D\atop\eta',C} \cr
&\hskip 3cm \lw f(\psi,\ze+\varphi+\varphi';\eta+\eta')\rw_{\varphi',D}
\,d\mu_{D}(\ze^{\prime\prime},\varphi')\,d\mu_{C}(\xi^{\prime\prime},\eta') \cr
}$$
We have, as in Proposition \propoverldecR

\proposition{\STM\propoverldecQ}{
Let $\bar K_1(\psi;\ze,\ze',\varphi;\xi,\xi',\eta),\cdots,
 \bar K_\ell(\psi;\ze,\ze',\varphi;\xi,\xi',\eta)$ be Grassmann functions with
$\bar K_i(\psi;\ze,\ze',\varphi;\xi,\xi',0)=0$. Then for any Grassmann function
$f(\psi,\ze,\xi)$ 
$$\eqalign{
Q(&\bar K_1,\cdots,\bar K_\ell)(f)(\psi,\ze;\xi) \cr
&=\ \lww \int \hskip-6pt\int  
\lW \bar K_1(\psi;\ze,\ze',\varphi;\xi,\xi',\eta) 
\rW_{\ze',\varphi;D\atop\xi',\eta;C}\cr 
&\hskip 1.5cm \lW\tilde Q(\bar K_2,\cdots,\bar K_\ell)(f)(\psi;\ze,\ze',\varphi;\xi,\xi',\eta)
\rW_{\ze',\varphi;D\atop\xi',\eta;C}
\,d\mu_{D}(\ze',\varphi)\,d\mu_{C}(\xi',\eta) \ \rww_{\xi,C} 
}$$

}

\lemma{\STM\normestQ}{
Let $\bar K_i(\psi;\ze,\ze',\varphi;\xi,\xi',\eta),\ i=1,\cdots,\ell$  
be Grassmann functions such that
$\bar K_i(\psi;\ze,\ze',\varphi;\xi,\xi',0)=0$. 
Furthermore let
$f(\psi,\ze,\xi)$ be any Grassmann function. Set
$$
\lw f_1(\psi,\ze,\xi)\rw_{\xi,C}\  
  =\sfrac{1}{\ell !}\,Q(\bar K_1,\cdots,\bar K_\ell)
(\lw f\rw_{\xi,C})(\psi,\ze,\xi) 
$$
Then
$$\eqalign{
N(f_1\cl \al) & \le \sfrac{1}{\al^\ell}\,N(f\cl 2\al) \,
                      \smprod_{i=1}^\ell N(\bar K_i\cl \al) \cr
N(f_1(\psi,0,\xi)\cl \al) & \le \sfrac{1}{\al^\ell}\,N(f\cl \al) \,
                      \smprod_{i=1}^\ell N(\bar K_i\cl \al) \cr
 N\big(\sfrac{1}{(\ell-1) !}\, \tilde Q(\bar K_2,\cdots,\bar K_\ell)
(\lw f\rw_{\xi,C})\cl \al) 
 & \le \sfrac{ 1}{\al^{\ell-1}}\,N(f\cl 4\al) \,
                      \smprod_{i=2}^\ell N(\bar K_i\cl 2\al) \cr
N_{\rm impr}\big(\sfrac{1}{(\ell-1) !}\, 
\tilde Q(\bar K_2,\cdots,\bar K_\ell)(\lw f\rw_{\xi,C})\cl \al) 
 & \le \sfrac{1}{\al^{\ell-1}}\,N_{\rm impr}(f\cl 4\al) \,
                      \smprod_{i=2}^\ell N_{\rm impr}(\bar K_i\cl 2\al) \cr
}$$
}
\prf  Set $\tilde f(\psi;\ze,\varphi',\xi) =  f(\psi;\ze+\varphi',\xi)$ and
$$\eqalign{
\lw\tilde f_1(\psi;\ze,\ze',\varphi,\varphi';\xi)\rw_{\xi,C}\ 
&=\
\sfrac{1}{\ell !}\,
\lW R_C(\lw\bar K_1\rw_{\ze',D},\cdots,\lw\bar K_\ell\rw_{\ze',D})
\big(\lw\tilde f\rw_{\xi,C} \big)\rW_{\varphi,\varphi';D}\cr
\lw\tilde f_2(\psi;\ze,\ze',\varphi,\varphi';\xi)\rw_{\xi,C}\ 
&=\
\sfrac{1}{\ell !}\,
 R_C(\bar K_1,\cdots,\bar K_\ell)
\big(\lw\tilde f\rw_{\xi,C} \big)\cr
}$$
where $R_C$ is considered as an operator over the Grassmann algebra with coefficients in $A$, generated by $\psi_i,\ze_i,\ze'_i,\varphi_i,\varphi'_i$. Then
$f_1(\psi,\ze,\xi) = 
\int \tilde f_1(\psi;\ze,\ze',\varphi,\varphi;\xi)\,d\mu_D(\ze',\varphi) $
, and consequently by Lemma \lemwicknorm\ (twice, with $C$ replaced by $D$), Proposition \propnormestR\ and Remark \remfunctnorm
$$
N(f_1)\ \le\ N(\tilde f_2) 
\ \le\ \sfrac{1}{\al^\ell} N(\tilde f) \smprod_{i=1}^\ell N(\bar K_i)
\ \le\ \sfrac{1}{\al^\ell}\,N(f\cl 2\al) \,
                      \smprod_{i=1}^\ell N(\bar K_i\cl \al) 
$$
When $\ze$ is set to zero,
$$
\lw\tilde f_2(\psi;0,\ze',\varphi,\varphi';\xi)\rw_{\xi,C}\ 
=\
\sfrac{1}{\ell !}\,
 R_C(\bar K_1,\cdots,\bar K_\ell)
\big(\lw \tilde f\rw_{\xi,C} \big)\Big|_{\ze=0}
=\
\sfrac{1}{\ell !}\,
 R_C(\bar K_1,\cdots,\bar K_\ell)
\big(\lw f\rw_{\xi,C} \big)\Big|_{\ze=0}
$$
so
$$
N\big(f_1(\psi,0,\xi)\big)\ \le\ N(\tilde f_2(\psi,0,\xi)) 
\ \le\ \sfrac{1}{\al^\ell} N( f) \smprod_{i=1}^\ell N(\bar K_i)
$$
The proofs of the other inequalities are similar.
\endproof

\vfill\eject

\chap{ Overlapping Loops created by the second Covariance}\PG\pgRIX

In this Section we prove Theorem \theoremVb\ in the general case.
We assume that $A$ is a superalgebra in which all elements have degree 
zero and that $\|\,\cdot\|$ and $\|\,\cdot\|_{\rm impr}$ are two systems of 
symmetric seminorms on the spaces $A\otimes V^{\otimes n}$ such that 
the covariances $(C,D)$ have improved integration constants $\cb,\ib,\imp$ 
for these families of seminorms. 

\sect{Implementing Overlapping Loops}\PG\pgRIXa

\proposition{\STM\propsplittwoQ}{ Let  
$ K(\psi;\ze,\ze',\varphi;\xi,\xi',\eta)$ be an even Grassmann function 
such that
$ K(\psi;\ze,\ze',\varphi;\xi,\xi',0)=0$. Decompose
$$
K = K' + K^{\prime\prime} 
$$
where $K'$ has degree at most two in the variables $\ze',\xi',\varphi,\eta$ and
$K^{\prime\prime}$ has degree at least three in these variables.
Let $f(\psi;\ze;\xi)$ be a Grassmann function, and set
$$
\lw g\rw_{\xi,C} \ =\ \cQ_{K}(\lw f\rw_{\xi,C})  \qquad\qquad
\lw g'\rw_{\xi,C} \ =\ \cQ_{K'}(\lw f\rw_{\xi,C})
$$
Then, if $\al \ge 2$ and $N(K\cl 2\al)_\0 < \al$
$$
N_{\rm impr}(g-g'\cl \al) 
 \le \sfrac{2^5\,\imp}{\al^6}\,
N(f\cl 4\al)\,\sfrac{N(K\cl 2\al)}{1-{1\over\al}N(K\cl 2\al)}
$$
}

The proof is analogous to the proof of Proposition \propsplittwo\ and is
given following

\lemma{\STM\lemsplittwoQ}{ Let  
$ K(\psi;\ze,\ze',\varphi;\xi,\xi',\eta)$ be an even Grassmann function that
satisfies $ K(\psi;\ze,\ze',\varphi;\xi,\xi',0)=0$. Decompose
$$
K = K' + K^{\prime\prime} 
$$
where $K'$ has degree at most two in the variables $\ze',\xi',\varphi,\eta$ and
$K^{\prime\prime}$ has degree at least three in these variables.
Let each of the functions $K^{(1)},\cdots,K^{(\ell)}$ be either 
$K'$ or $K^{\prime\prime}$, and assume that at least one of them is equal to 
$K^{\prime\prime}$. Let $f(\psi;\ze;\xi)$ be a Grassmann function, and set
$$
\sfrac{1}{\ell !}\, Q(K^{(1)},\cdots,K^{(\ell)})(\lw f\rw)(\psi;\ze;\xi) = \ \lw f'(\psi;\ze;\xi)\rw_{\xi,C}
$$
where$\lw f\rw$ is shorthand for $\lw f(\psi;\ze;\xi)\rw_{\xi,C}$.
Then, if $\al\ge 2$ 
$$
N_{\rm impr}(f'\cl \al) \le \sfrac{2^5}{\ell\al^{\ell+5}}\,\imp\,
 N(f\cl 4\al)\,N(K\cl 2\al)^\ell
$$
}

\prf
We may assume that $K^{(1)}=K^{\prime\prime}$. Set
$$
g(\psi;\ze,\ze',\varphi;\xi,\xi',\eta) = \tilde Q(K^{(2)},\cdots, K^{(\ell)})
(\lw f\rw)
$$
By Lemma \normestQ
$$
\sfrac{1}{(\ell-1)!} \,N(g) 
\le \sfrac{1}{\al^{\ell-1}}\, N(f\cl 4\al)\,N(K\cl 2\al)^{\ell-1} $$
By Proposition \propoverldecQ
$$\eqalign{
f'(\psi;\ze;\xi) = \sfrac{1}{\ell !} \int \hskip-6pt\int  
\lW  K^{\prime\prime}(\psi;\ze,\ze',\varphi;\xi,\xi',\eta) 
&\rW_{\ze',\varphi;D\atop\xi',\eta;C}  \cr
 \lW g(\psi;\ze,\ze',\varphi&;\xi,\xi',\eta)
\rW_{\ze',\varphi;D\atop\xi',\eta;C}
\,d\mu_{D}(\ze',\varphi)\,d\mu_{C}(\xi',\eta) \cr 
}$$
Proposition \propimprnormest\ implies that
$$\eqalign{
N_{\rm impr} (f') &\le  \sfrac{27\imp}{\ell!\,\al^6}\,N(K^{\prime\prime})\,N(g)  \cr
&\le   \sfrac{2^5}{\ell\,\al^{\ell+5}}\,\imp\, N(f\cl 4\al)\,N(K\cl 2\al)^\ell  \cr
}$$
\endproof

\proof{ of Proposition \propsplittwoQ}
By Definition \defQ ii
$$
g = \smsum_{\ell=1}^\infty g_\ell \qquad  \qquad 
g' = \smsum_{\ell=1}^\infty g'_\ell
$$
where
$$\eqalign{
\lw g_\ell\rw\  &= \ \sfrac{1}{\ell !}\,Q(K,\cdots,K)(\lw f\rw) \cr
\lw g'_\ell\rw\ &= \ \sfrac{1}{\ell !}\,Q(K',\cdots,K')(\lw f\rw)  \cr
}$$
Since
$$\eqalign{
Q&(K,\cdots,K) - Q(K',\cdots,K') \cr
&=  Q(K-K',K,\cdots,K)
 + Q(K',K-K',K,\cdots,K) + \cdots +Q(K',\cdots,K-K') \cr
&=  Q(K^{\prime\prime},K,\cdots,K)
 + Q(K',K^{\prime\prime},K,\cdots,K) + \cdots +Q(K',\cdots,K^{\prime\prime}) \cr
}$$ 
it follows from Lemma \lemsplittwoQ\ that
$$
N_{\rm impr} (g_\ell-g'_\ell) \le \sfrac{2^5}{\al^{\ell+5}}\,\imp\, N(f\cl 4\al)\,N(K\cl 2\al)^\ell $$
Therefore
$$
N_{\rm impr}(g-g') \le \smsum_{\ell=1}^\infty N_{\rm impr}(g_\ell-g'_\ell)
 \le \sfrac{2^5\,\imp}{\al^6}\,
N(f\cl 4\al)\,\sfrac{N(K\cl 2\al)}{1-{1\over\al}N(K\cl 2\al)}
$$
\endproof

\proposition{\STM\propimprQQ}{
Let $ K(\psi;\ze,\ze',\varphi;\xi,\xi',\eta)$ be an even Grassmann function 
of degree at most two in the variables $\ze',\xi',\varphi,\eta$ and of degree 
at least one in $\eta$. Write
$$
K(\psi;\ze,\ze',\varphi;\xi,\xi',\eta)
 = \sum_{n_0,n_1,n_2,n_3\atop p_1,p_2,p_3} 
K_{n_0}\tailind{p_1}{p_2}{p_3}{n_1}{n_2}{n_3}
(\psi;\ze,\ze',\varphi;\xi,\xi',\eta)
$$ with
$K_{n_0}\tailind{p_1}{p_2}{p_3}{n_1}{n_2}{n_3}
 \in A[n_0,p_1,p_2,p_3, n_1,n_2,n_3]$. Let
$$
K_{\cdot,1} = \smsum_{n_0,n_1,p_1}
K_{n_0}\big({\scriptstyle p_1\,0\,0\atop \scriptstyle n_1\,0\,1}\big) 
$$
be the part of $K$ that has degree precisely one in $\ze',\xi',\varphi,\eta$.

\noindent
Furthermore let $T(\psi;\varphi;\et)$ be a Grassmann function that has 
degree two in the variables $\varphi,\eta$ and degree at least one in 
the variable $\eta$. Write $T=T_{11}+T_{02}$ where $T_{11}$ has degree one in $\varphi$ and $\eta$, and $T_{02}$ has degree two in $\eta$. Set
$$\eqalign{
T_{\rm mix}(\psi;\varphi',\varphi^{\prime\prime};\eta',\eta^{\prime\prime}) 
&= T_{11}(\psi;\varphi';\et'') +
T_{11}(\psi;\varphi'';\et') \cr
&\ +\big[ T_{02}(\psi;0;\eta'+\eta^{\prime\prime}) - T_{02}(\psi;0;\eta')
-T_{02}(\psi;0;\eta^{\prime\prime}) \big] \cr
}$$
Furthermore set
$$\eqalign{
&\tilde K_1(\psi;\ze,\ze',\varphi;\xi,\xi',\eta) \cr
& \hskip .4cm = \int \hskip -6pt \int 
\lw T(\psi;\varphi';\et')\rw_{\eta',C}\
\lW K(\psi;\ze+\varphi',\ze',\varphi;\xi+\eta',\xi',\eta)
\rW_{\varphi',D\atop\eta',C}\,d\mu_D(\varphi')\,d\mu_C(\eta') \cr
&\tilde K_2(\psi;\ze,\ze',\varphi;\xi,\xi',\eta) 
  = \int  T_{11}(\psi;\varphi;\et')\,
\lw K(\psi;\ze,\ze',\varphi;\xi+\eta',\xi',\eta)
\rw_{\eta',C}\,d\mu_C(\eta') \cr
}
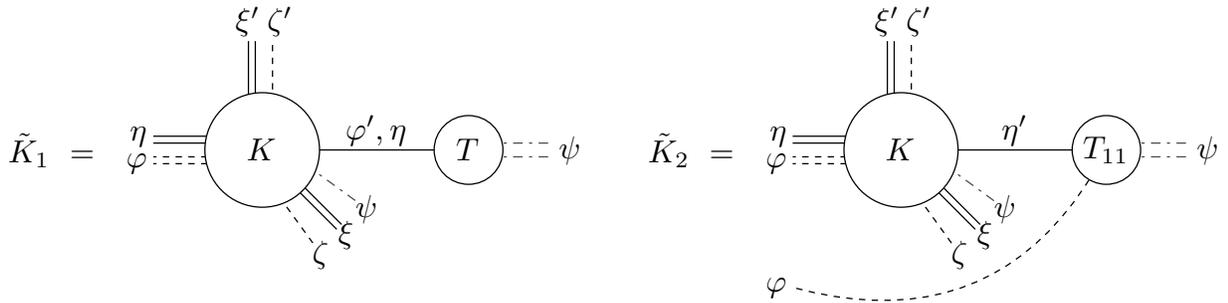
$$ 
$$
\hskip.2in\figput{figK1}\quad \figput{figVII2}
$$
and $\tilde K = \tilde K_1+\tilde K_2$, 
$$\eqalign{
&\hat K(\psi;\ze,\varphi;\xi,\eta) 
= \int\hskip-6pt \int 
T_{\rm mix}(\psi;\varphi',\varphi^{\prime\prime};\eta',\eta^{\prime\prime}) \,\lW K_{\cdot,1}(\psi;\ze+\varphi',\ze',\varphi;\xi+\eta',\xi',\eta)
\rW_{\varphi',D\atop\eta',C} \cr
&\hskip 2.2cm \lW K_{\cdot,1}(\psi;\ze+\varphi^{\prime\prime},\ze',\varphi;
\xi+\eta^{\prime\prime},\xi',\eta)
\rW_{\varphi^{\prime\prime},D\atop\eta^{\prime\prime},C}\,
d\mu_D(\ze',\varphi',\varphi^{\prime\prime})\,
d\mu_C(\xi',\eta',\eta^{\prime\prime}) \cr
}
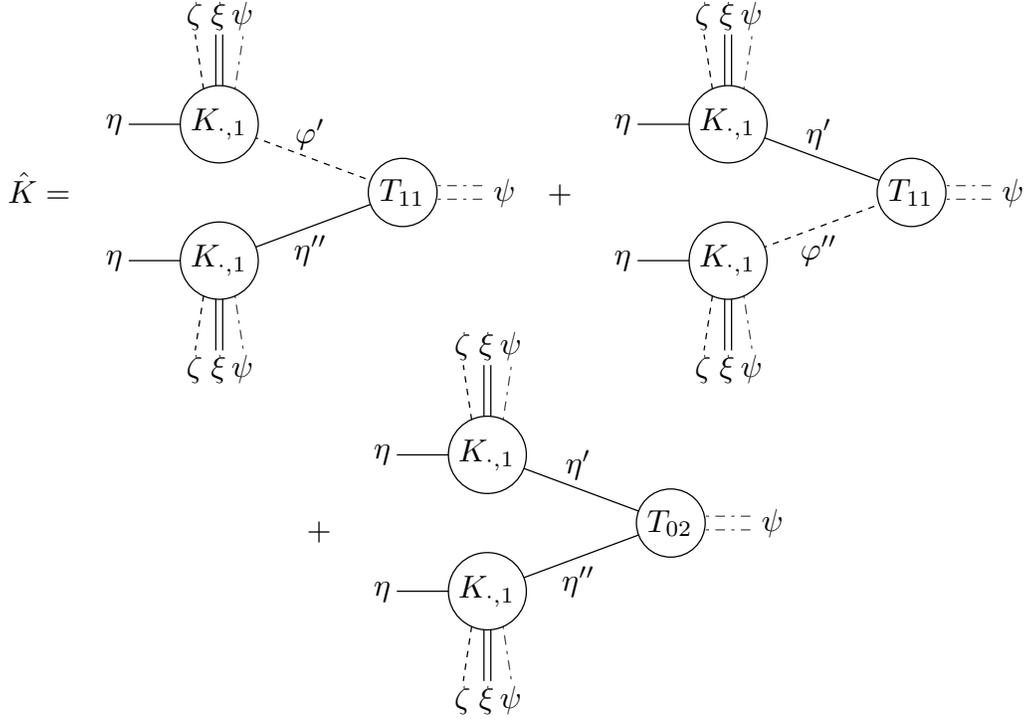
$$
$$\eqalign{
\hat K=&\figplace{figKhat1}{0in}{-0.65in}
+\figplace{figKhat2}{0in}{-0.65in}\cr
&\hskip1in+\figplace{figKhat3}{0in}{-0.82in}
}$$
Finally let $f(\psi;\ze;\xi)$ be a Grassmann function and set
$$\eqalign{
f'(\psi;\ze,\varphi;\xi) &= \tilde Q(T)\, Q(K,K)(\lw f\rw_{\xi,C})
(\psi;\ze,0,\varphi;0,0,\xi) \cr
\lw \tilde f(\psi;\ze,\varphi;\xi)\rw_{\xi,C} 
&= 2\,Q(\tilde K,K)(\lw f\rw_{\xi,C})(\psi;\ze+\varphi;\xi) \cr
\lw\hat f(\psi;\ze,\varphi;\xi)\rw_{\xi,C} 
&= Q(\hat K)(\lw f\rw_{\xi,C})(\psi;\ze+\varphi;\xi) \cr
}$$
Then, if $\al\ge 2$,
$$
N_{\rm impr}(f'-(\tilde f+\hat f)\cl \al) \le \sfrac{2^{9}\,\imp}{\al^{10}}\,
N(f\cl 4\al)\,N(K\cl 4\al)^2\,\,N(T\cl 2\al)
$$
}

The proof is similar to the proof of Proposition \propimprRR\ and is given
following

\lemma{\STM\lemimprRRCD}{ Let 
$B(\psi;\varphi',\varphi^{\prime\prime};\eta',\eta^{\prime\prime})$ 
be a Grassmann function that has degree one in the variables $\varphi',\eta'$,
degree one in the variables $\varphi^{\prime\prime},\eta^{\prime\prime}$ and degree at least one in the variables $\eta',\eta^{\prime\prime}$. Furthermore
let 
$H(\psi;\ze,\ze',\gamma;\xi,\xi',\eta)$, $K(\psi;\ze,\ze',\gamma;\xi,\xi',\eta)$ be even Grassmann functions that vanish for $\eta =0$. Assume that $H$ or $K$ has degree at least two in the variables 
$\ze',\gamma,\xi',\eta$.
Let $f(\psi;\ze;\xi) $ be any Grassmann function and set
$$\eqalign{
g(\psi;\ze,\varphi;\xi) = &\int \hskip -6pt \int 
\lww \Big[ \int\hskip -6pt\int  
B(\psi;\varphi',\varphi^{\prime\prime};\eta',\eta^{\prime\prime})\ 
 \lW H(\psi;\ze+\varphi+\varphi',\ze',\gamma;\xi+\eta',\xi',\eta)
\rW_{\ze',\varphi';D\atop\eta',\xi';C} \cr
& \ 
\lW K(\psi;\ze\!+\!\varphi\!+\!\varphi'',\ze',\gamma;
    \xi\!+\!\eta^{\prime\prime},\xi',\eta)
\rW_{\ze',\varphi^{\prime\prime};D\atop\xi',\eta^{\prime\prime};C}\,
d\mu_D(\varphi',\varphi^{\prime\prime},\ze')\,d\mu_C(\eta',\eta^{\prime\prime},\xi') \Big]
\rww_{\gamma,D\atop\eta,C}    \cr
& \hskip 5cm  \lW f(\psi;\ze+\varphi+\gamma;\eta)\rW_{\gamma,D\atop\eta,C}
\,d\mu_D(\gamma)\,d\mu_C(\eta) \cr
}$$
Then, if $\al \ge 2$
$$
N_{\rm impr}(g\cl \al)  \le \  \sfrac{2^5\,\imp}{\al^{10}} \,N(B\cl \al)\,N(H\cl 4\al)\,N(K\cl 4\al)
\,N(f\cl 4\al)$$
}

\prf In the proof we suppress the variable $\psi$. 
We assume that $K$ has degree at least two in the variables 
$\ze',\gamma,\xi',\eta$.

\noindent
First we discuss the case that $B$ has degree one in the variable $\eta'$.
Set 
$$\eqalign{
h(\ze,\varphi,\gamma,\ze',\varphi^{\prime\prime};&\xi,\eta,\xi',\eta^{\prime\prime}) \cr
&= \int\hskip -6pt\int 
B(\varphi',\varphi^{\prime\prime};\eta',\eta^{\prime\prime}) \
\lW H(\ze+\varphi+\varphi',\ze',\ze'';\xi+\eta',\xi',\xi'')
\rW_{\varphi',\ze'';D\atop\eta',\xi'';C}\cr
& \hskip 3cm\lW f(\ze+\varphi+\gamma+\ze'';\et+\xi'')
\rW_{\ze^{\prime\prime},D\atop\xi^{\prime\prime},C} 
\,d\mu_D(\varphi',\ze^{\prime\prime})\,d\mu_C(\eta',\xi^{\prime\prime})
}$$
By Lemma \:\lemBV, twice
$$\eqalign{
&\int\hskip -7pt\int 
 \lW H({\sst\cdots},\gamma{\sst\cdots},\eta) K({\sst\cdots},\gamma{\sst\cdots},\eta)
        \rW_{\gamma,D\atop\et,C}\ \lW f({\sst\cdots},\gamma;\et) \rW_{\gamma,D\atop\et,C}
        \ d\mu_C(\et)d\mu_D(\ga)\cr
&=\!\!\int\hskip -7pt\int \!\!
 \lW H({\sst\cdots},\ze''{\sst\cdots},\xi'') \rW_{\!\ze'',D\atop\!\xi'',C}\!
 \lW K({\sst\cdots},\gamma{\sst\cdots},\eta) \rW_{\!\gamma,D\atop\!\et,C}\!
  \lW f({\sst\cdots},\gamma+\ze'';\et+\xi'') \rW_{\!\gamma,\ze'';D\atop\!\et,\xi'';C}
        d\mu_C(\et,\xi'')d\mu_D(\gamma,\ze'')\cr
}\EQN\eqnVIIIi$$
so
$$\eqalign{
g(\ze,\varphi;\xi) 
= \int \hskip -6pt \int 
&\lW h(\ze,\varphi,\gamma,\ze',\varphi^{\prime\prime};\xi,\eta,\xi',\eta^{\prime\prime}) \rW_{\ze',\gamma;D\atop\xi',\eta;C}\cr
&\lW K(\ze+\varphi+\varphi^{\prime\prime},\ze',\gamma;\xi+\eta^{\prime\prime},\xi',\eta)
\rW_{\varphi^{\prime\prime},\ze',\gamma;D\atop\eta^{\prime\prime},\xi',\eta;C}\,
d\mu_D(\gamma,\varphi^{\prime\prime},\ze')
\,d\mu_C(\eta,\eta^{\prime\prime},\xi')    \cr
}$$
By iterated application of Lemma \lemwicknorm\ combined with Proposition \propestCconnect,
first integrating $\int\ \cdot\ d\mu_C(\et')d\mu_D(\varphi')$
and then integrating  $\int\ \cdot\ d\mu_C(\xi'')d\mu_D(\ze'')$,
and Remark \remfunctnorm, several times,
$$\eqalign{
N(h) &\le \sfrac{1}{\al^4} N(B)\,N(H({\sst\ze+\varphi+\varphi',\ze',\ze'';\xi+\eta',\xi',\xi''})\cl \al)
\,N(f({\sst \ze+\varphi+\gamma+\ze'';\et+\xi''})\cl \al)\cr
&\le \sfrac{1}{\al^4} N(B)\,N(H(\ze+\varphi,\ze',\ze'';\xi,\xi',\xi'')\cl 2\al)
\,N(f(\ze+\gamma;\et)\cl 2\al)\cr
&\le \sfrac{1}{\al^4} N(B)\,N(H\cl 4\al)\,N(f\cl 4\al)\cr
}$$
Since $h$, through $B$,  has degree one in $\varphi'',\et''$,
only the part of 
$ K(\ze+\varphi+\varphi^{\prime\prime},\ze',\gamma;\xi+\eta^{\prime\prime},\xi',\eta)$
that has degree at least one in in $\varphi'',\et''$ can contribute to
$g(\ze,\varphi;\xi)$. As $K$ also has degree at least two in the variables 
$\ze',\gamma,\xi',\eta$ and degree at least one in $\eta$, Proposition \propimprnormest\ implies that
$$\eqalign{
N_{\rm impr}(g)
&\le \sfrac{27\,\imp}{\al^6}\, N(h)\,N(K\cl 4\al) \cr
&\le \  \sfrac{2^5\,\imp}{\al^{10}} \,N(B)\,N(H\cl 4\al)\,N(f\cl 4\al)\,
N(K\cl 4\al)
}$$

Next we discuss the situation that $B$ has degree zero in the variable $\eta'$.
This implies that $B$ has degree one in $\varphi'$ and degree one in $\et''$. Set
$$\eqalign{
h(\ze,\varphi,\gamma,\ze',\varphi';\xi,\xi',\eta,\eta') 
&= \int\hskip -6pt\int 
\lW H(\ze+\varphi+\varphi',\ze',\ze'';\xi+\eta',\xi',\xi'')
\rW_{\ze'';D\atop\xi'';C}\cr
& \hskip 2.5cm\lW f(\ze+\varphi+\gamma+\ze'';\et+\xi'')
\rW_{\ze'';D\atop\xi'';C} 
\,d\mu_D(\ze'')\,d\mu_C(\xi'') \cr
k(\ze,\varphi,\gamma,\ze',\varphi';\xi,\xi',\eta,\eta') 
&= \int\hskip -6pt\int 
B(\varphi',\varphi^{\prime\prime};\eta',\eta^{\prime\prime})
\lW K(\ze+\varphi+\varphi^{\prime\prime},\ze',\gamma;\xi+\eta^{\prime\prime},\xi',\eta)
\rW_{\varphi^{\prime\prime};D\atop\eta^{\prime\prime};C}\cr
&\hskip 7.3cm \,d\mu_D(\varphi^{\prime\prime})\,d\mu_C(\eta^{\prime\prime}) \cr
}$$
Again by (\eqnVIIIi)
$$\eqalign{
g = \int\hskip -6pt\int
\lW h(\ze,\varphi,\ga,\ze'&,\varphi';\xi,\xi',\eta,\eta')
  \rW_{\varphi',\ze',\gamma;D\atop\eta',\xi',\eta;C}\cr 
&\lW k(\ze,\varphi,\ga,\ze',\varphi';\xi,\xi',\eta,\eta')
  \rW_{\varphi',\ze',\gamma;D\atop\eta',\xi',\eta;C}
\,d\mu_D(\varphi',\ze',\gamma)\,d\mu_C(\eta',\xi',\eta) \cr  
}$$
By  Lemma \lemwicknorm, Proposition \propestCconnect\ and Remark \remfunctnorm
$$\eqalign{
N(h) &\le \sfrac{1}{\al^2} N(H\cl 4\al)\,N(f\cl 4\al)\cr
N(k) &\le \sfrac{1}{\al^2} N(B\cl \al)\,N(K\cl 4\al)\cr
}$$
Since $k$ has degree at least one in $\eta$ and degree at least three in 
$\varphi',\ze',\gamma,\eta',\xi',\eta$ we have
$$
N_{\rm impr}(g)\ \le\ \sfrac{27\imp}{\al^6}\,N(h)\,N(k) \
\le \  \sfrac{2^5\,\imp}{\al^{10}} \,N(H\cl 4\al)\,N(f\cl 4\al)\,
N(K\cl 4\al)\,N(B)
$$
\endproof

\proof{of Proposition \propimprQQ}
We again suppress $\psi$ in the proof. By definition
$$
f'(\ze,\varphi;\et)
= \int\hskip -6pt\int
\lw T(\varphi',\eta')\rw_{\eta',C}\,
\lw Q(K,K)(\lw f\rw_{\xi,C})(\ze+\varphi+\varphi';\eta+\eta')\rw_{\varphi',D}\,
d\mu_D(\varphi')\,d\mu_C(\eta')
$$
so that, by Lemma \:\lemBVI, recalling that $T(\varphi',\et')$ is of degree two
in $\varphi',\et'$,
$$\eqalign{
&f'(\ze,\varphi;\xi)
= \int\hskip -6pt\int
\lw T(\varphi',\eta')\rw_{\eta',C}\,
\lw Q(K,K)(\lw f\rw)(\ze+\varphi+\varphi';\xi+\eta')\rw_{\varphi',D}\,
d\mu_D(\varphi')\,d\mu_C(\eta')  \cr
\hskip.5cm&\hskip.5cm=\ \int\hskip -6pt\int
\lw T(\varphi',\eta')\rw_{\eta',C}\,
\lww \Big[ \lW \Big(
\lW K(\ze+\varphi+\varphi',\ze',\ga;\xi+\eta',\xi',\eta)\rW_{\ze',D\atop\xi',C} 
\cr
& \hskip 5cm 
\lW K(\ze+\varphi+\varphi',\ze',\ga;\xi+\eta',\xi',\eta)\rW_{\ze',D\atop\xi',C} 
\Big)\rW_{\ga,D\atop\eta,C} \cr
&\hskip 2cm 
\lW f(\ze+\varphi+\varphi'+\ga;\eta)\rW_{\ga,D\atop\eta,C} 
\Big]\rww_{\varphi',D\atop\eta',C}\ 
d\mu_D(\ze',\varphi',\ga)\,d\mu_C(\xi',\eta,\eta') \cr
}$$
$$\eqalign{
\hskip.5cm&\hskip.5cm=\  2\int\hskip -6pt\int
\lw T(\varphi',\eta')\rw_{\eta',C}\
\lww\Big[ \lW \Big(
\lW K(\ze+\varphi+\varphi',\ze',\ga;\xi+\eta',\xi',\eta)
\rW_{\ze',\varphi';D\atop\xi',\eta';C} \cr
& \hskip 5cm 
\lW K(\ze+\varphi,\ze',\ga;\xi,\xi',\eta)\rW_{\ze',D\atop\xi',C} 
\Big)\rW_{\ga,D\atop\eta,C} \cr
&\hskip 3cm 
\lW f(\ze+\varphi+\varphi'+\ga;\eta)\rW_{\ga,D\atop\eta,C} \Big]\rww_{\varphi',D}\, 
d\mu_D(\ze',\varphi',\ga)\,d\mu_C(\xi',\eta,\eta') \cr
\hskip.5cm&\hskip.5cm\ \ + \int\hskip -6pt\int \lww \Big[
T_{\rm mix}(\varphi',\varphi^{\prime\prime};\eta',\eta^{\prime\prime})\,
\lW K(\ze+\varphi+\varphi',\ze',\ga;\xi+\eta',\xi',\eta)
\rW_{\varphi',\ze';D\atop\eta',\xi';C} \cr
& \hskip 4cm
\lW K(\ze+\varphi+\varphi^{\prime\prime},\ze',\ga;
   \xi+\eta^{\prime\prime},\xi',\eta)
      \rW_{\varphi^{\prime\prime},\ze';D\atop\eta^{\prime\prime},\xi';C} 
 \Big]\rww_{\ga,D\atop\eta,C} \cr
& \hskip 4cm
\lW f(\ze+\varphi+\ga;\eta)\rW_{\ga,D\atop\eta,C} \,
d\mu_D(\varphi',\varphi^{\prime\prime},\ze',\ga)\,
d\mu_C(\eta',\eta^{\prime\prime},\xi',\eta)\hskip.6cm \cr
}\EQN\eqnVIIIii$$
The $T_{\rm mix}$ term in (\eqnVIIIii) above differs from
$$
\int\hskip -6pt\int \lW  
\hat K(\ze+\varphi,\ga;\xi,\eta) \rW_{\ga,D\atop\eta,C}\
\lW  f(\ze+\varphi+\ga;\eta)  \rW_{\ga,D\atop\eta,C} d\mu_D(\ga) \,d\mu_C(\eta)
=\hat f(\ze,\varphi;\xi)
$$
by
$$\eqalign{
&\ \ \int\hskip -6pt\int \lww \Big[
T_{\rm mix}(\varphi',\varphi^{\prime\prime};\eta',\eta^{\prime\prime})\,
\lW H(\ze+\varphi+\varphi',\ze',\ga;\xi+\eta',\xi',\eta)
\rW_{\varphi',\ze';D\atop\eta',\xi';C} \cr
& \hskip 4cm
\lW H(\ze+\varphi+\varphi^{\prime\prime},\ze',\ga;
   \xi+\eta^{\prime\prime},\xi',\eta)
      \rW_{\varphi^{\prime\prime},\ze';D\atop\eta^{\prime\prime},\xi';C} 
 \Big]\rww_{\ga,D\atop\eta,C} \cr
& \hskip 4cm
\lW f(\ze+\varphi+\ga;\eta)\rW_{\ga,D\atop\eta,C} \,
d\mu_D(\varphi',\varphi^{\prime\prime},\ze',\ga)\,
d\mu_C(\eta',\eta^{\prime\prime},\xi',\eta) \cr
&\ \ \ + 2\int\hskip -6pt\int \lww \Big[
T_{\rm mix}(\varphi',\varphi^{\prime\prime};\eta',\eta^{\prime\prime})\,
\lW H(\ze+\varphi+\varphi',\ze',\ga;\xi+\eta',\xi',\eta)
\rW_{\varphi',\ze';D\atop\eta',\xi';C} \cr
& \hskip 4cm
\lW K_{\cdot,1}(\ze+\varphi+\varphi^{\prime\prime},\ze',\ga;
   \xi+\eta^{\prime\prime},\xi',\eta)
      \rW_{\varphi^{\prime\prime},\ze';D\atop\eta^{\prime\prime},\xi';C} 
 \Big]\rww_{\ga,D\atop\eta,C} \cr
& \hskip 4cm
\lW f(\ze+\varphi+\ga;\eta)\rW_{\ga,D\atop\eta,C} \,
d\mu_D(\varphi',\varphi^{\prime\prime},\ze',\ga)\,
d\mu_C(\eta',\eta^{\prime\prime},\xi',\eta) \cr
}$$
where $K=K_{\cdot,1}+H$.
By Lemma \lemimprRRCD, the improved norm of this difference is bounded by
$$\eqalign{
3\,&\sfrac{2^5\,\imp}{\al^{10}} \,N(T_{\rm mix}\cl \al)\,
N(H\cl 4\al)\,N(K\cl 4\al)
\,N(f\cl 4\al) \cr
&\le\, \sfrac{2^9\,\imp}{\al^{10}} \,
N(T\cl 2\al)\, N(K\cl 4\al)^2 \,N(f\cl 4\al) \cr
}$$
We apply Lemma \:\lemBV\ to the variable $\varphi'$ in the 
$2\ \lw T(\varphi',\et')\rw_{\et',C}$ term
of (\eqnVIIIii). We get
$$\eqalign{
 &2\int\hskip -6pt\int
\lw T(\varphi'+\varphi'';\eta')\rw_{\eta',C}\,
 \lww \Big( \lW K(\ze+\varphi+\varphi',\ze',\ga;\xi+\eta',\xi',\eta)
     \rW_{\ze',\varphi';D\atop\xi',\eta';C} \cr
& \hskip 5cm 
\lW K(\ze+\varphi,\ze',\ga;\xi,\xi',\eta)\rW_{\ze',D\atop\xi',C} 
\Big)\rww_{\ga,D\atop\eta,C} \cr
&\hskip 4cm 
\lW f(\ze+\varphi+\varphi^{\prime\prime}+\ga;\eta)
\rW_{\varphi^{\prime\prime},\ga;D\atop\eta,C} \, 
d\mu_D(\ze',\varphi',\varphi^{\prime\prime},\ga)\,d\mu_C(\xi',\eta,\eta') \cr
}$$
As $T$ has degree at most one in $\varphi'+\varphi^{\prime\prime}$, this is equal to
$$\eqalign{
 &2\int\hskip -6pt\int \lww  \Big(
\lw T(\varphi';\eta')\rw_{\eta',C}\ 
  \lW K(\ze+\varphi+\varphi',\ze',\ga;\xi+\eta',\xi',\eta)
     \rW_{\ze',\varphi';D\atop\xi',\eta';C} \cr
& \hskip 5cm 
\lW K(\ze+\varphi,\ze',\ga;\xi,\xi',\eta)\rW_{\ze',D\atop\xi',C} 
\Big)\rww_{\ga,D\atop\eta,C} \cr
&\hskip 4cm 
\lW f(\ze+\varphi+\ga;\eta)\rW_{\ga,D\atop\eta,C} \, 
d\mu_D(\ze',\varphi',\ga)\,d\mu_C(\xi',\eta,\eta') \cr
&\ \ +2\int\hskip -6pt\int
\lw T_{11}(\varphi'';\eta')\rw_{\eta',C}\,
 \lww\Big( \lW K(\ze+\varphi,\ze',\ga;\xi+\eta',\xi',\eta)
     \rW_{\ze',D\atop\xi',\eta';C} \cr
& \hskip 5cm 
\lW K(\ze+\varphi,\ze',\ga;\xi,\xi',\eta)\rW_{\ze',D\atop\xi',C} 
\Big)\rww_{\ga,D\atop\eta,C} \cr
&\hskip 4cm 
\lW f(\ze+\varphi+\varphi^{\prime\prime}+\ga;\eta)
\rW_{\varphi^{\prime\prime},\ga;D\atop\eta,C} \, 
d\mu_D(\ze',\varphi^{\prime\prime},\ga)\,d\mu_C(\xi',\eta,\eta') \cr
}$$
Applying Lemma \:\lemBV, with $\xi$ replaced by $\ga$ and $\xi'$ replaced by
$\varphi''$, to the second term, this is equal to
$$\eqalign{
 &2\int\hskip -6pt\int \lww  \Big(
\lw T(\varphi';\eta')\rw_{\eta',C}\ 
  \lW K(\ze+\varphi+\varphi',\ze',\ga;\xi+\eta',\xi',\eta)
     \rW_{\ze',\varphi';D\atop\xi',\eta';C} \cr
& \hskip 5cm 
\lW K(\ze+\varphi,\ze',\ga;\xi,\xi',\eta)\rW_{\ze',D\atop\xi',C} 
\Big)\rww_{\ga,D\atop\eta,C} \cr
&\hskip 4cm 
\lW f(\ze+\varphi+\ga;\eta)\rW_{\ga,D\atop\eta,C} \, 
d\mu_D(\ze',\varphi',\ga)\,d\mu_C(\xi',\eta,\eta') \cr
&\ \ +2\int\hskip -6pt\int \lww\Big(
T_{11}(\ga;\eta')\,
 \lW K(\ze+\varphi,\ze',\ga;\xi+\eta',\xi',\eta)
     \rW_{\ze',D\atop\xi',\eta';C} \cr
& \hskip 5cm 
\lW K(\ze+\varphi,\ze',\ga;\xi,\xi',\eta)\rW_{\ze',D\atop\xi',C} 
\Big)\rww_{\ga,D\atop\eta,C} \cr
&\hskip 4cm 
\lW f(\ze+\varphi+\ga;\eta)
\rW_{\ga;D\atop\eta,C} \, 
d\mu_D(\ze',\ga)\,d\mu_C(\xi',\eta,\eta') \cr
&=   2\int\hskip -6pt\int \lww  \Big( 
  \lW \tilde K_1(\ze+\varphi,\ze',\ga;\xi,\xi',\eta)
     \rW_{\ze',D\atop\xi',C} \ 
\lW K(\ze+\varphi,\ze',\ga;\xi,\xi',\eta)\rW_{\ze',D\atop\xi',C} 
\Big)\rww_{\ga,D\atop\eta,C} \cr
&\hskip 5cm 
\lW f(\ze+\varphi+\ga;\eta)\rW_{\ga,D\atop\eta,C} \, 
d\mu_D(\ze',\ga)\,d\mu_C(\xi',\eta) \cr
&\ \ +2\int\hskip -6pt\int \lww  \Big( 
  \lW  \tilde K_2(\ze+\varphi,\ze',\ga;\xi,\xi',\eta)
     \rW_{\ze',D\atop\xi',C} \ 
\lW K(\ze+\varphi,\ze',\ga;\xi,\xi',\eta)\rW_{\ze',D\atop\xi',C} 
\Big)\rww_{\ga,D\atop\eta,C} \cr
&\hskip 5cm 
\lW f(\ze+\varphi+\ga;\eta)\rW_{\ga,D\atop\eta,C} \, 
d\mu_D(\ze',\ga)\,d\mu_C(\xi',\eta) \cr
}$$
So the $2\ \lw T(\varphi',\et')\rw_{\et',C}$ term of (\eqnVIIIii) equals
$$
\tilde f_1(\ze,\varphi;\xi) + \tilde f_2(\ze,\varphi;\xi)
$$
where
$$\eqalign{
\lw \tilde f_1(\ze,\varphi;\xi)\rw_{\xi,C} 
& \ =\ 2 \, Q(\tilde K_1,K)(\lw f\rw)(\ze+\varphi;\xi) \cr
\lw \tilde f_2(\ze,\varphi;\xi)\rw_{\xi,C} 
& \ =\ 2 \, Q(\tilde K_2,K)(\lw f\rw)(\ze+\varphi;\xi) \cr
}$$
As $\tilde f =\tilde f_1+\tilde f_2$, the Proposition follows.
\endproof

\sect{ Tails}\PG\pgRIXb

In this subsection let $K(\psi;\ze,\ze',\varphi;\xi,\xi',\eta)$ be an even Grassmann function that has {\bf degree at least four} in the variables
$\psi,\ze,\ze',\varphi,\xi,\xi',\eta$ such that
 $K(\psi;\ze,\ze',\varphi;\xi,\xi',0)=0$ . We always write
$$
K = \sum_{n_0,n_1,n_2,n_3\atop p_1,p_2,p_3} 
K_{n_0}\tailind{p_1}{p_2}{p_3}{n_1}{n_2}{n_3}
\qquad {\rm with}\qquad
K_{n_0}\tailind{p_1}{p_2}{p_3}{n_1}{n_2}{n_3}
\in A[n_0,p_1,p_2,p_3, n_1,n_2,n_3]
$$

\goodbreak
\definition{\STM\deftailsec}{
\Item{(i)} An $n$--legged tail with at least $e$ external legs
is a Grassmann function
$$
T(\psi;\varphi;\et) \in \bigoplus_{d\ge e} 
\bigoplus_{ n_1+n_2=n \atop n_2\ge 1} A[d,n_1,n_2]
$$
A $n$--legged tail is a $n$--legged tail with at least two external legs.
\Item{(ii)} If $T$ is a two--legged tail we define the two--legged tail 
$T\circ K$ by
$$\eqalign{
(T\circ K)(\psi;\varphi;\et)& = 
\int \hskip -6pt \int \lw T(\psi;\varphi';\et')\rw_{\eta',C}\cr 
& \lww \smsum_{p_1,n_1 \atop p_3+n_3 =2}K_0\tailind{p_1}{0}{p_3}{n_1}{0}{n_3} (\psi;\varphi',\ze',\varphi;\eta',\xi',\eta)\rww_{\varphi',D\atop\eta',C}
d\mu_D(\varphi')\,d\mu_C(\eta')
}$$

}

\remark{\STM\remtailendsec}{
Again, a two--legged tail with two external legs is an end in the sense of Definition \defladders. If $K(\psi;\ze,\ze',\varphi;\xi,\xi',\eta) = U(\psi+\ze+\ze'+\varphi;\xi+\xi'+\eta) - U(\psi+\ze+\ze'+\varphi;\xi+\xi')$
for some Grassmann function $U(\psi;\xi)$ then $T\circ K$ agrees with 
$T\circ {\rm Rung}(U)$ of Definitions \defladders\ and \deftails.
}

Recall that we are interested in the two-- and four--legged contributions to
the Grassmann function $f'(\psi)$ of
Theorem \theoremVb. As in Definition \defP\ 
 such contributions are extracted by 
\definition{\STM\defPII}{
The operator $P$ maps 
$\lw f(\psi;\ze;\xi)\rw_{\xi,C}$ to $f_{4,0,0}(\psi;0;0) + f_{2,0,0}(\psi;0;0)$,
when
$f = \smsum_{n_0,n_1,n_2} f_{n_0,n_1,n_2}$ with
$f_{n_0,n_1,n_2} \in A[n_0,n_1,n_2]$.
}

\definition{\STM\defeffnorm}{ Let $T(\psi;\varphi;\et)$ be an 
$n$--legged tail.
\Item i)
An element $\nu$ of the norm domain $\fN_d$ 
is said to be an effective bound for $T$ if
$$\eqalign{
N\Big(\int \lw T(\psi;\varphi;\et) \rw_{\et,C}
h\big(\psi;\varphi;\et;\xi^{(1)},\cdots,\xi^{(r)}\big)\ d\mu_C(\et)\ 
\cl \al\Big)
&\le \sfrac{\nu}{\al^2} N(h\cl \al)\cr
N_{\rm impr}\Big(\int \lw T(\psi;\varphi;\et) \rw_{\et,C}
h\big(\psi;\varphi;\et;\xi^{(1)},\cdots,\xi^{(r)}\big)\ d\mu_C(\et)\ 
\cl \al\Big)
&\le \sfrac{\nu}{\al^2} N_{\rm impr}(h\cl \al)\cr
}$$
for all Grassmann functions $h\big(\psi;\varphi;\et;\xi^{(1)},\cdots,\xi^{(r)}\big)$.
\Item ii) For $\nu\in\fN_d$, we write
$$
N_{\rm eff}(T\cl \al)\le\nu
\qquad{\Longleftrightarrow}\qquad
\hbox{$\nu$ is an effective bound for $T$}
$$
}

\remark{\STM\remeffnorm}{ 
Proposition \propestCconnect\ and the fact that 
$N_{\rm impr}(T\cl \al)\le N(T\cl \al)$ imply
$$
N_{\rm eff}(T\cl \al)\le N(T\cl \al)
$$
}

\lemma{\STM\lemtailIsec}{ Assume that $K(\psi;\ze,\ze',\varphi;\xi,\xi',\eta)$ 
has degree at most two in the variables $\ze',\varphi,\xi',\eta$ and that 
$\al \ge 2$. Let $T$ be a two--legged tail with at least $e$ external legs.
If $e\ge 4$, then
$$
 P Q(T)\cQ_K(\lw f\rw) =P Q(T\circ K)(\lw f\rw)
$$ 
where $\lw f\rw$ is shorthand for $\lw f\rw_{\xi,C}$.
More generally, if $e\ge2$, there exists a one--legged tail $t_1$ with at least $e+1$ external legs, a two--legged tail $t_2$ with at least $e+1$ external legs and a two--legged tail $\tau(\psi;\varphi;\et)$ with at least $e+2$ external legs and degree
 two\footnote{$^{(1)}$}{Hence $\tau$ is independent of $\varphi$.} 
in $\et$ such that for any Grassmann function
$f(\psi;\ze;\xi)$ the following holds:

\noindent
Set
$$
f'(\psi) = P\Big[ Q(T)\cQ_K(\lw f\rw) - Q(T\circ K)(\lw f\rw)
 - Q\big(T\circ K,T_1(K)\big)(\lw f\rw) - Q(t_1+t_2+\tau)(\lw f\rw) \Big]
$$
where
$$
T_1(K) = K_{2}\tailind{0}{0}{0}{0}{0}{2}
+ K_{2}\tailind{0}{0}{1}{0}{0}{1}
 $$
Then, 
$$
N_{\rm impr}(f'\cl \al) \le \sfrac{2^{9}\,\imp}{\al^8}\,
N(f\cl 4\al)\,\,N(T\cl 2\al)\,N(K\cl 4\al)
\Big(1+\sfrac{1}{\al^2} N(K\cl 4\al) \Big)
$$
Furthermore
$$\eqalign{
N(t_1) & \le \sfrac{1}{\al^2}\, N(T)\,N(K) \cr
N(t_2) & \le \sfrac{2}{\al^2}\, N(T)\,N(K)\cr
N_{\rm eff}(\tau) & \le \sfrac{4}{\al^4}\, N(T)\,N(K)^2 \cr
}$$
}

\prf The proof is similar to that of Lemma \lemtailI.
If $\lw f_\ell\rw_{\xi,C}=Q(T)Q(K,\cdots,K)(\lw f\rw)$, with $\ell$ $K$'s,
then, as $K$ is of degree at least two in $\psi,\ze,\xi$ and $T$ is 2--legged
with $e$ external legs, $f_\ell$ has degree at least $e+2\ell-2$. Hence
$$\deqalign{
 P\big[ Q(T)\cQ_K(\lw f\rw)\big] 
&= P \big[ Q(T)Q(K)(\lw f\rw)\big]
& \hbox{if $e\ge 4$}\cr
 P\big[ Q(T)\cQ_K(\lw f\rw)\big] 
&= P \big[ Q(T)Q(K)(\lw f\rw)\big]
+ \sfrac{1}{2} P \big[ Q(T)Q(K,K)(\lw f\rw)\big]
\hskip.5in& \hbox{if $e\ge 2$}\cr
}$$
The contribution $PQ(T)Q\Big(K_{n_0}\tailind{p_1}{p_2}{p_3}{n_1}{n_2}{n_3}\Big)$
vanishes unless
\item{$\bullet$} $p_1+n_1\le 2$, since otherwise 
$Q(T)Q\Big(K_{\cdot}\tailind{p_1}{p_2}{p_3}{n_1}{n_2}{n_3}\Big)$ is of
degree at least one in $\ze$ and $P$ sets $\ze$ to zero.
\item{$\bullet$} $n_1\in\{1,2\}$ since these fields must connect to $\et$
fields of $T$ and $T$ has degree one or two in $\et$
\item{$\bullet$} $p_2=n_2=0$ since there is only a single $K$ in 
$PQ(T)Q\Big(K_{n_0}\tailind{p_1}{p_2}{p_3}{n_1}{n_2}{n_3}\Big)$
\item{$\bullet$} $p_3+n_3\le 2$ because $K$ is of degree at most two in
 $\ze',\varphi,\xi',\et$.
\item{$\bullet$} $n_3\ge 1$ because $K$ is of degree at least one in $\et$.
\item{$\bullet$} $n_0+n_1+p_1+n_3+p_3\ge 4$ because $K$ is of degree at 
least four overall.

\noindent Hence
$$\eqalign{
P \, Q(T)Q(K) 
&= P \, Q(T)\,Q\Big( \smsum_{p_1+n_1=2 \atop p_3+n_3 =2} K_{\cdot}\tailind{p_1}{0}{p_3}{n_1}{0}{n_3}\Big) + P \, Q(T)\,Q\Big( \smsum_{ p_3+n_3 =2} 
K_{\cdot}\tailind{0}{0}{p_3}{1}{0}{n_3}\Big) \cr
&+ P \, Q(T)\,Q\Big( \smsum_{p_1+n_1=2 } 
K_{\cdot}\tailind{p_1}{0}{0}{n_1}{0}{1}\Big) 
+ P \, Q(T)\,Q\Big( 
K_{\cdot}\tailind{0}{0}{0}{1}{0}{1}\Big) \cr
&= P \big[ Q(T\circ K) + Q( k) + Q(t_{1}+t_{21}+t_{22}) \big]
}$$
where
$$\eqalign{
k(\psi;\varphi;\eta) &=  \int T_{11}(\psi;\varphi;\eta')  \big( \smsum_{ p_3+n_3 =2 \atop n_0\ge 1} 
K_{n_0}\tailind{0}{0}{p_3}{1}{0}{n_3}
(\psi;0,0,\varphi;\eta',0,\eta) \big)\, d\mu_C(\eta') \cr
t_{1}(\psi;\varphi;\eta) &= \int\hskip-6pt \int \lw T(\psi;\varphi';\eta')\rw_{\eta'}\, \big(
\smsum_{ p_1+n_1 =2 \atop n_0\ge 1} 
K_{n_0}\tailind{p_1}{0}{0}{n_1}{0}{1}\big)
(\psi;\varphi',0,0;\eta',0,\eta) \big)\,
d\mu_D(\varphi')\, d\mu_C(\eta') \cr
&\phantom{=}\hbox{is a one--legged tail with
at least $e+1$ external legs}\cr
t_{21}(\psi;\varphi;\eta) &= \int\hskip-6pt \int \hskip -2pt \lw T(\psi;\varphi';\eta')\rw_{\eta'}  \big(\hskip -4pt
\smsum_{n_0\ge 1} \smsum_{ p_1+n_1 =2 \atop p_3+n_3=2}\hskip-7pt 
K_{n_0}\tailind{p_1}{0}{p_3}{n_1}{0}{n_3}\big)
(\psi;\varphi',0,\varphi;\eta',0,\eta) \big)\,
d\mu_D(\varphi')\, d\mu_C(\eta') \cr
&\phantom{=}\hbox{is a two--legged tail with
at least $e+1$ external legs}\cr
t_{22}(\psi;\varphi;\eta) &=  \int T_{11}(\psi;\varphi;\eta')  
\big(\smsum_{n_0\ge 2} K_{n_0}\tailind{0}{0}{0}{1}{0}{1}
(\psi;0,0,0;\eta',0,\eta) \big)\, d\mu_C(\eta') \cr
&\phantom{=}\hbox{is a two--legged tail with
at least $e+2$ external legs}\cr
}$$
If $e\ge 4$,
$$
PQ(k)=PQ(t_{1})=PQ(t_{21})=PQ(t_{22})=0
$$
If $e\ge 2$, by Lemma \lemwicknorm\ and Proposition \propestCconnect,
$$
N(t_{1}),\,N(t_{21}),\,N(t_{22}) \ \le \sfrac{1}{\al^2} \,N(T)\,N(K)
$$
Furthermore $k$ has degree at least one in the 
variable $\eta$ and degree three in the variables $\varphi,\eta$. By 
Lemma \lemsplittwoQ\ and Proposition \propestCconnect
$$\eqalign{
N_{\rm impr}\big( P\,Q( k)(\lw f\rw) \big) 
&\le \sfrac{2^5}{\al^6}\, \imp\,N(k\cl 2\al)\,N(f\cl 4\al)\cr
&\le \sfrac{2^3}{\al^8}\, \imp\,N(T\cl 2\al)
N(K\cl 2\al)\,N(f\cl 4\al)\cr
}$$

\vskip1cm
We define the projection $P'$ by
$$
P'\Big( f(\psi;\ze;\xi) \Big) = f_4(\psi;0;0) + f_2(\psi;0;0)
$$
Observe that
$$\eqalign{
P \big[ Q(T)Q(K,K)(\lw f\rw)\big] 
&= P' \big[ \tilde Q(T)Q(K,K)(\lw f\rw)(\psi;\ze,0,0;\xi,0,0)\big]\cr
&= P' \big[ \tilde Q(T)Q(K,K)(\lw f\rw)(\psi;0,0,0;0,0,0)\big]\cr
}$$
so that we can apply Proposition \propimprQQ. Modulo a term whose improved norm 
$N_{\rm impr}$ is bounded by 
$\sfrac{2^9\,\imp}{\al^{10}}\,
N(f\cl 4\al)\,N(K\cl 4\al)^2\,\,N(T\cl 2\al)$
$$\eqalign{
&P' \big[ \tilde Q(T)Q(K,K)(\lw f\rw)(\psi;\ze,0,0;0,0,\xi)\big]\cr
&\hskip.5in= 2\,P \big[ Q(\tilde K_1,K)(\lw f\rw)\big] + 2\,P \big[ Q(\tilde K_2,K)(\lw f\rw)\big]
 + P \big[ Q(\hat K)(\lw f\rw)\big]
}$$
where $\tilde K_1, \tilde K_2, \hat K$ are as in Proposition \propimprQQ.
Since $K_{\cdot,1}$ has degree at least three in $\psi,\ze,\xi$, the tail 
$\hat K$ has at least six external legs, so that
$P \big[ Q(\hat K)(\lw f\rw)\big]=0$. Similarly, as
$\tilde K_2$ has degree at least three in $\psi,\ze,\xi$ and 
$K$ has degree at least two in $\psi,\ze,\xi$,
$ P \big[ Q(\tilde K_2,K)(\lw f\rw)\big] = 0$.
As
$\tilde K_1$ and $K$ have degree at most two in $\ze',\varphi,\xi',\et$ and 
degree at least four overall,
$$\eqalign{
P \big[ Q(\tilde K_1,K)(\lw f\rw)\big] 
&= P \big[ Q \big(T_1(\tilde K_1),T_1(K)\big)(\lw f\rw)\big]
    +P \big[ Q(\tau)(\lw f\rw)\big] \cr
&= P \big[ Q(T\circ K,T_1(K))(\lw f\rw)\big]
   +P \big[ Q(\tau)(\lw f\rw)\big] \cr
}$$
where 
$$\eqalign{
\tau&(\psi;\varphi;\eta)=\cr
& \smsum_{p_2+n_2=1}
\int \hskip-6pt \int 
 (\tilde K_1)_2\tailind{0}{p_2}{0}{0}{n_2}{1}
(\psi;0,\ze',0;0,\xi',\eta) \
K_2\tailind{0}{p_2}{0}{0}{n_2}{1}
(\psi;0,\ze',0;0,\xi',\eta) \,
d\mu_D(\ze')\,d\mu_C(\xi')\cr
}$$
is a two--legged tail with at least $e+2$ external legs and degree two in $\et$.
 By Proposition \propestCconnect
$$\eqalign{
N\Big( &\smsum_{p_2+n_2=1} 
(\tilde K_1)_2\tailind{0}{p_2}{0}{0}{n_2}{1}(\psi;0,\ze',0;0,\xi',\eta) \Big)\cr
& = N\bigg( \int \hskip -6pt \int 
\lw T_{11}(\psi;\varphi';\et')\rw_{\eta',C}\
\Big(\smsum_{p_2+n_2=1} K_2\tailind{1}{p_2}{0}{1}{n_2}{1}(\psi;\varphi',\ze',0;\eta',\xi',\eta)\Big)
\,d\mu_D(\varphi')\,d\mu_C(\eta')\bigg) \cr
& \le\sfrac{1}{\al^2} N(T)\,N(K)
}$$
Therefore, by Proposition \propestCconnect, for any Grassmann function $h$,
$$\eqalign{
N\Big(\int\hskip-6pt\int \lw \tau(\psi;\varphi;\et) \rw_{\et,C}&
\ h\big(\psi;\varphi;\et;\xi^{(1)},\cdots,\xi^{(r)}\big)\ d\mu_C(\et) \Big) \cr
&\le \sfrac{4}{\al^4}\, N\Big(\smsum_{p_2+n_2=1} 
(\tilde K_1)_2\tailind{0}{p_2}{0}{0}{n_2}{1}(\psi;0,\ze',0;0,\xi',\eta) \Big)\,
N(K)\,N(h) \cr
&\le \sfrac{4}{\al^6}\,N(T)\,N(K)^2\,N(h) \cr
}$$
Similarly,
$$\eqalign{
N_{\rm impr}\Big(\hskip-2pt\int\hskip-6pt\int \lw \tau(\psi;\varphi;\et) \rw_{\et,C}
\,h\big(\psi;\varphi;\et;\xi^{(1)},\cdots,\xi^{(r)}\big)\ d\mu_C(\et) 
\hskip-2pt\Big) 
&\le \sfrac{4}{\al^6}N_{\rm impr}(T)\,N_{\rm impr}(K)^2\,N_{\rm impr}(h) \cr
&\le \sfrac{4}{\al^6}\,N(T)\,N(K)^2\,N_{\rm impr}(h) \cr
}$$
 Therefore
$$
N_{\rm eff}(\tau) \le \sfrac{4}{\al^4}\, N(T)\, N(K)^2
$$
\endproof

\lemma{\STM\lemtailIIsec}{  
Assume that $K(\psi;\ze,\ze',\varphi;\xi,\xi',\eta)$ has degree at 
most two in the variables $\ze',\varphi,\xi',\eta$. Let $T_1,T_2$ be 
two--legged tails. Then there exists a one--legged tail $t_1$, a two--legged 
tail $t_2$ and a two--legged tail $\tau(\psi;\varphi;\et)$ of degree two 
in $\et$, each with at least four external legs, such that for any 
Grassmann function $f(\psi;\ze;\xi) $ the following holds:

\noindent
Set
$$
f'(\psi) = P\Big[ Q(T_1,T_2)\cQ_K(\lw f\rw) - Q(T_1\circ K,T_2\circ K)(\lw f\rw)
 - Q(t_1+t_2+\tau)(\lw f\rw) \Big]
$$
Then, if $\al \ge 2$
$$
N_{\rm impr}(f') \le \sfrac{2^{9}\,\imp}{\al^{10}}\,
N(f\cl 4\al)\,
N(T_1\cl 2\al)\,N(T_2\cl 2\al)\,
\,N(K\cl 4\al)\,\big(1+\sfrac{N(K\cl 4\al)}{\al^2}\big)
$$
and
$$\eqalign{
N(t_1) & \le \sfrac{4}{\al^4}\, N(T_1)\,N(T_2)\,N(K) \cr
N(t_2) & \le \sfrac{12}{\al^4}\, N(T_1)\,N(T_2)\,N(K) \cr
N_{\rm eff}(\tau) & \le \sfrac{8}{\al^6}\, N(T_1)\,N(T_2)\,N(K)^2 \cr
}$$

}

\prf The proof is similar to that of Lemma \lemtailII.
Again, if 
$$
\lw f_\ell\rw_{\xi,C}=Q(T_1,T_2)Q(K,\cdots,K)(\lw f\rw)
$$
with $\ell$ $K$'s,
then, as $K$ is of degree at least two in $\psi,\ze,\xi$ and $T_1, \ T_2$ are 2--legged
with at least two external legs, $f_\ell$ has degree at least $4+2\ell-4$.
Hence
$$
P\big[ Q(T_1,T_2)\cQ_K(\lw f\rw)\big] 
= P\big[ Q(T_1,T_2)Q(K)(\lw f\rw)\big]  + 
\sfrac{1}{2}\, P\big[ Q(T_1,T_2)Q(K,K)(\lw f\rw)\big] 
$$
The contribution $PQ(T_1,T_2)Q\Big(K_{n_0}\tailind{p_1}{p_2}{p_3}{n_1}{n_2}{n_3}\Big)$
vanishes unless
\item{$\bullet$} $n_0=0$ because $T_1$ and $T_2$ are each of degree at
least two in $\psi$. 
\item{$\bullet$} $p_1+n_1\le 4$, since otherwise 
$Q(T_1,T_2)Q\Big(K_0\tailind{p_1}{p_2}{p_3}{n_1}{n_2}{n_3}\Big)$ is of
degree at least one in $\ze$ and $P$ sets $\ze$ to zero.
\item{$\bullet$} $n_1\ge 2$ since these fields must connect to $\et$
fields of $T_1,\ T_2$ which have combined degree at least two in $\et$
\item{$\bullet$} $p_2=n_2=0$ since there is only a single $K$ in 
$PQ(T_1,T_2)Q\Big(K_{n_0}\tailind{p_1}{p_2}{p_3}{n_1}{n_2}{n_3}\Big)$
\item{$\bullet$} $p_3+n_3\le 2$ because $K$ is of degree at most two in
 $\ze',\varphi,\xi',\et$.
\item{$\bullet$} $n_3\ge 1$ because $K$ is of degree at least one in $\et$.
\item{$\bullet$} $n_1+p_1+n_3+p_3\ge 4$ because $K$ is of degree at 
least four overall.

\noindent Hence

$$
P\big[ Q(T_1,T_2)Q(K)(\lw f\rw)\big] 
= P\big[ Q(t_{11}+t_{21}+t_{22}+t_{23})(\lw f\rw)\big]
+ P\big[ Q(k)(\lw f\rw)\big]
$$
where
$$\eqalign{
t_{11}(\psi;\varphi;\eta) &= \int  \hskip -6pt \int
\lW T_1(\psi;\varphi';\eta')T_2(\psi;\varphi';\eta')\rW_{\varphi',D\atop\eta',C}\,
 \Big( \smsum_{p_1+n_1 =4} 
K_0\tailind{p_1}{0}{0}{n_1}{0}{1}
(0;\varphi',0,0;\eta',0,\eta) \Big) \cr
& \hskip 9.5cm d\mu_D(\varphi')\,d\mu_C(\eta')\cr
t_{21}(\psi;\varphi;\eta) 
&= \int  \hskip -6pt \int
\lW T_1(\psi;\varphi';\eta')T_2(\psi;\varphi';\eta')\rW_{\varphi',D\atop\eta',C}
\, \Big( \smsum_{p_1+n_1 =4\atop p_3+n_3=2} 
K_0\tailind{p_1}{0}{p_3}{n_1}{0}{n_3}(0;\varphi',0,\varphi;\eta',0,\eta) \Big) \cr
& \hskip 9.5cm d\mu_D(\varphi')\,d\mu_C(\eta')\cr
t_{22}(\psi;\varphi;\eta) 
&= \int  \hskip -6pt \int
\lW T_1(\psi;\varphi;\eta')T_2(\psi;\varphi';\eta')\rW_{\eta',C}
\, \Big( \smsum_{p_1+n_1 =3} 
K_0\tailind{p_1}{0}{0}{n_1}{0}{1}(0;\varphi',0,\varphi;\eta',0,\eta) \Big) \cr
& \hskip 9.5cm d\mu_D(\varphi')\,d\mu_C(\eta')\cr
t_{23}(\psi;\varphi;\eta) 
&= \int  \hskip -6pt \int
\lW T_1(\psi;\varphi';\eta')T_2(\psi;\varphi;\eta')\rW_{\eta',C}
\, \Big( \smsum_{p_1+n_1 =3} 
K_0\tailind{p_1}{0}{0}{n_1}{0}{1}(0;\varphi',0,\varphi;\eta',0,\eta) \Big) \cr
& \hskip 9.5cm d\mu_D(\varphi')\,d\mu_C(\eta')\cr
k(\psi;\varphi;\eta) 
&= \int  \hskip -6pt \int
\lW T_1(\psi;\varphi;\eta')T_2(\psi;\varphi;\eta')\rW_{\eta',C}
\, \, \Big( \smsum_{p_3+n_3 =2\atop n_3\ge 1} \hskip-2pt
K_0\tailind{0}{0}{p_3}{2}{0}{n_3}(0;0,0,\varphi;\eta',0,\eta) \Big) d\mu_C(\eta')\cr
}$$
Here, each $t_{ij}$ is a $i$--legged tail with at least four external legs,
that, by Lemma \lemwicknorm\ and Proposition \propestCconnect, with $\ell=2$, fulfills
$$
N(t_{ij})\ \le  \ \sfrac{4}{\al^4}\, N(T_1)\,N(T_2)\,N(K)
$$
Furthermore $k$ has degree at least one in the 
variable $\eta$ and degree three in the variables $\varphi,\eta$. By 
Lemma \lemsplittwoQ\ and Proposition \propestCconnect\ with $\ell=2$ and
$\al$ replaced by $2\al$,
$$\eqalign{
N_{\rm impr}\big( P\,Q( k)(\lw f\rw) \big) 
&\le \sfrac{2^5}{\al^6}\, \imp\,N(k\cl 2\al)\,N(f\cl 4\al)\cr
&\le \sfrac{2^3}{\al^{10}}\, \imp\,N(T_1\cl 2\al)N(T_2\cl 2\al)
N(K\cl 2\al)\,N(f\cl 4\al)\cr
}$$

\vskip1cm 
By Proposition \propoverldecQ 
$$
Q(T_1,T_2)Q(K,K)(\lw f\rw)
=\ \lww \int \hskip -6pt \int 
\lw T_1(\psi;\varphi;\et)\rw_{\eta,C}\,
\lW g(\psi;\ze,0,\varphi;\xi,0,\eta)\rW_{\varphi,D\atop\eta,C} 
\,d\mu_D(\varphi)\,d\mu_C(\eta) \rww_{\xi,C} 
$$
where $g = \tilde Q(T_2)Q(K,K)(\lw f\rw)$. In particular
$$
P\,Q(T_1,T_2)\,Q(K,K)(\lw f\rw)
=P'\int \hskip -6pt \int 
\lw T_1(\psi;\varphi;\et)\rw_{\eta,C}\,
\lW g(\psi;0,0,\varphi;0,0,\eta)\rW_{\varphi,D\atop\eta,C}
\,d\mu_D(\varphi)\,d\mu_C(\eta) 
$$
By Proposition \propimprQQ 
$$
g(\psi;0,0,\varphi;0,0,\eta)
=\tilde g_1(\psi;\varphi;\et) + \tilde g_2(\psi;\varphi;\et)
 +\hat g(\psi;\varphi;\et) + h(\psi;\varphi;\et)
$$
where
$$\eqalign{
\tilde g_1(\psi;\varphi;\et) &= 2Q(\tilde K_1,K)(\lw f\rw)(\psi;\varphi;\et) \cr
\tilde g_2(\psi;\varphi;\et) &= 2Q(\tilde K_2,K)(\lw f\rw)(\psi;\varphi;\et) \cr
\hat g(\psi;\varphi;\et) &= Q(\hat K)(\lw f\rw)(\psi;\varphi;\et) \cr
}$$
and
$$
N_{\rm impr}(h) \le 
\sfrac{2^{9}\,\imp}{\al^{10}}\,
N(f\cl 4\al)\,N(K\cl 4\al)^2\,N(T_2\cl 2\al)
$$
By Lemma \lemwicknorm\ and Proposition \propestCconnect, the improved norm of
$$
\int \hskip -4pt \int 
\lw T_1(\psi;\varphi;\et)\rw_{\eta,C}\,
\lW h(\psi;\varphi;\et)\rW_{\varphi,D\atop\eta,C} 
\,d\mu_D(\varphi)\,d\mu_C(\eta)
$$
is bounded by
$$
\sfrac{1}{\al^2}N_{\rm impr}(T_1)N_{\rm impr}(h)
\le \sfrac{2^9\,\imp}{\al^{12}}\,
N(f\cl 4\al)\,N(K\cl 4\al)^2\,
N(T_1\cl \al)\,N(T_2\cl 2\al)
$$
Observe that $\hat g$ has degree at least six, so that
$$
P'\int \hskip -6pt \int 
\lw T_1(\psi;\varphi;\et)\rw_{\eta,C}\,
\lW \hat g(\psi;\varphi;\et)\rW_{\varphi,D\atop\eta,C} 
\,d\mu_D(\varphi)\,d\mu_C(\eta')  = 0
$$
Similarly, $\tilde g_2$ has degree at least five, so that
$$
P'\int \hskip -6pt \int 
\lw T_1(\psi;\varphi;\et)\rw_{\eta,C}\,
\lW \tilde g_2(\psi;\varphi;\et)\rW_{\varphi,D\atop\eta,C} 
\,d\mu_D(\varphi)\,d\mu_C(\eta')  = 0
$$
Finally, the contribution to $\tilde g_1$ with $\psi$--degree two and overall
degree four is
$\tilde g_{11}+\tilde g_{12}$, where
$$\eqalign{
\tilde g_{11}
& = 2Q \Big( T_2\circ K, \smsum_{p_1+n_1=2 \atop p_3+n_3=2}
K_0\tailind{p_1}{0}{p_3}{n_1}{0}{n_3} \Big)(\lw f\rw)
\cr
\tilde g_{12}
& = 2Q \Big( p, \smsum_{p_1+n_1=2 \atop p_2+n_2=1}
K_0\tailind{p_1}{p_2}{0}{n_1}{n_2}{1} \Big)(\lw f\rw)
\cr
}$$
where
$$\eqalign{
p(\psi;\ze',\varphi;\xi',\eta)
&= \int \hskip -6pt \int
\lw T_2(\psi;\varphi';\et')\rw_{\eta',C} \ 
\Big( \smsum_{p_1+n_1=2 \atop p_2+n_2=1}
K_0\tailind{p_1}{p_2}{0}{n_1}{n_2}{1}
(\psi;\varphi',\ze',\varphi;\eta',\xi',\eta) \Big) \cr  
&\hskip9cm \,d\mu_D(\varphi')\,d\mu_C(\eta')  \cr
}$$
Then
$$
P'\int \hskip -6pt \int 
\lw T_1(\psi;\varphi;\et)\rw_{\eta,C}\,
\lW \tilde g_{11}(\psi;\varphi;\et)\rW_{\varphi,D\atop\eta,C} 
\,d\mu_D(\varphi)\,d\mu_C(\eta) 
=2P\, Q(T_1\circ K,T_2\circ K)(\lw f\rw)
$$
while
$$
P'\int \hskip -6pt \int 
\lw T_1(\psi;\varphi;\et)\rw_{\eta,C}\,
\lW \tilde g_{12}(\psi;\varphi;\et)\rW_{\varphi,D\atop\eta,C} 
\,d\mu_D(\varphi)\,d\mu_C(\eta) 
=P\,Q(\tau)(\lw f\rw)
$$
with the two--legged tail, with four external legs and degree two in $\et$,
$$\eqalign{
\tau &= 2\int\,
p(\psi;\ze',\varphi;\xi',\eta)\ p'(\psi;\ze',\varphi;\xi',\eta) 
\ d\mu_D(\ze')\,d\mu_C(\xi')
}$$
where $p$ was defined above and
$$\eqalign{
p'(\psi;\ze',\varphi;\xi',\eta)
&= \int \hskip -6pt \int
 \lw T_1(\psi;\varphi'',\eta'')\rw_{\eta'',C}
\Big( \smsum_{p_1+n_1=2 \atop p_2+n_2=1}
K_0\tailind{p_1}{p_2}{0}{n_1}{n_2}{1}
(\psi;\varphi^{\prime\prime},\ze',\varphi;\eta^{\prime\prime},\xi',\eta)\Big) \cr  
&\hskip8.5cm \,d\mu_D(\varphi'')\,d\mu_C(\eta'')  \cr
}$$
By Lemma \lemwicknorm\ and  Proposition \propestCconnect,
$$\eqalign{
N(p)&\le\sfrac{1}{\al^2}N(T_2)N(K) \cr
N(p')&\le\sfrac{1}{\al^2}N(T_1)N(K) \cr
}$$
By Lemma \lemwicknorm\ and  Proposition \propestCconnect,  with $\ell=2$,
$$\eqalign{
N\Big(\int\hskip-6pt\int \lw \tau(\psi;\varphi;\et) \rw_{\et,C}\ 
h\big(\psi;\varphi;\et;\xi^{(1)},\cdots,\xi^{(r)}\big)\ d\mu_C(\et) \Big) \ 
&\le\ \sfrac{8}{\al^4}N(p)N(p')N(h)\cr
&\le\ \sfrac{8}{\al^8}N(T_1)N(T_2)N(K)^2N(h)\cr
}$$
for all Grassmann functions $h$. A similar estimate applies for the $N_{\rm impr}$ norm so that
$$
N_{\rm eff}(\tau)\ \le\  \sfrac{8}{\al^6}N(T_1)N(T_2)N(K)^2
$$
\endproof

\lemma{\STM\lemtailIIIsec}{  Assume that $K$ has degree at most two in the variables $\ze',\varphi,\xi',\eta$.
Let $T$ be a one--legged tail with at least three external legs. Then there is a
two--legged tail $t_2$ with at least four external legs such that for all Grassmann
functions $f(\psi;\ze,\xi) $
$$
P\big[ Q(T)\cQ_K(\lw f\rw)\big] = P\big[Q(t_2)(\lw f\rw)\big]
$$
and $$
N(t_2) \le  \sfrac{2}{\al^2} \,N(T)\,N(K)
$$
}

\prf The proof is similar to that of Lemma \lemtailIII.
Since $T$ is one--legged and $K$ has degree at least four
$$\eqalign{
P\,Q(T)\,\cQ_K & = P\,Q(T)\,Q(K) \cr
& = P\,Q(T)\,
Q\Big(\smsum_{p_3+n_3=2}
  K_1\tailind{0}{0}{p_3}{1}{0}{n_3} \Big) \ =\ P\,Q(t_2)
}$$
with 
$$
t_2(\psi;\varphi;\et) = \int T(\psi;\varphi;\eta')\Big(
 K_1\tailind{0}{0}{1}{1}{0}{1}
        (\psi;0,0,\varphi;\eta',0,\eta)
+ K_1\tailind{0}{0}{0}{1}{0}{2}
        (\psi;0,0,0;\eta',0,\eta)
\Big)\,d\mu_C(\eta')
$$
\endproof

\definition{\STM\defeffecttailsec}{ The two--legged tails $T_\ell(K)$ are recursively defined as follows: $T_1(K)$ was defined in Lemma \lemtailIsec, and
$$
T_{\ell+1}(K) = T_\ell \circ K \qquad \qquad {\rm for\ } \ell \ge 1  
$$
}

\remark{\STM\remeffecttailsec}{
\Item{(i)}
Clearly $T_\ell(K)$ has two external legs. Using Lemma \lemwicknorm\ and Proposition \propestCconnect\ one proves by induction that for $\al \ge 2$
$$
N\big( T_\ell(K) \big) \le \sfrac{1}{\al^{2\ell-2}}\, N(K)^\ell
$$
\Item{(ii)} If $K(\psi;\ze,\ze',\varphi;\xi,\xi',\eta) = U(\psi+\ze+\ze'+\varphi;\xi+\xi'+\eta) - U(\psi+\ze+\ze'+\varphi;\xi+\xi')$
for some even Grassmann function $U(\psi;\xi)$ then, by Remark \remtailendsec,
$T_\ell(K) = T_\ell(U)$. Recall that $T_\ell(U)$ was defined in Definition
\deftails.
}

\sect{ Proof of Theorem \theoremVb\ in the general case}\PG\pgRIXc

First, we prove the analog of Proposition \propprodtails\ for the enlarged
algebra.

\proposition{\STM\propprodtailssec}{
Let $K(\psi;\ze;\xi,\xi',\eta)$ be an even Grassmann function that vanishes for $\eta=0$ and has degree at least four overall. Set
$$
\bar K(\psi;\ze,\ze',\varphi;\xi,\xi',\eta) =
K(\psi;\ze+\ze'+\varphi;\xi,\xi',\eta)
$$
Assume that $\al \ge 8$ and $N(\bar K\cl 4\al)_\0< \sfrac{2\al}{3}$.
Furthermore let $ f(\psi;\ze;\xi) $ be a Grassmann function.

\noindent
For each $n \ge 1$ there exists a Grassmann function $h_n(\psi;f)$, a 
one--legged tail $t_1$, a two--legged tail $t_2$, each with at least three external legs
and a two--legged tail $\tau(\psi;\varphi;\et)$ with at least four
 external legs and of degree two in $\et$ such that 
$$\eqalign{
P\hskip -3pt \int \hskip -2pt{\rm Ev}\ \cR^n_{\lw K \rw_{\ze,D},C}
   \big( \lw  f \rw_{\xi,C \atop\ze,D} \big)\, d\mu_D(\ze)
= P \Big[ Q\big(T_n(\bar K)\big)(\lw  f \rw_{\xi,C} ) 
 &+\sfrac{1}{2}\hskip -.3cm\smsum_{\ell,\ell'\ge 1 \atop \max\{\ell,\ell'\}=n} 
   \hskip -.4cm   
   Q\big(T_\ell(\bar K),T_{\ell'}(\bar K)\big)(\lw  f \rw_{\xi,C}) \cr
 &\hskip-.5cm+Q(t_1+t_2+\tau)(\lw  f\rw_{\xi,C}) \Big] +h_n(\psi;f) \cr 
}$$ 
and
$$\eqalign{
N(t_1),\,N(t_2)\  &\le\ 
4\big(\sfrac{8}{\al^2}\big)^{n-1}\,\sfrac{N(\bar K\cl \al)^n}
{1-{2\over\al^2} N(\bar K\cl \al)} 
\qquad\qquad
N_{\rm eff}(\tau)\le\ \sfrac{8}{\al^{2n}}\,
\sfrac{N(\bar K\cl \al)^{n+1}}{1-{1\over\al^2}N(\bar K\cl \al)}
\cr
N_{\rm impr}\big(h_n(\psi;f)\cl \al\big) 
&\le \sfrac{2^{10}\,\imp}{\al^5}\, \,\smsum_{m=0}^{n-1}
\sfrac{1}{\al^{n-m}}\ 
\sfrac{N(\bar K\cl 4\al)^{n-m}}{1-{3\over 2\al}N(\bar K\cl 4\al) } \ 
N\big( F_m(\psi;f)\cl 4\al\big)
\cr
}$$
where
$$
\lw F_m(\psi;f) \rw_{\xi,C \atop\ze,D} = {\rm Ev}\, \cR_{\lw K \rw_{\ze,D},C}^m
 \big( \lw  f \rw_{\xi,C \atop\ze,D} \big)
$$
}

\prf 
Decompose
$$
\bar K = K' + K^{\prime\prime} 
$$
where $K'$ has degree at most two in the variables $\ze',\xi',\varphi,\eta$ and
$K^{\prime\prime}$ has degree at least three in these variables.

\noindent
We perform induction on $n$. For $n=1$, by Lemma \lemQWick.ii
$$
{\rm Ev}\,\cR_{\lw K \rw_{\ze,D},C}\big(\lw  f \rw_{\xi,C \atop \ze,D} \big)
= \lw \cQ_{\bar K}(\lw  f \rw_{\xi,C}) \rw_{\ze,D}
$$
and therefore
$$
\int {\rm Ev}\,\cR_{\lw K \rw_{\ze,D},C}\big(\lw  f \rw_{\xi,C \atop\ze,D} \big)\,d\mu_D(\ze)
=  \cQ_{\bar K}(\lw  f \rw_{\xi,C})(\psi,0;\xi)
$$
Set
$$
h_1(\psi;f) = P\,(\cQ_{\bar K}-\cQ_{K'})(\lw  f \rw_{\xi,C})
$$
By Proposition \propsplittwoQ
$$
N_{\rm impr}\big(h_1(\psi;f)\big)
\le \sfrac{2^5\,\imp}{\al^6}\,N\big( f\cl 4\al\big)\,
\sfrac{N(\bar K\cl 2\al)}{1-{1\over\al}N(\bar K\cl 2\al)}
$$
Since $K'$  has degree at most two in the variables $\ze',\varphi,\xi',\eta$ and degree at least four overall
$$\eqalign{
P\, \cQ_{K'}
& = P\,Q \Big(
    K_{\cdot}\tailind{0}{0}{0}{0}{0}{1}
  + K_{\cdot}\tailind{0}{0}{0}{0}{0}{2}
  + K_{\cdot}\tailind{0}{0}{1}{0}{0}{1}
  \Big)\cr 
&  +\sfrac{1}{2}\, P\,Q \Big(
    K_2\tailind{0}{0}{0}{0}{1}{1}
  + K_2\tailind{0}{1}{0}{0}{0}{1}\ ,\ 
    K_2\tailind{0}{0}{0}{0}{1}{1}
  + K_2\tailind{0}{1}{0}{0}{0}{1} 
  \Big)\cr 
&  +\sfrac{1}{2}\, P\,Q \Big(
    K_2\tailind{0}{0}{0}{0}{0}{2}
  + K_2\tailind{0}{0}{1}{0}{0}{1}\ ,\ 
    K_2\tailind{0}{0}{0}{0}{0}{2}
  + K_2\tailind{0}{0}{1}{0}{0}{1} 
  \Big)\cr 
& = P\,Q\big(T_1(\bar K)\big) 
  + P\,Q(t_1+t_2)
  +\sfrac{1}{2}\, P\,Q\big(T_1(\bar K),T_1(\bar K)\big) \cr
}$$
with
$$\eqalign{
t_1 &=  K_3\tailind{0}{0}{0}{0}{0}{1} 
  + K_4\tailind{0}{0}{0}{0}{0}{1} \cr
t_2 &=  K_3\tailind{0}{0}{0}{0}{0}{2} 
  + K_4\tailind{0}{0}{0}{0}{0}{2} 
      + K_3\tailind{0}{0}{1}{0}{0}{1} 
  + K_4\tailind{0}{0}{1}{0}{0}{1}  
+ \sfrac{1}{2}\,\int 
    K_2\tailind{0}{0}{0}{0}{1}{1}\, 
    K_2\tailind{0}{0}{0}{0}{1}{1}
    \,d\mu_C(\xi')\cr
\tau &=   \sfrac{1}{2}\,\int 
    K_2\tailind{0}{1}{0}{0}{0}{1}\, 
    K_2\tailind{0}{1}{0}{0}{0}{1}
    \,d\mu_D(\ze')\cr
}$$
In particular 
$$\deqalign{
N(t_1) & \le N(\bar K) \cr
N(t_2) & \le N(\bar K)+\sfrac{1}{2\al^2}\,N(\bar K)^2
&\le \sfrac{N(\bar K)}{1-{1\over 2\al^2}N(\bar K)}\cr
N_{\rm eff}(\tau) & \le\sfrac{1}{2}\sfrac{4}{\al^2}\,N(\bar K)^2
& \le\sfrac{2}{\al^2}\,N(\bar K)^2 \cr
}$$
Proposition, \propestCconnect, with $\ell=1$, was used to bound the last
term of $t_2$. Lemma \lemwicknorm\ and Proposition \propestCconnect, 
with $\ell=2$, were used to bound 
$\int \lw \tau(\psi;\varphi;\et) \rw_{\et,C}\ 
h\big(\psi;\varphi;\et;\xi^{(1)},\cdots,\xi^{(r)}\big)\ d\mu_C(\et)$
as in Definition \defeffnorm.

Now assume that the statement of the Lemma is true for $n$.
By Remark \remEv.ii
$$
{\rm Ev}\,\cR^{n+1}_{\lw K \rw_{\ze,D},C}
  \big(\lw  f \rw_{\xi,C \atop\ze,D} \big)
= {\rm Ev}\ \cR^n_{\lw K \rw_{\ze,D},C}\,
\Big(\lw F_1(\psi;f) \rw_{\xi,C \atop\ze,D} \Big) 
$$
By Lemma \lemQWick.ii
$$
\lw F_1(\psi;f) \rw_{\xi,C } = Q_{\bar K}(\lw  f \rw_{\xi,C})
$$
The induction hypothesis therefore implies that
$$\eqalign{
&P\hskip -3pt \int \hskip -2pt{\rm Ev}\ \cR^{n+1}_{\lw K \rw_{\ze,D},C}
   \big( \lw  f \rw_{\xi,C \atop\ze,D} \big)\, d\mu_D(\ze)
=P \int {\rm Ev}\  \cR^n_{\lw K \rw_{\ze,D},C}
   \big( \lw \cQ_{\bar K}(\lw  f \rw_{\xi,C}) \rw_{\ze,D} \big)\, d\mu_D(\ze)\cr
&\hskip.3cm= \ h_n(\psi;F_1(\psi;f))  
+ P Q\big(T_n(\bar K)\big) 
\big(\cQ_{\bar K}( \lw  f \rw_{\xi,C})\big)
 + \sfrac{1}{2} \hskip -.5cm 
\smsum_{\ell,\ell'\ge 1 \atop \max\{\ell,\ell'\}=n} 
 \hskip -.4cm P\,Q\big(T_\ell(\bar K),T_{\ell'}(\bar K)\big)
\big(\cQ_{\bar K}(\lw  f \rw_{\xi,C}) \big)\cr
&\hskip8cm +P\,Q(t_1+t_2+\tau)\big(\cQ_{\bar K}(\lw  f \rw_{\xi,C})\big)  
}$$
with
$$
N(t_1),\,N(t_2)\ 
\le\ 4\big(\sfrac{8}{\al^2}\big)^{n-1}\,\sfrac{N(\bar K)^n}
{1-{2\over\al^2} N(\bar K)}
\qquad\qquad
N_{\rm eff}(\tau)\ 
\le\ \sfrac{8}{\al^{2n}}\,
\sfrac{N(\bar K)^{n+1}}{1-{1\over\al^2}N(\bar K)}
$$
and
$$
N_{\rm impr}\big( h_n(\psi;F_1(\psi;f))\big)
\le \sfrac{2^{10}\,\imp}{\al^5}\, \,\smsum_{m=0}^{n-1}
\sfrac{1}{\al^{n-m}}\ 
\sfrac{N(\bar K\cl 4\al)^{n-m}}{1-{3\over 2\al}N(\bar K\cl 4\al) }\ 
N\Big( F_m\big(\psi;F_1(f;\psi)\big)\cl 4\al\Big)
$$
By Remark \remEv.ii
$$\eqalign{
\lW F_m\big(\psi;F_1(\psi;f)\big)\rW_{\xi,C \atop\ze,D} 
&= {\rm Ev}\ \cR^m_{\lw K \rw_{\ze,D},C}\,
\Big(\lW F_1(\psi;f) \rW_{\xi,C \atop\ze,D} \Big) \cr
&= {\rm Ev}\ \cR^m_{\lw K \rw_{\ze,D},C}\,
\Big({\rm Ev}\ \cR_{\lw K \rw_{\ze,D},C} \big(\lW f \rW_{\xi,C \atop\ze,D}\big) \Big) \cr
& = {\rm Ev}\ \cR^{m+1}_{\lw K \rw_{\ze,D},C}\,
\big(\lW f \rW_{\xi,C \atop\ze,D}\big) \cr
& = \lW F_{m+1}(\psi;f)\rW_{\xi,C \atop\ze,D}
}$$
Therefore
$$
N_{\rm impr}\big( h_n(\psi;F_1(\psi;f))\big)
\le \sfrac{2^{10}\,\imp}{\al^5}\, \,\smsum_{m=1}^{n}
\sfrac{1}{\al^{n+1-m}}
\sfrac{N(\bar K\cl 4\al)^{n+1-m}}{1-{3\over 2\al}N(\bar K\cl 4\al) } 
N\big( F_{m}(\psi;f)\cl 4\al\big)
$$

Let $ g_0$ be the difference between
$$\eqalign{
&P\Big[ Q\big(T_n(\bar K)\big) \big(\cQ_{\bar K}(\lw  f \rw_{\xi,C})\big)
+ \sfrac{1}{2} \hskip -.5cm \smsum_{\ell,\ell'\ge 1 \atop \max\{\ell,\ell'\}=n} 
 \hskip -.4cm Q\big(T_\ell(\bar K),T_{\ell'}(\bar K)\big)
\big(\cQ_{\bar K}(\lw  f \rw_{\xi,C})\big)\cr
\noalign{\vskip-15pt}
&\hskip9cm +Q(t_1+t_2+\tau)\big(\cQ_{\bar K}(\lw  f \rw_{\xi,C})\big) \Big]
}$$
and 
$$\eqalign{
&P\Big[ Q\big(T_n(\bar K)\big) \big(\cQ_{K'}(\lw  f \rw_{\xi,C})\big)
+ \sfrac{1}{2} \hskip -.5cm \smsum_{\ell,\ell'\ge 1 \atop \max\{\ell,\ell'\}=n} 
 \hskip -.4cm Q\big(T_\ell(\bar K),T_{\ell'}(\bar K)\big)
 \big(\cQ_{K'}(\lw  f \rw_{\xi,C})\big)\cr
\noalign{\vskip-15pt}
&\hskip9cm +Q(t_1+t_2+\tau)\big(\cQ_{K'}(\lw  f \rw_{\xi,C})\big) \Big]
}$$
Let $ \lw f''\rw_{\xi,D} = \cQ_{\bar K}(\lw  f \rw_{\xi,C}) -\cQ_{K'}(\lw  f \rw_{\xi,C})$.
By Lemma \normestQ\ and  Definition \defeffnorm
$$\eqalign{
&N_{\rm impr}(g_0)
\le \sfrac{1}{\al} \,
N_{\rm impr}\big( f''\cl \al\big) 
\Big[ N\big( T_n(\bar K)\big) + \sfrac{1}{\al} \hskip -.4cm
\smsum_{\ell,\ell'\ge 1 \atop \max\{\ell,\ell'\}=n} \hskip -.4cm
  N\big( T_\ell(\bar K)\big)\,N\big( T_{\ell'}(\bar K)\big) \cr
\noalign{\vskip-15pt}
&\hskip9cm+N(t_1)+N(t_2)+\nu\Big]
\cr
}$$
if $\nu$ is any effective bound for $\tau$.
By Proposition \propsplittwoQ, Remark \remeffecttailsec\ and the induction hypothesis,
$$\eqalign{
N_{\rm impr}(g_0)
& \le \sfrac{2^5\imp}{\al^7}\,N( f\cl 4\al)  \,
\sfrac{N(\bar K\cl 2\al)}{1-{1\over\al}N(\bar K\cl 2\al)}\,
 \Big[ \sfrac{1}{\al^{2n-2}} N(\bar K)^n 
\Big(1+ 2\smsum_{\ell=1}^n \sfrac{N(\bar K)^\ell }{\al^{2\ell-1}} \Big) \cr
&\hskip2in
+8\big(\sfrac{8}{\al^2}\big)^{n-1}\,\sfrac{N(\bar K)^n}
{1-{2\over\al^2} N(\bar K)}+\sfrac{8}{\al^{2n}}\,
\sfrac{N(\bar K)^{n+1}}{1-{1\over\al^2}N(\bar K)} \Big]\cr
}$$
Since
$$\eqalign{
1+ 2\smsum_{\ell=1}^n \sfrac{N(\bar K)^\ell }{\al^{2\ell-1}} 
&=1+ 2\al\smsum_{\ell=1}^n \sfrac{N(\bar K)^\ell }{\al^{2\ell}} 
\le 1+2\al \sfrac{{1\over\al^2} N(\bar K)}{1-{1\over\al^2} N(\bar K)}
=\sfrac{1-{1\over\al^2} N(\bar K)+{2\over\al} N(\bar K)}
{1-{1\over\al^2} N(\bar K)}\cr
&\le 8\sfrac{1-{1\over 8\al^2} N(\bar K)+{1\over4\al} N(\bar K)}
{1-{1\over\al^2} N(\bar K)}
\le 8\sfrac{1}{1-{1\over\al^2} N(\bar K)}
\sfrac{1}{1+{1\over 8\al^2} N(\bar K)-{1\over4\al} N(\bar K)}\cr
&\le 8\sfrac{1}{1-\big({1\over4\al}+{7\over8\al^2} \big)N(\bar K)}
\le 8\sfrac{1}{1-{1\over2\al}N(\bar K)}
}$$
we have
$$\eqalign{
N_{\rm impr}(g_0)& \le \sfrac{2^5\imp}{\al^7}\,N( f\cl 4\al)  \,
\sfrac{N(\bar K\cl 2\al)^{n+1}}{1-{1\over\al}N(\bar K\cl 2\al)}\,
 \Big[ \sfrac{8}{\al^{2n-2}}\sfrac{1}{1-{1\over2\al}N(\bar K)}
+8\big(\sfrac{8}{\al^2}\big)^{n-1}\,\sfrac{1}{1-{2\over\al^2} N(\bar K)}
\,\sfrac{1}{1-{1\over\al^2} N(\bar K)}\Big]\cr
& \le \sfrac{2^5\imp}{\al^7}\,N( f\cl 4\al)  \,
\sfrac{N(\bar K\cl 2\al)^{n+1}}{1-{1\over\al}N(\bar K\cl 2\al)}\,
 \Big[ \sfrac{8}{\al^{2n-2}}\sfrac{1}{1-{1\over2\al}N(\bar K)}
+8\big(\sfrac{8}{\al^2}\big)^{n-1}\,\sfrac{1}{1-{1\over2\al} N(\bar K)}\Big]\cr
& \le \sfrac{2^5\imp}{\al^{n+6}}\,N( f\cl 4\al)  \,
\sfrac{N(\bar K\cl 2\al)^{n+1}}{1-{3\over 2\al}N(\bar K\cl 2\al)}\,[8+8]\cr
& \le \sfrac{2^9\imp}{\al^{n+6}}\,N( f\cl 4\al)  \,
\sfrac{N(\bar K\cl 2\al)^{n+1}}{1-{3\over 2\al}N(\bar K\cl 2\al)}\cr
}$$

By Lemma \lemtailIsec\ and Remark \remfunctnorm, there exists a one--legged tail $t_{11}$, a two--legged tail $t_{21}$, each with at least three external legs, and a two--legged tail $\tau_1$ with at least four
 external legs and of degree two in $\et$ such that
$$\eqalign{
P\Big[ Q\big(T_n(\bar K)\big)
\big(\cQ_{K'}(\lw  f \rw_{\xi,C})\big)\Big]
&= P\big[ Q\big(T_{n+1}(\bar K)\big)(\lw  f \rw_{\xi,C}) \big]
 + P\big[ Q\big(T_{n+1}(\bar K),T_1(\bar K)\big)(\lw  f \rw_{\xi,C}) \big] \cr
&+ P\big[Q(t_{11}+t_{21}+\tau_1)(\lw  f \rw_{\xi,C})\big]+g_1(\psi)
}$$
with
$$\eqalign{
N_{\rm impr}(g_1) 
&\le \sfrac{2^{9}\,\imp}{\al^8}\,N( f\cl 4\al)\, 
N(T_n(\bar K)\cl 2\al) \,N(\bar K\cl 4\al)
\big(1+\sfrac{1}{\al^2}N(\bar K\cl 4\al)\big) \cr
&\le \sfrac{2^{9}\,\imp}{\al^{2n+6}}\,N( f\cl 4\al)\, 
N(\bar K\cl 2\al)^{n}
\sfrac{N(\bar K\cl 4\al)}{1-{1\over\al^2}N(\bar K\cl 4\al)} \cr
&\le \sfrac{2^{9}\,\imp}{\al^{2n+6}}\,N( f\cl 4\al)\,  
\sfrac{N(\bar K\cl 4\al)^{n+1}}{1-{1\over\al^2}N(\bar K\cl 4\al)} \cr
}$$
and
$$\deqalign{
N(t_{11}),\ N(t_{21})\ 
&\le\ \sfrac{2}{\al^2} N\big(T_n(\bar K)\big)\, N(\bar K) 
&\le\ \sfrac{2}{\al^{2n}} N(\bar K)^{n+1} \cr
N_{\rm  eff}(\tau_1)\ 
&\le\ \sfrac{4}{\al^4} N\big(T_n(\bar K)\big)\, N(\bar K)^2
&\le\ \sfrac{4}{\al^{2n+2}} N(\bar K)^{n+2}
\cr
}$$
Similarly, for $1\le \ell\le n$, by Lemma \lemtailIIsec, there exists a one--legged tail $t_{12}^{(\ell)}$ and a two--legged tail $t_{22}^{(\ell)}$, each with at least three external legs, and 
a two--legged tail $\tau^{(\ell)}$ with at least four
 external legs and of degree two in $\et$
such that
$$\eqalign{
P\Big[ Q\big(T_\ell(\bar K),T_n(\bar K)\big)
\big(\cQ_{K'}(\lw  f \rw_{\xi,C})\big) \Big]
&= P\Big[ Q\big(T_{\ell+1}(\bar K),T_{n+1}(\bar K)\big)
(\lw  f \rw_{\xi,C}) \Big] \cr
&+ P\big[Q(t_{12}^{(\ell)}+t_{22}^{(\ell)}+\tau^{(\ell)})(\lw  f \rw_{\xi,C})\big]
+g_{2,\ell}(\psi) \cr
}$$
with
$$\eqalign{
N_{\rm impr}(g_{2,\ell}) 
&\le \sfrac{2^{9}\,\imp}{\al^{10}}\,N( f\cl 4\al)\,
\,N(T_n(\bar K)\cl 2\al) \,N(T_{\ell}(\bar K)\cl 2\al)
\,N(\bar K\cl 4\al) \big(1+\sfrac{1}{\al^2}N(\bar K\cl 4\al)\big) \cr
&\le \sfrac{2^{9}\,\imp}{\al^{2n+6}}\,N(f\cl 4\al)\,
N(\bar K\cl 2\al)^{n} 
\big( \sfrac{1}{\al^2}N(\bar K\cl 2\al)\big)^\ell 
\sfrac{N(\bar K\cl 4\al)}{1-{1\over\al^2}N(\bar K\cl 4\al)}\cr
&\le \sfrac{2^{9}\,\imp}{\al^{2n+6}}\,N(f\cl 4\al)\,
\big( \sfrac{1}{\al^2}N(\bar K\cl 2\al)\big)^\ell 
\sfrac{N(\bar K\cl 4\al)^{n+1}}{1-{1\over\al^2}N(\bar K\cl 4\al)}\cr
}$$
and
$$\deqalign{
N\big(t_{12}^{(\ell)}\big),\ N\big(t_{22}^{(\ell)}\big)
&\le \sfrac{12}{\al^4} N(\bar K) \,N\big(T_n(\bar K)\big) 
\,N\big(T_\ell(\bar K)\big)
&\le \sfrac{12}{\al^{2(n+\ell)}} N(\bar K)^{n+\ell+1} \cr
N_{\rm eff}\big(\tau^{(\ell)}\big)
&\le \sfrac{8}{\al^6} N(\bar K)^2 \,N\big(T_n(\bar K)\big) 
\,N\big(T_\ell(\bar K)\big)
&\le \sfrac{8}{\al^{2(n+\ell)+2}} N(\bar K)^{n+\ell+2} \cr
}$$
In particular
$$\eqalign{
N_{\rm impr}\big( g_1 \big)+
 \smsum_{\ell=1}^n N_{\rm impr}\big( g_{2,\ell} \big)
&\le \sfrac{2^{9}\imp}{\al^{2n+6}}\,N( f\cl 4\al)\ 
\sfrac{N(\bar K\cl 4\al)^{n+1}}{1-{1\over\al^2}N(\bar K\cl 4\al)}
\ \sfrac{1}{1-{1\over\al^2}N(\bar K\cl 2\al)} \cr&\le \sfrac{2^{9}\imp}{\al^{2n+6}}\,N( f\cl 4\al)\ 
\sfrac{N(\bar K\cl 4\al)^{n+1}}{1-{2\over\al^2}N(\bar K\cl 4\al)} \cr
N\big( t_{11} \big)+\smsum_{\ell=1}^n N\big( t_{12}^{(\ell)} \big),\ 
N\big( t_{21} \big)+ \smsum_{\ell=1}^n N\big( t_{22}^{(\ell)} \big)
&\le \sfrac{12}{\al^{2n}}\,
\sfrac{N(\bar K)^{n+1}}{1-{1\over\al^2}N(\bar K)} \cr
N_{\rm eff}\big(\tau_1+ \smsum_{\ell=1}^{n-1}\tau^{(\ell)}+\half\tau^{(n)}  \big) 
&\le \sfrac{8}{\al^{2n+2}}\,\sfrac{N(\bar K)^{n+2}}{1-{1\over\al^2}N(\bar K)}  \cr
}$$
By Lemma \lemtailIIIsec, there exists a two--legged tail $t_{23}$ with at least four external legs such that
$$
P\Big[Q(t_1)\big(\cQ_{K'}(\lw  f \rw_{\xi,C})\big) \Big]
= P\Big[Q(t_{23})(\lw  f \rw_{\xi,C}) \Big]
$$
and, by the induction hypothesis,
$$
N(t_{23})\ \le\ \sfrac{2}{\al^2}N(t_1)\,N(\bar K)\
$$
Also, by Lemma \lemtailIsec
$$
P\Big[Q(t_2)\big(\cQ_{K'}(\lw  f \rw_{\xi,C})\big) \Big]
= P\Big[Q(t_{14}+t_{24}+t_{25})(\lw  f \rw_{\xi,C}) \Big]+g_3
$$
where $t_{14},t_{24}$ are one-- resp. two--legged tails with at least three external legs, fulfilling
$$
N(t_{14}),\ N(t_{24})\ 
\le\ \sfrac{2}{\al^2} N(t_2)\, N\big(\bar K\big)
$$
$t_{25}=t_2\circ K'$ is a two--legged tail with at least four external legs
fulfilling
$$
N(t_{25})\ 
\le\ \sfrac{1}{\al^2} N(t_2)\, N\big(\bar K\big)
$$
and the $g_3$ term obeys
$$\eqalign{
N_{\rm impr}\big( g_3 \big)
&\le \sfrac{2^{9}\imp}{\al^{8}}\,N( f\cl 4\al)
\,N( t_2\cl 2\al)\,
N(\bar K\cl 4\al) \big(1+\sfrac{1}{\al^2}N(\bar K\cl 4\al)\big) \cr
&\le \sfrac{2^{9}\imp}{\al^8}\,N( f\cl 4\al)\,
4\big(\sfrac{2}{\al^2}\big)^{n-1}\,\sfrac{N(\bar K\cl 2\al)^n}
{1-{1\over2\al^2} N(\bar K\cl 2\al)}
\,\sfrac{N(\bar K\cl 4\al)}
{1-{1\over\al^2} N(\bar K\cl 4\al)} \cr
&\le \sfrac{2^{n+10}\imp}{\al^{2n+6}}\,N( f\cl 4\al)\,
\sfrac{N(\bar K\cl 4\al)^{n+1}}
{1-{2\over\al^2} N(\bar K\cl 4\al)}\cr
}$$
Here we have used that $Q\big(t_2\circ K,T_1(K)\big)(\lw f\rw)$ has at least five external legs so that 
$$
PQ\big(t_2\circ K',T_1(K)\big)(\lw f\rw)=0
$$
For the same reason, the term ``$PQ(\tau)(\lw f\rw)$'' of Lemma
\lemtailIsec\ also vanishes.

\noindent
Finally, by Lemma \lemtailIsec,
$$
P\Big[Q(\tau)\big(\cQ_{K'}(\lw  f \rw_{\xi,C})\big) \Big]
= P\Big[Q(t_{26})(\lw  f \rw_{\xi,C}) \Big]
$$
where 
$t_{26}=\tau\circ K'$ is a two--legged tail with at least four external legs.
By Definition \defeffnorm,
$\ 
N(t_{26})\ 
\le\ \sfrac{1}{\al^2} \nu\, N\big(\bar K\big)\ 
\ $
for any effective bound $\nu$ for $\tau$. Hence, by the induction hypothesis
$$
N(t_{26})\ 
\le\ \sfrac{1}{\al^2} \ \sfrac{8}{\al^{2n}}\,
\sfrac{N(\bar K)^{n+1}}{1-{1\over\al^2}N(\bar K)}\ 
N\big(\bar K\big)\ 
=\  \sfrac{8}{\al^{2n+2}}\,
\sfrac{N(\bar K)^{n+2}}{1-{1\over\al^2}N(\bar K)}
$$

Combining the results above, we see that
$$\eqalign{
P \int {\rm Ev}\ \cR^{n+1}_{\lw K \rw_{\ze,D},C}
   &\big( \lw  f \rw_{\xi,C \atop\ze,D} \big)\, d\mu_D(\ze)
= P \, Q\big(T_{n+1}(\bar K)\big)(\lw  f \rw_{\xi,C} ) \cr
 &\hskip-1cm+ \sfrac{1}{2}
\hskip -.5cm\smsum_{\ell,\ell'\ge 1 \atop \max\{\ell,\ell'\}=n+1} 
   \hskip -.5cm   
  P\, Q\big(T_\ell(\bar K),T_{\ell'}(\bar K)\big)(\lw  f \rw_{\xi,C})\Big] 
 +P\,Q(t'_1+t'_2+\tau')(\lw  f\rw_{\xi,C})  +h_{n+1}(\psi;f) \cr 
}$$ 
with
$$
h_{n+1}(\psi;f) = h_n(\psi;F_1(\psi;f)) +  g_0(\psi) +g_1(\psi) 
+\sfrac{1}{2} g_{2,n}(\psi) + \smsum_{\ell=1}^{n-1} g_{2,\ell}(\psi)
+g_3
$$ 
 one-- resp. two--legged tails with at least three external legs
$$\eqalign{
t_1'&=t_{11}+\sfrac{1}{2}t_{12}^{(n)}+\smsum_{\ell=1}^{n-1}t_{12}^{(\ell)} +t_{14}\cr
t_2'&=t_{21}+\sfrac{1}{2}t_{22}^{(n)}+\smsum_{\ell=1}^{n-1}t_{22}^{(\ell)} 
+t_{23}+t_{24}+t_{25}+t_{26}\cr
}$$
and the two--legged tail with four external legs and degree two in $\et$
$$
\tau'=\tau_1+\sfrac{1}{2}\tau^{(n)}+\smsum_{\ell=1}^{n-1}\tau^{(\ell)} 
$$
By the estimates obtained above
$$\eqalign{
N(t_1'),N(t_2') \ &\le \ \sfrac{12}{\al^{2n}}\,
\sfrac{N(\bar K)^{n+1}}{1-{1\over\al^2}N(\bar K)}+
\sfrac{1}{\al^2}N(\bar K)\Big[2N(t_1)+2N(t_2)+N(t_2) \Big] 
+\sfrac{8}{\al^{2n+2}}\,
\sfrac{N(\bar K)^{n+2}}{1-{1\over\al^2}N(\bar K)}\cr
\ &\le \ \sfrac{12}{\al^{2n}}\,
\sfrac{N(\bar K)^{n+1}}{1-{1\over\al^2}N(\bar K)}
\Big(1+\sfrac{2}{3\al^2}N(\bar K)\Big)+
\sfrac{5}{\al^2}N(\bar K)
\ 4\big(\sfrac{8}{\al^2}\big)^{n-1}\,\sfrac{N(\bar K\cl \al)^n}
{1-{2\over\al^2} N(\bar K\cl \al)} \cr
\ &\le \ \sfrac{12}{\al^{2n}}\,
\sfrac{N(\bar K)^{n+1}}{1-{2\over\al^2}N(\bar K)}+
\sfrac{5}{2}
\ \big(\sfrac{8}{\al^2}\big)^{n}\,\sfrac{N(\bar K\cl \al)^{n+1}}
{1-{2\over\al^2} N(\bar K\cl \al)} \cr
\ &\le \ \Big(\sfrac{12}{8^n}+\sfrac{5}{2}\Big)
\ \big(\sfrac{8}{\al^2}\big)^{n}\,\sfrac{N(\bar K\cl \al)^{n+1}}
{1-{2\over\al^2} N(\bar K\cl \al)} \cr
\ &\le \ 4
\ \big(\sfrac{8}{\al^2}\big)^{n}\,\sfrac{N(\bar K\cl \al)^{n+1}}
{1-{2\over\al^2} N(\bar K\cl \al)} \cr
N_{\rm eff}(\tau')\ &\le\ 
\sfrac{8}{\al^{2n+2}}\,\sfrac{N(\bar K)^{n+2}}{1-{1\over\al^2}N(\bar K)}
}$$
and
$$\eqalign{
& N_{\rm impr}( g_0) +  N_{\rm impr}(g_1) 
   + \smsum_{\ell'=1}^n N_{\rm impr}(g_{2,\ell})+  N_{\rm impr}(g_3)  \cr
&\le \sfrac{2^9\imp}{\al^{n+6}}\,N( f\cl 4\al)  \,
\sfrac{N(\bar K\cl 4\al)^{n+1}}{1-{3\over 2\al}N(\bar K\cl 4\al)}
 \Big[ 1 + \sfrac{1}{\al^n} +\sfrac{2^{n+1}}{\al^n}\Big]   \cr
&\le \sfrac{2^{10}\imp}{\al^{n+6}}\,N( f\cl 4\al)  \,
\sfrac{N(\bar K\cl 4\al)^{n+1}}{1-{3\over 2\al}N(\bar K\cl 4\al)}
}$$
so that
$$\eqalign{
N&_{\rm impr}\big(h_{n+1}(\psi;f)\big)\cr
&\le N_{\rm impr}\big(h_n(\psi;F_1(\psi;f))\big) 
     + N_{\rm impr}( g_0) +  N_{\rm impr}(g_1) 
   + \smsum_{\ell'=1}^n N_{\rm impr}(g_{2,\ell})
   +  N_{\rm impr}(g_3) \cr
& \le  \sfrac{2^{10}\,\imp}{\al^5}\, \,\smsum_{m=1}^{n}
\sfrac{1}{\al^{n+1-m}}
\sfrac{N(\bar K\cl 4\al)^{n+1-m}}{1-{3\over 2\al}N(\bar K\cl 4\al)} 
N\big( F_{m}(\psi;f)\cl 4\al\big)\cr
&\hskip7cm+ \sfrac{2^{10}\imp}{\al^{n+6}}\,N( f\cl 4\al)  \,
\sfrac{N(\bar K\cl 4\al)^{n+1}}{1-{3\over 2\al}N(\bar K\cl 4\al)}
 \cr
& \le  \sfrac{2^{10}\,\imp}{\al^5}\, \,\smsum_{m=0}^{n}
\sfrac{1}{\al^{n+1-m}}
\sfrac{N(\bar K\cl 4\al)^{n+1-m}}{1-{3\over 2\al}N(\bar K\cl 4\al)} 
N\big( F_{m}(\psi;f)\cl 4\al\big)\cr}$$
since $F_0(\psi;f)=f$.
\endproof

\corollary{\STM\corprodtailssec}{
Let $K(\psi;\ze;\xi,\xi',\eta)$ be a Grassmann function that vanishes for $\eta=0$ and has degree at least four overall. Furthermore let $f(\psi;\ze;\xi)$ be a Grassmann function of degree at least four in the variables $\psi,\ze,\xi$. Set
$$\eqalign{
h(\psi) = P \Big[ \int {\rm Ev}\frac{1}{\bbbone - \cR_{\lw K\rw_{\ze,D},C}}
  (\lw f\rw_{\xi,C \atop \ze,D}) \,d\mu_D(\ze)
     &-\lw f(\psi;0;\xi)\rw_{\xi,C}
     - \smsum_{\ell=1}^\infty Q\big(T_\ell(\bar K)\big)(\lw f\rw_{\xi,C}) \cr
     & - \sfrac{1}{2} \smsum_{\ell,\ell'\ge 1 } 
  Q\big(T_\ell(\bar K),T_{\ell'}(\bar K)\big)
              (\lw f\rw_{\xi,C}) \Big] \cr
}$$
If $\al \ge 8$ and $N(K\cl 16\al)_\0 < \sfrac{1}{3} \al$, then
$$
N_{\rm impr}(h)  
\le \sfrac{2^{10}\,\imp}{\al^6}\,N(f\cl 16\al)\,
\sfrac{N( K\cl 16\al)}{1-{3\over\al}N( K\cl 16\al)}
$$
If 
$\ 
K(\psi;\ze;\xi,\xi',\eta) = 
\hat U(\psi+\ze;\xi+\xi'+\eta) - 
\hat U(\psi+\ze;\xi+\xi')$,
$\al \ge 8$ and $N(\hat U\cl 32\al)_\0 < \sfrac{1}{3} \al$, then
$$
N_{\rm impr}(h) 
\le \sfrac{2^{10}\,\imp}{\al^6}\,N(f\cl 16\al)\, 
\sfrac{N(\hat U\cl 16\al) }{1-{3\over\al}N(\hat U\cl 32\al)} 
$$

}
\prf 
By Proposition \propprodtailssec
$$\eqalign{
P  \int \Big[{\rm Ev}\frac{1}{\bbbone - \cR_{\lw K\rw_{\ze,D},C}}
  &(\lw f\rw_{\xi,C \atop \ze,D})- \lw f\rw_{\xi,C \atop \ze,D}\Big] \,d\mu_D(\ze) 
= \smsum_{\ell=1}^\infty P \Big[\int {\rm Ev}
  \,\cR_{\lw K\rw_{\ze,D},C}^\ell(\lw f\rw_{\xi,C \atop \ze,D})\,d\mu_D(\ze)\Big] \cr
&= \smsum_{\ell=1}^\infty 
  P \big[ Q\big(T_\ell(\bar K)\big)(\lw f\rw_{\xi,C})\big] 
  + \half \smsum_{\ell,\ell'\ge 1} 
     P \big[Q\big( T_\ell(\bar K),T_{\ell'}(\bar K) \big)(\lw f\rw_{\xi,C})\big] \cr
& \hskip 4.5cm  + \smsum_{\ell=1}^\infty h_\ell(\psi;f) 
  + P\big[ Q(\bar t_1+\bar t_2)(\lw f\rw_{\xi,C})\big] \cr   
}$$
with a one--legged tail $\bar t_1$ and a two--legged tail $\bar t_2$, each with at least three external legs. As the degree of $f$ is at least four, 
$P\big[ Q(\bar t_1+\bar t_2)(\lw f\rw_{\xi,C})\big]=0$. Set 
$h(\psi)=\smsum_{\ell=1}^\infty h_\ell(\psi;f)$. Then, 
$$\eqalign{
N_{\rm impr}(h) 
&\le \sfrac{2^{10}\,\imp}{\al^5}\, \,\smsum_{\ell=1}^\infty\,\smsum_{m=0}^{\ell-1}
\sfrac{1}{\al^{\ell-m}}
\sfrac{N\big(\bar K\cl 4\al\big)^{\ell-m} }{1-{3\over2\al}N(\bar K\cl 4\al)}
N\big( F_m(\psi;f)\cl 4\al\big)\cr
}$$
Define $\tilde F_m(\psi;f)$ by
$$
\lw \tilde F_m(\psi;f) \rw_{\xi,C } = {\rm Ev}\, \cR_{\lw K\rw_{\ze,D},C}^m
 \big( \lw  f \rw_{\xi,C \atop\ze,D} \big)
$$
Then, by 
Remark \remfunctnorm, followed by Lemma \lemestEv\ and Lemma \lemnormestcalR
$$\eqalign{
N\big( F_m(\psi;f)\cl 4\al\big)
&\le N\big(\tilde  F_m(\psi;f)\cl 8\al\big)\cr
&\le\Big(\sfrac{1}{32\al^{2}}
\sfrac{N(\lw K\rw_{\ze,D} \cl 8\al)}
{1-{1\over 32\al^2}N(\lw K\rw_{\ze,D} \cl 8\al)}\Big)^{m} 
N\big( \lw f\rw_{\ze,D}\cl 8\al\big)\cr
}$$
so that
$$\eqalign{
N_{\rm impr}(h) 
&\le \sfrac{2^{10}\,\imp}{\al^5}\, \,\smsum_{\ell=1}^\infty\,\smsum_{m=0}^{\ell-1}
\sfrac{1}{\al^{\ell-m}}
\sfrac{N(\bar K\cl 4\al)^{\ell-m} }{1-{3\over2\al}N(\bar K\cl 4\al)} 
\Big(\sfrac{{1\over 32\al^2}N( \lw K\rw_{\ze,D} \cl 8\al)}
{1-{1\over 32\al^2}N(\lw K\rw_{\ze,D} \cl 8\al)}\Big)^{m} 
N\big(f\cl 16\al\big)\cr
&= \sfrac{2^{10}\,\imp}{\al^5}\, \,\smsum_{p=1}^\infty\,\smsum_{m=0}^\infty
\sfrac{1}{\al^p}
\sfrac{N(\bar K\cl 4\al)^{p} }{1-{3\over2\al}N(\bar K\cl 4\al)} 
\Big(\sfrac{{1\over 32\al^2}N( \lw K\rw_{\ze,D} \cl 8\al)}
{1-{1\over 32\al^2}N(\lw K\rw_{\ze,D} \cl 8\al)}\Big)^{m} 
N\big(f\cl 16\al\big)\cr
&= \sfrac{2^{10}\,\imp}{\al^6}\, 
\sfrac{N(\bar K\cl 4\al) }{1-{1\over\al}N(\bar K\cl 4\al)} 
\sfrac{1 }{1-{3\over2\al}N(\bar K\cl 4\al)} 
\frac{1}{1-\sfrac{{1\over 32\al^2}N(\lw K\rw_{\ze,D} \cl 8\al)}{1-{1\over 32\al^2}N(\lw K\rw_{\ze,D} \cl 8\al)}}
N\big(f\cl 16\al\big)\cr
&\le \sfrac{2^{10}\,\imp}{\al^6}\, 
\sfrac{N(\bar K\cl 4\al) }{1-{1\over\al}N(\bar K\cl 4\al)} 
\sfrac{1 }{1-{3\over2\al}N(\bar K\cl 4\al)} 
\sfrac{1}{1-{1\over 16\al^2}N(\lw K\rw_{\ze,D} \cl 8\al)}N\big(f\cl 16\al\big)\cr
&\le \sfrac{2^{10}\,\imp}{\al^6}\, 
\sfrac{N(K\cl 16\al) }{1-{1\over\al}N(K\cl 16\al)} 
\sfrac{1 }{1-{3\over2\al}N(K\cl 16\al)} 
\sfrac{1}{1-{1\over 16\al^2}N(K\cl 16\al)}N\big(f\cl 16\al\big)\cr
&\le \sfrac{2^{10}\,\imp}{\al^6}\, 
\sfrac{N(K\cl 16\al) }{1-{3\over\al}N(K\cl 16\al)} N\big(f\cl 16\al\big)\cr
}$$
If 
$$
K(\psi;\ze;\xi,\xi',\eta) = 
\hat U(\psi+\ze;\xi+\xi'+\eta) - 
\hat U(\psi+\ze;\xi+\xi')
$$ 
so that
$$\eqalign{
\bar K(\psi;\ze,\ze',\varphi;\xi,\xi',\eta) 
&=K(\psi;\ze+\ze'+\varphi;\xi,\xi',\eta) \cr
&=\hat U(\psi+\ze+\ze'+\varphi;\xi+\xi'+\eta) - 
\hat U(\psi+\ze+\ze'+\varphi;\xi+\xi') \cr
}$$
and
$$
\lw K(\psi;\ze;\xi,\xi',\eta)\rw_{\ze,D}
=\int\big[\hat U(\psi+\ze+\ze';\xi+\xi'+\eta) - 
\hat U(\psi+\ze+\ze';\xi+\xi') \big]d\mu_{-D}(\ze')
$$
then, by Lemma \lemwicknorm\  and Remark \remfunctnorm
$$\eqalign{
N(\bar K\cl 4\al)&\le N(\hat U\cl 16\al)\cr
N(\lw K\rw_{\ze,D} \cl 8\al)&\le N(\hat U\cl 32\al)\cr
}$$
Consequently
$$\eqalign{
N_{\rm impr}(h) 
&\le \sfrac{2^{10}\,\imp}{\al^6}\, 
\sfrac{N(\bar K\cl 4\al) }{1-{1\over\al}N(\bar K\cl 4\al)} 
\sfrac{1 }{1-{3\over2\al}N(\bar K\cl 4\al)} 
\sfrac{1}{1-{1\over 16\al^2}N(\lw K\rw_{\ze,D} \cl 8\al)}N\big(f\cl 16\al\big)\cr
&\le \sfrac{2^{10}\,\imp}{\al^6}\, 
\sfrac{N(\hat U\cl 16\al) }{1-{1\over\al}N(\hat U\cl 32\al)} 
\sfrac{1 }{1-{3\over2\al}N(\hat U\cl 32\al)} 
\sfrac{1}{1-{1\over 16\al^2}N(\hat U\cl 32\al)}N\big(f\cl 16\al\big)\cr
&\le \sfrac{2^{10}\,\imp}{\al^6}\, 
\sfrac{N(\hat U\cl 16\al) }{1-{3\over\al}N(\hat U\cl 32\al)} 
}$$
\endproof

\proof{of Theorem \theoremVb}
Set
$$
K(\psi;\ze;\xi,\xi',\eta) = 
\hat U(\psi+\ze;\xi+\xi'+\eta) - 
\hat U(\psi+\ze;\xi+\xi')
$$ 
and
$$\eqalign{
\bar K(\psi;\ze,\ze',\varphi;\xi,\xi',\eta) 
&=K(\psi;\ze+\ze'+\varphi;\xi,\xi',\eta) \cr
&=\hat U(\psi+\ze+\ze'+\varphi;\xi+\xi'+\eta) - 
\hat U(\psi+\ze+\ze'+\varphi;\xi+\xi') \cr
\tilde f(\psi;\ze;\xi) &=  \hat f(\psi+\ze;\xi)
}$$
By Proposition \propalgtwoWick  
$$\eqalign{
\cS_{U,C}(f)\ 
= \cS_{U,C} \big( \lw \tilde f(\psi;0;\xi)\rw_{\xi,C\atop\psi,D} \big)
=\ \lW \int\hskip-6pt \int {\rm Ev} \sfrac{1}{\bbbone -\cR_{\lw K\rw_{\ze,D},C}}
 (\lw \tilde f \rw_{\xi,C \atop \ze,D})\,d\mu_C(\xi)\,d\mu_D(\ze)\rW_{\psi,D} 
}$$
so that
$$
P\,f' = P\,\int {\rm Ev} \sfrac{1}{\bbbone -\cR_{\lw K\rw_{\ze,D},C}}
 (\lw \tilde f \rw_{\xi,C \atop \ze,D})\,d\mu_D(\ze)
$$
Therefore, by Corollary \corprodtailssec\ and Remark \remfunctnorm,
there is $g(\psi)$ with
$$
Pf'=P \Big[ \hat f(\psi;0) + 
\smsum_{\ell=1}^\infty Q\big(T_\ell(\bar K)\big)(\lw\tilde f\rw_{\xi,C}) 
      + \sfrac{1}{2}  \smsum_{\ell,\ell'\ge 1 }
Q\big(T_\ell(\bar K),T_{\ell'}(\bar K)\big)(\lw\tilde f\rw_{\xi,C}) \Big] 
                 \ +\ g
$$
and
$$\eqalign{
N_{\rm impr}(g) 
&\le\sfrac{2^{10}\,\imp}{\al^6}\, 
\sfrac{N(\hat U\cl 32\al) }{1-{3\over\al}N(\hat U\cl 32\al)}
N\big(\tilde f\cl 16\al\big)\, \cr
&\le\sfrac{2^{10}\,\imp}{\al^6}\, 
\sfrac{N(\hat U\cl 32\al) }{1-{3\over\al}N(\hat U\cl 32\al)}
N\big(\hat f\cl 32\al\big)\, \cr
}$$
Since $\hat f$ and $K$ have degree at least four overall,
$$
P\,Q\big(T_\ell(\bar K)\big)(\lw \tilde f \rw_{\xi,C \atop \ze,D}) = T_\ell(\bar K) \circ T_1(\hat f)
$$
and by Remark \remformladder
$$
P Q\big(T_\ell(\bar K),T_{\ell'}(\bar K)\big)(\lw \tilde f \rw_{\xi,C \atop \ze,D}) = 
T_\ell(\bar K) \circ {\rm Rung}(\hat f) \circ T_{\ell'}(\bar K)
$$
Furthermore, by part (ii) of Remark \remeffecttailsec,\ 
$T_\ell(\bar K)=T_\ell(\hat U)$. Thus
$$
Pf'=P \hat f(\psi,0) + 
\smsum_{\ell=1}^\infty T_\ell(\hat U) \circ T_1(\hat f) 
      + \sfrac{1}{2}  \smsum_{\ell,\ell'\ge 1} 
   T_\ell(\hat U) \circ {\rm Rung}(\hat f) \circ T_{\ell'}(\hat U)
 \ +\ g
$$
\endproof

\vfill\eject

\chap{ Example: A Vector Model}\PG\pgRX

We consider a model, which while simple, still captures the main features of 
one scale of a many fermion model. Let $\cF$ be a finite set of at least two ``colours''
(Farben) and $\cX$ be a finite set of points in ``space--time''. Let
$V$ be the complex vector space with basis $\set{\xi_{c,x}}{c\in\cF,\ x\in\cX}$
and $V'$ be the complex vector space with basis 
$\set{\psi_{c,x}}{c\in\cF,\ x\in\cX}$. Let $C(x,x'),\ x,x'\in\cX$ be a skew 
symmetric matrix and define the covariance
$$
C\big(\xi_{c,x}, \xi_{c',x'}\big)=\de_{c,c'}C(x,x')
$$
An antisymmetric function $W$ on $(\cF\times\cX)^n$, with $n$ even,
 is said to be colour preserving if it is of the form
$$
W\big((c_1,x_1),\cdots(c_n,x_n)\big)
={\rm Ant}\big[\de_{c_1,c_2}\cdots\de_{c_{n-1},c_n}w(x_1,\cdots,x_n)\big]
\EQN\eqnColourpreserving$$
where Ant means antisymmetrization.
An example, with $n=2$, is the function $\de_{c,c'}C(x,x')$. An even element
$\cW$ of the Grassmann algebra $\bigwedge V$ is said to be colour preserving if it is of the form 
$$
\cW= \sum_{n\in 2\bbbn} \sum_{c_1,\cdots,c_n\in\cF\atop x_1,\cdots,x_n\in\cX}
W_n\big((c_1,x_1),\cdots(c_n,x_n)\big)\ \xi_{c_1,x_1}\cdots\xi_{c_n,x_n}
$$
with each coefficient function $W_n$ colour preserving.

For $p$ odd, we define the (0--dimensional) norm $\|\varphi\|_p$ of a 
complex valued function $\varphi$ on $(\cF\times\cX)^n$ by
$$\eqalign{
\| \varphi\|_p&=\max_{1\le i_1<\cdots<i_p\le n\atop 1\le k \le n}\ 
\sup_{c_{i_1},\cdots,c_{i_p}\in\cF}\sum_{c_i\in\cF\atop i\ne i_1,\cdots,i_p}
\sup_{x_k\in\cX}\ 
\sum_{x_j \in\cX\atop j\ne k}
 \big|\varphi\big((c_1,x_1),\cdots(c_n,x_n)\big)\big|\cr
}$$
Also, for a colour preserving function $W$ on $(\cF\times\cX)^n$, we define
$$
\tn W\tn =\inf\Big\{\max_{ 1\le k \le n}\,\sup_{x_k\in\cX}\ 
\smsum_{x_j \in\cX\atop j\ne k} \big|w(x_1,\cdots,x_n)\big|\,\Big|\,
w\hbox{ satisfies (\eqnColourpreserving)}\,\Big\}
$$
Then
$$\eqalign{
\| W\|_p&\le |\cF|^{n-p-1\over 2}\ \tn W\tn\cr
 \tn W\tn&\le (n-1)!!\,\|W\|_{n-1}\qquad\hbox{if}\quad|\cF|\ge\sfrac{n}{2}\cr
}\EQN\eqncptripnorm$$
In particular, if $n=4$,
$$
\| W \|_1\le |\cF|\,\tn W\tn \le 3|\cF|\,\|W\|_3
$$
Every element $f\in V^{\otimes n}$ has a unique representation
$$
f=\sum_{c_1,\cdots,c_n\in\cF\atop x_1,\cdots,x_n\in\cX}
\varphi\big((c_1,x_1),\cdots(c_n,x_n)\big)\ \xi_{c_1,x_1}\otimes\cdots\otimes\xi_{c_n,x_n}
$$
with $\varphi$ a complex valued function on $(\cF\times\cX)^n$. We set
$$
\|f\|_p=\|\varphi\|_p
$$
Observe that, for each odd $p$,  $\|\ \cdot\ \|_p$ is a family of symmetric 
norms on the spaces $V^{\otimes n}$, in the sense of Definition \defsymnorm.

\proposition{\STM\propvectorintconst}{
Suppose that $\cX$ is a disjoint union of two subsets $\cX_a$ and $\cX_c$ such that
$$
C(x,x')=0 \qquad {\rm if\ both}\ x,x'\in \cX_a\ \   {\rm or\ both}\ 
   x,x'\in \cX_c
$$
Assume furthermore that there is a Hilbert space $\cH$ and vectors $w_x\in \cH,\ x\in\cX$ such that
$$
C(x,x') = \<w_x,w_{x'}\>_\cH \qquad {\rm for\ all\ \ } x\in\cX_a,\ x'\in\cX_c
$$
Set
$$\eqalign{
\ib&= 2\sup_{x\in\cX}\|w_x\|  \cr
\cb&= \sup_{x\in\cX}\sum_{x'\in\cX}|C(x,x')|  \cr
}$$
Then  $(C,0)$ has integration constants $\cb,\ib$ for the 
configuration $\|\,\cdot\,\|_p$ of seminorms, in the sense of Definition
\defimprconf.

}
\prf
We first verify that $\ib$ is an integral bound for $C$ with respect to 
each family  $\|\ \cdot\ \|_p$ of seminorms. Set
$$\eqalign{
\cH'&= L^2(\cF)\otimes\cH\cr
w'_{c,x}&=\de_c \otimes w_x\cr
}$$
where $\de_c$ is the function on $\cF$ which vanishes except at $c$ and takes the value one there. Then, for all $c,c'\in\cF$,
$$
C(\xi_{c,x},\xi_{c',x'})=0 \qquad {\rm if\ both}\ x,x'\in \cX_a\ \   
{\rm or\ both}\    x,x'\in \cX_c
$$
and 
$$
C(\xi_{c,x},\xi_{c',x'}) = \de_{c,c'}\<w_x,w_{x'}\>_\cH
= \<w'_{c,x},w'_{c',x'}\>_{\cH'}\qquad 
{\rm for\ all\ \ } x\in\cX_a,\ x'\in\cX_c
$$
Furthermore
$$
\|w'_{c,x}\|_{\cH'}\le\| w_x\|_{\cH}\le \sfrac{\ib}{2}
$$
Let $V_c$ (respectively $V_a$) be the subspace of $V$ generated by
$\set{\xi_{c,x}}{x\in\cX_c,\ c\in\cF}$ (respectively
$\set{\xi_{c,x}}{x\in\cX_a,\ c\in\cF}$). By Proposition \propGII,
$$
\Big| \int \xi_{c_1,x_1} \cdots \xi_{c_m,x_m}\, d\mu_C(\xi)  \Big| 
\le \big(\sfrac{\ib}{2}\big)^m
$$
As in Example \egIIcompatnorm, $\ib$ is an integral bound for $C$ with 
respect to each family  $\|\ \cdot\ \|_p$ of seminorms.

Also
$$
\sup_{c,c'\in\cF\atop x,x'\in\cX}\big|C(\xi_{c,x},\xi_{c',x'})\big|
=\sup_{x,x'\in\cX}|C(x,x')|
\le \sup_{x,x'\in\cX}\| w_x\|\,\|w_{x'}\|
\le\sfrac{\ib^2}{4}
\EQN\eqnvectorCsup$$

We now verify the contraction estimates of Definition \defimprconf.
Let
$$\deqalign{
f&=\sum_{c_1,\cdots,c_n\in\cF\atop x_1,\cdots,x_n\in\cX}
\varphi\big((c_1,x_1),\cdots(c_n,x_n)\big)\ \xi_{c_1,x_1}\otimes\cdots\otimes\xi_{c_n,x_n}&\in V^{\otimes n}\cr
f'&=\sum_{c_1,\cdots,c_{n'}\in\cF\atop x_1,\cdots,x_{n'}\in\cX}
\varphi'\big((c_1,x_1),\cdots(c_{n'},x_{n'})\big)\ \xi_{c_1,x_1}\otimes\cdots\otimes\xi_{c_{n'},x_{n'}}&\in V^{\otimes {n'}}\cr
}$$
and, for $1\le k\le n$ and $1\le k'\le n'$,
$$\eqalign{
\Phi_k(c_1,\cdots,c_n)&=\sup_{x_k\in\cX}\sum_{x_i\in\cX\atop i\ne k}
\big|\varphi\big((c_1,x_1),\cdots(c_n,x_n)\big)\big|\cr
\Phi'_{k'}(c'_1,\cdots,c'_{n'})&=\sup_{x'_{k'}\in\cX}
\sum_{x'_{i}\in\cX\atop i\ne k'}
\big|\varphi'\big((c'_1,x'_1),\cdots(c'_{n'},x'_{n'})\big)\big|\cr
}$$
Observe that, for all $1\le k\le n$, all $1\le i_1<\cdots<i_p\le n$
and all $c_{i_1},\cdots,c_{i_p}$
$$
\sum_{c_i\in\cF\atop i\ne i_1,\cdots,i_p}\Phi_k(c_1,\cdots,c_n)
\le \|\varphi\|_p= \|f\|_p
$$
and similarly for $\Phi'$.

For the first contraction estimate of Definition \defimprconf,
let $1\le i\le n$ and $1\le j\le n'$. By the symmetry of the norms, it suffices to consider $i=j=1$. 
Set
$$
\ga({\sst (c_2,x_2),\cdots(c_n,x_n),(c'_2,x'_2),\cdots(c'_{n'},x'_{n'})})
=\!\!\!\sum_{c_1,c'_1\in\cF\atop x_1,x'_1\in\cX}\!\!\!
\varphi({\sst (c_1,x_1),\cdots,(c_n,x_n)})\ \de_{c_1,c'_1}C({\sst x_1,x'_1})\ 
\varphi'({\sst(c'_1,x'_1),\cdots,(c'_{n'},x'_{n'})})
$$
Then
$$
\Cont{1}{n+1}{C} (f\otimes f')
=\sum_{c_1,\cdots,c_{n+n'-2}\in\cF\atop x_1,\cdots,x_{n+n'-2}\in\cX}\!\!
\ga\big((c_1,x_1),\cdots,(c_{n+n'-2},x_{n+n'-2})\big)
\ \xi_{c_1,x_1}\otimes\cdots\otimes\xi_{c_{n+n'-2},x_{n+n'-2}}
$$
In particular
$\ 
\big\|\Cont{i}{n+j}{C} (f\otimes f')\big\|_p=\|\ga\|_p
\ .$

\noindent
Fix any $1\le i_1<\cdots<i_p\le n+n'-2$ and colours 
$c_{i_1},\cdots,c_{i_p}$. 
Set 
$ 
q=\max\set{\nu}{i_\nu\le n-1}
\ $
with the convention that if $\set{\nu}{i_\nu\le n-1}=\emptyset$, then $q=0$. 
Set $j_1=i_1+1,\ \cdots,\ j_q=i_q+1;\ j'_{1}=i_{q+1}-n+2,\ \cdots,\ j'_{p-q}=i_p-n+2$ and 
$$\meqalign{
d_{j_\nu}&=c_{i_\nu}&&\nu&=1,\cdots,q\cr
d'_{j'_\nu}&=c_{i_{q+\nu}}&&\nu&=1,\cdots,p-q\cr
}$$
Also fix $1\le k\le n+n'-2$. 
First assume that  $k\le n-1$.  We have,
for all $c_1,\cdots,c_{n+n'-2}\in\cF$
$$\eqalign{
&\sup_{x_k}\sum_{x_i, i\ne k}
\big|\ga\big((c_1,x_1),\cdots,(c_{n+n'-2},x_{n+n'-2})\big)\big|\cr
&\hskip1in\le \sum_{c,c'\in\cF}\Phi_{k+1}(c,c_1,\cdots,c_{n-1})
\ \de_{c,c'}\cb\ 
\Phi'_1(c',c_n,\cdots,c_{n+n'-2})
}$$
Now assume that $q$ is odd. Then $p-q$ is even and
$$\eqalign{
&\sum_{c_i\in\cF\atop i\ne i_1,\cdots,i_p}\sup_{x_k}\sum_{x_i, i\ne k}
\big|\ga\big((c_1,x_1),\cdots,(c_{n+n'-2},x_{n+n'-2})\big)\big|\cr
&\hskip.5in\le  \cb
\sum_{d_j\in\cF\atop j\ne 1,j_1,\cdots,j_q}
\sum_{d'_{\ell}\in\cF\atop \ell\ne 1,j'_1,\cdots,j'_{p-q}}
\sum_{c,c'\in\cF}\Phi_{k+1}(c,d_2,\cdots,d_n)
\ \de_{c,c'}\ 
\Phi'_1(c',d'_2,\cdots,d'_{n'})\cr
&\hskip.5in\le  \cb
\sum_{d_j\in\cF\atop j\ne j_1,\cdots,j_q} \Phi_{k+1}(d_1,\cdots,d_n)
\ \ \ \sup_{d'_1}\hskip-10pt\sum_{d'_{\ell}\in\cF\atop \ell\ne 1,j'_1,\cdots,j'_{p-q}}
\Phi'_1(d'_1,d'_2,\cdots,d'_{n'})\cr
&\hskip.5in\le  \cb\ \|f\|_q\ \|f'\|_{p-q+1}
}$$
In the event that $q$ is even, one obtains in a similar way that
$$
\sum_{c_i\in\cF\atop i\ne i_1,\cdots,i_p}\sup_{x_k}\sum_{x_i, i\ne k}
\big|\ga\big((c_1,x_1),\cdots,(c_{n+n'-2},x_{n+n'-2})\big)\big|
\le  \cb\ \|f\|_{q+1}\ \|f'\|_{p-q}
$$
The case that $k\ge n$ follows by interchanging the roles of $f$ and $f'$.

For the second contraction estimate of Definition \defimprconf\ 
 it suffices to consider, by the symmetry of the norms,
$\big\|\Cont{1}{n+1}{C}\,\Cont{2}{n+2}{C}\,\Cont{3}{n+3}{C}\,
 (f\otimes f')\big\|_p$. 
Set
$$\eqalign{
&\tilde\ga({\sst (c_4,x_4),\cdots(c_n,x_n),(c'_4,x'_4),\cdots(c'_{n'},x'_{n'})})
\cr
&=\hskip-8pt\sum_{c_1,c'_1,c_2,c'_2,c_3,c'_3\in\cF\atop 
x_1,x'_1,x_2,x'_2,x_3,x'_3\in\cX}\hskip-20pt
\varphi({\sst (c_1,x_1),\cdots,(c_n,x_n)})
\ \de_{c_1,c'_1}C({\sst x_1,x'_1})
\de_{c_2,c'_2}C({\sst x_2,x'_2})
\de_{c_3,c'_3}C({\sst x_3,x'_3})\ 
\varphi'({\sst(c'_1,x'_1),\cdots,(c'_{n'},x'_{n'})})
}$$
Then
$$\eqalign{
&\Cont{1}{n+1}{C}\,\Cont{2}{n+2}{C}\,\Cont{3}{n+3}{C}\, (f\otimes f')\cr
&\hskip.5in=\sum_{c_1,\cdots,c_{n+n'-6}\in\cF\atop x_1,\cdots,x_{n+n'-6}\in\cX}
\tilde\ga({\sst(c_1,x_1),\cdots,(c_{n+n'-6},x_{n+n'-6})})
\ \xi_{c_1,x_1}\otimes\cdots\otimes\xi_{c_{n+n'-6},x_{n+n'-6}}\cr
}$$

\noindent
Fix any $1\le i_1<\cdots<i_p\le n+n'-6$ and colours 
$c_{i_1},\cdots,c_{i_p}$. 
Set 
$ 
q=\max\set{\nu}{i_\nu\le n-3}
\ $. 
Set $j_1=i_1+3,\ \cdots,\ j_q=i_q+3;\ j'_{1}=i_{q+1}-n+6,\ \cdots,\ j'_{p-q}=i_p-n+6$ and 
$$\meqalign{
d_{j_\nu}&=c_{i_\nu}&&\nu&=1,\cdots,q\cr
d'_{j'_\nu}&=c_{i_{q+\nu}}&&\nu&=1,\cdots,p-q\cr
}$$
Also fix $1\le k\le n+n'-6$. 
First assume that  $k\le n-3$.  We have,
for all $c_1,\cdots,c_{n+n'-6}\in\cF$
$$\eqalign{
&\sup_{x_k}\sum_{x_i, i\ne k}
\big|\tilde\ga\big((c_1,x_1),\cdots,(c_{n+n'-6},x_{n+n'-6})\big)\big|\cr
&\hskip.5in\le \sum_{d_1,d_2,d_3\in\cF\atop d'_1,d'_2,d'_3\in\cF}
\Phi_{k+1}(d_1,d_2,d_3,c_1,\cdots,c_{n-3})
\ \ \de_{d_1,d'_1}\de_{d_2,d'_2}\de_{d_3,d'_3}
\cb\big[\sup_{y,y'\in\cX}|C(y,y')|\big]^2\cr
&\hskip3.4in
\Phi'_1(d'_1,d'_2,d'_3,c_{n-2},\cdots,c_{n+n'-6})\cr
&\hskip.5in\le \ib^4\cb\sum_{d_1,d_2,d_3\in\cF}
\Phi_{k+1}(d_1,d_2,d_3,c_1,\cdots,c_{n-3})
\Phi'_1(d_1,d_2,d_3,c_{n-2},\cdots,c_{n+n'-6})\cr
}$$
by (\eqnvectorCsup).  Now assume that $q$ is odd. Then
$$\eqalign{
&\sum_{c_i\in\cF\atop i\ne i_1,\cdots,i_p}\sup_{x_k}\sum_{x_i, i\ne k}
\big|\tilde\ga\big((c_1,x_1),\cdots,(c_{n+n'-6},x_{n+n'-6})\big)\big|\cr
&\hskip.5in\le  \ib^4\cb
\sum_{d_j\in\cF\atop j\ne j_1,\cdots,j_q} \Phi_{k+1}(d_1,\cdots,d_n)
\ \ \ \sup_{d'_1,d'_2,d'_3}\hskip-10pt\sum_{d'_{\ell}\in\cF\atop \ell\ne 1,2,3,j'_1,\cdots,j'_{p-q}}
\Phi'_1(d'_1,d'_2,\cdots,d'_{n'})\cr
&\hskip.5in\le  \ib^4\cb\ \|f\|_q\ \|f'\|_{p-q+3}
}$$
The remaining cases are similar.
\endproof

\theorem{\STM\thmExample}{Let $W$ be a colour preserving function on
$(\cF\times\cX)^n$ and
$$
\cW=\sum_{c_1,\cdots,c_4\in\cF\atop x_1,\cdots,x_4\in\cX}
W\big((c_1,x_1),\cdots(c_4,x_4)\big)\ \psi_{c_1,x_1}\cdots\psi_{c_4,x_4}
$$
be the associated interaction. Let
$$
\cW'(\psi)=\Om_C\big(\lw\cW\rw\big)
=\log \sfrac{1}{Z}\int e^{\lw\cW\rw(\psi +\xi)} d\mu_C(\xi) 
\qquad{\rm where}\quad Z= \int e^{\lw\cW\rw(\xi)} d\mu_C(\xi)
$$
and $C$ is the covariance of Proposition \propvectorintconst.
Since $C$ and $W$ are colour preserving, we can write
$$
\cW'=\sum_{n\in 2\bbbn} \sum_{c_1,\cdots,c_n\in\cF\atop x_1,\cdots,x_n\in\cX}
W'_n\big((c_1,x_1),\cdots(c_n,x_n)\big)\ \psi_{c_1,x_1}\cdots\psi_{c_n,x_n}
$$
with colour preserving coefficients $W'_n$. If 
$\tn W\tn\le \sfrac{1}{2^{38}\ib^2\cb\,|\cF|} $, then
$$
\sum_{n=6}^\infty \al_0^{n-6}\ib^{n-6} \| W'_n\|_1
 \le 2^{48}\cb\,|\cF|^2\ \tn W\tn^2
$$
with $\al_0=\sfrac{1}{\root {3} \of {2^{29}\ib^2\cb \,|\cF|\,\tn W\tn}}$, and
$$\eqalign{
\tn W'_2\tn&\le 2^{61}\ib^4\cb\, |\cF|\ \tn W\tn^2 \cr
\TN W'_4-W+\sfrac{1}{4}\smsum_{r=1}^\infty
(-12)^{r+1} W\circ(\cC\circ W)^r \TN&\le 2^{57}\ib^2\cb\  \tn W\tn^2 \cr
}$$
where $W\circ(\cC\circ W)^r$ is the ladder

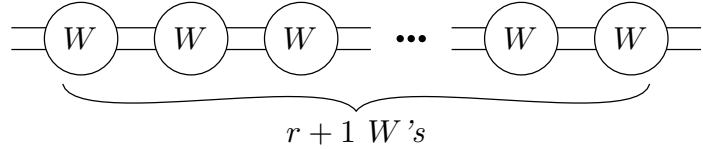
\centerline{\figput{figX1}}

\noindent
with $r+1$  rungs $W$ and propagator $C$, defined in Appendix C.
}
\prf Set $\imp=\sfrac{1}{|\cF|}$.  For $f\in V^{\otimes n}$, set
$$\eqalign{
\|f\| &= \|f\|_1+|\cF|\, \|f\|_3+|\cF|^2 \|f\|_5 \cr
\|f\|_{\rm impr} &= \|f\|_1+|\cF|\, \|f\|_3 \cr
}$$ 
By Lemma \lemimprconfig, $(C,0)$ have improved integration constants $\cb,\ib,\imp$ for the 
families $\|\,\cdot\|$ and $\|\,\cdot\|_{\rm impr}$ of seminorms.
By (\eqncptripnorm)
$$
\| W\|\le 2\,|\cF|\,\tn W\tn 
$$
Let $\al_1=8$ and $\al\in\{\al_0,\al_1\}$.  Then $\al\ge 8$ and, 
using the notation of Definition \deffunctnorm,
$$
N(W\cl64\al)=2^{24}\al^4\ib^2\cb \|W\|\le 2^{25}\al^4\ib^2\cb \,|\cF|\,\tn W\tn
\le \sfrac{1}{16}\al
$$
since $\ib^2\cb \,|\cF|\,\tn W\tn\le\sfrac{1}{2^{38}} $. Therefore 
the hypotheses of Theorem \theoremVa\ are fulfilled. By part (i) of 
Theorem \theoremVa,
$$
N(\cW'-\cW\cl \al)
\le  \sfrac{1}{2\al^2}\,
     \sfrac{N( \cW\cl 32\al)^2}{1-{1\over\al^2}N( \cW\cl 32\al)}
\le 2^{48}\al^6\ib^4\cb^2|\cF|^2\ \tn W\tn^2
$$
In particular
$$
\sfrac{\cb}{\ib^2}\sum_{n=6}^\infty \al^n\ib^n \| W'_n\|_1
\le 2^{48}\al^6\ib^4\cb^2|\cF|^2\ \tn W\tn^2
$$
so that
$$
\sum_{n=6}^\infty \al^{n-6}\ib^{n-6} \| W'_n\|_1
 \le 2^{48}\cb\,|\cF|^2\ \tn W\tn^2
$$
By part (ii) of Theorem \theoremVa\ and Proposition \:\propAIV, both
$\,N_{\rm impr}\big( W'_2\cl \al\big) \,$ and \hfill\break
$\,N_{\rm impr}\big( W'_4-W+\sfrac{1}{4}\smsum_{r=1}^\infty
(-12)^{r+1} W\circ(\cC\circ W)^r\cl \al\big) \,$ are bounded by
$$
\sfrac{2^{10}\,\imp}{\al^6}\,\sfrac{N(W\cl 64\al)^2}
{1-{8\over\al}N(W\cl 64\al)}
\le 2^{61}\al^2\ib^4\cb^2|\cF|\ \tn W\tn^2 
$$
Using (\eqncptripnorm)
$$
\tn W'_2\tn \le \|W'_2\|_{\rm impr} 
= \sfrac{1}{\al^2\cb}\,N_{\rm impr}\big( W'_2\cl \al\big)
\le 2^{61}\ib^4\cb\,|\cF|\ \tn W\tn^2 
$$
Similarly, setting $F=W'_4-W+\sfrac{1}{4}\smsum_{r=1}^\infty
(-12)^{r+1} W\circ(\cC\circ W)^r$
$$
\tn F \tn \le 3 \|F\|_3 \le \sfrac{3}{|\cF|}\| F\|_{\rm impr} 
= \sfrac{3}{\al^4 \ib^2\cb\,|\cF|}\,N_{\rm impr}\big(F\cl \al\big)
\le 2^{63}\sfrac{1}{\al^2}\ib^2\cb\  \tn W\tn^2 
$$
With $\al=\al_1=8$ , we get the desired bound.
\endproof

\vskip1cm

The $j^{\rm th}$ shell, $j\ge 1$, of the many fermion model of [FKTf1]
behaves qualitatively like the vector model that we have just discussed.
The covariance for the $j^{\rm th}$ shell of that many fermion model is
$$
C^{(j)}(k) = \sfrac{\nu^{(j)}(k)}{\imath k_0 -e(\k)}
$$
where $\nu^{(j)}(k)$ is approximately the characteristic function with support
$$
S_j = \set{k=(k_0,\k)\in\bbbr\times\bbbr^d}
{ \sfrac{1}{M^j} \le |\imath k_0 -e(\k)| <\sfrac{1}{M^{j-1}}}
$$
To define seminorms in position space that mimic well, in our context,
the supremum in momentum space, we introduce sectorizations. We choose a 
projection $k \mapsto \pi_F(k)$ onto the Fermi curve $F$ and a decomposition of the Fermi curve into disjoint intervals $I_1,\cdots, I_N$ 
each of length $\fl$ between $\sfrac{1}{M^j}$ and $\sfrac{1}{M^{j/2}}$. The sectorization $\Si$ is the collection of ``sectors''
$s_\ell=\pi_F^{-1}(I_\ell)\cap S_j$, $\ell=1,\cdots, N$.

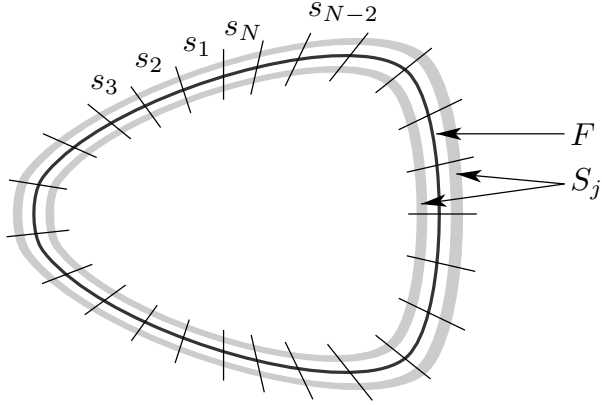
\centerline{\figput{nsphsectors2}}

\noindent
Let $V$ be the complex vector space with basis
$\set{\psi_{s,x},\ \bar\psi_{s,x}}{s\in\Si,\ x\in\bbbr\times\bbbr^2}$
and define the covariance\footnote{$^{(1)}$}{We are deliberately ignoring many technical fine points; in particular, we ignore spin, smoothness of partitions of unity, factors of $2\pi$.} 
$$
C(\psi_{s,x},\psi_{s',x'})
=C(\bar\psi_{s,x},\bar\psi_{s',x'})
=0\qquad
C(\psi_{s,x},\bar\psi_{s',x'})=\de_{s,s'}\int_s d^3 k\  
e^{\imath k\cdot x}C^{(j)}(\k) 
$$
The norms on $V^{\otimes n}$ are similar to those of the vector model.
In this case
$$
\ib^2 = \const\sfrac{\fl}{M^j}
\qquad
\cb = \const M^j
\qquad
\imp = \sfrac{1}{|\Si|}=\const\fl
$$
and the conclusions of Theorem \thmExample\ become
$$\eqalign{
\tn W'_2\tn&\le \const\sfrac{\fl}{M^j}\ \tn W\tn^2 \cr
\TN W'_4-W+\sfrac{1}{4}\smsum_{r=1}^\infty
(-12)^{r+1} W\circ(\cC\circ W)^r \TN&\le \const\fl\  \tn W\tn^2 \cr
}
$$
for all four--legged interactions $\cW$ whose norm  $\tn W\tn $
is bounded by a sufficiently small, but $j$--independent, constant.

\vfill\eject

\appendix{C}{Ladders expressed in terms of kernels}\PG\pgRC

Let $V$ be a complex vector space and $\{\xi_i\}$ a system of generators
for $V$ that is indexed by $i$ in a set $\fX$. 

\definition{\STM\defAI}{
\Item{i)} Depending on the context, a complex valued function on $\fX^4$ is called a four legged kernel over $\fX$ or a bubble propagator over $\fX$.
\Item{ ii)} To a four legged kernel $f$ over $\fX$ we associate the Grassmann function
$$
Gr(\xi;f) = \sum_{i_1,i_2,i_3,i_4\in\fX} f({\sst i_1,i_2,i_3,i_4})\,
\xi_{i_1}\xi_{i_2}\xi_{i_3}\xi_{i_4}
$$
in the complex Grassmann algebra $\bigwedge V$.
\Item{ iii)} If $A$ and $B$ are covariances on $V$ then the tensor product
$$
A\otimes B(i_1,i_2,i_3,i_4)=A(\xi_{i_1},\xi_{i_3})B(\xi_{i_2},\xi_{i_4})
$$
is a bubble propagator over $\fX$. In particular we shall consider the bubble propagator
$$
\cC = C\otimes C + C\otimes D + D\otimes C
$$
where $C$ and $D$ are the covariances of Section V.2.
\Item{ iv)} Let $f_1$ and $f_2$ be functions on $\fX^4$. The operator product of $f_1$ and $f_2$ is
$$
(f_1\fcirc f_2)(i_1,i_2,i_3,i_4)
 = \smsum_{j_1,j_2\in\fX}
f_1(i_1,i_2,j_1,j_2)\ f_2(j_1,j_2,i_3,i_4)
$$
whenever the sum is well--defined.
\Item {v)} Let $f$ be a four legged kernel over $\fX$. The antisymmetrization of $f$ is the four legged kernel
$$
\big({\rm Ant} f\big) (i_1,i_2,i_3,i_4) = 
\sfrac{1}{4!} \sum_{\pi\in S_4} 
\sgn(\pi)\,f(i_{\pi(1)},i_{\pi(2)},i_{\pi(3)},i_{\pi(4)})
$$
Clearly $Gr(\xi;f)=Gr(\xi;{\rm Ant} f)$.
The kernel $f$ is called antisymmetric if $f={\rm Ant} f$. 
}

\lemma{\STM\lemAII}{
Let
$$
E = \sum_{i_1,i_2,i_3,i_4\in\fX} \Big( 
  e_{202}({\sst i_1,i_2,i_3,i_4})\,\psi_{i_1}\psi_{i_2}\xi'_{i_3}\xi'_{i_4}
 +2e_{211}({\sst i_1,i_2,i_3,i_4})\,\psi_{i_1}\psi_{i_2}\ze'_{i_3}\xi'_{i_4} \Big)
$$
be an end as in Definition \defladders, where $e_{202}$ and $e_{211}$ are 
four legged kernels over $\fX$ that are antisymmetric under permutations
of their four arguments.
\Item{ i)} If 
$$
E' = \sum_{i_1,i_2,i_3,i_4\in\fX} \Big( 
  e'_{202}({\sst i_1,i_2,i_3,i_4})\,\psi_{i_1}\psi_{i_2}\xi'_{i_3}\xi'_{i_4}
 +2e'_{211}({\sst i_1,i_2,i_3,i_4})\,\psi_{i_1}\psi_{i_2}\ze'_{i_3}\xi'_{i_4} \Big)
$$
is another end with antisymmetric four legged kernels $e'_{202}$ and $e'_{211}$ then
$$
E \circ E' = -2\,Gr\big(\psi;\ e_{202} \circ (C\otimes C) \circ e'_{202} 
+ e_{211} \circ (C\otimes D+D\otimes C) \circ e'_{211} \big)
$$
\Item{ ii)}
If
$$\eqalign{
\rho = \sum_{i_1,i_2,i_3,i_4\in\fX} 
  \Big( &\rho_{0202}({\sst i_1,i_2,i_3,i_4})\xi_{i_1}\xi_{i_2}\xi'_{i_3}\xi'_{i_4}
    +2\,\rho_{1102}({\sst i_1,i_2,i_3,i_4})\ze_{i_1}\xi_{i_2}\xi'_{i_3}\xi'_{i_4} \cr
    +&2\,\rho_{0211}({\sst i_1,i_2,i_3,i_4})\xi_{i_1}\xi_{i_2}\ze'_{i_3}\xi'_{i_4}
    +4\,\rho_{1111}({\sst i_1,i_2,i_3,i_4})\ze_{i_1}\xi_{i_2}\ze'_{i_3}\xi'_{i_4} \Big) \cr
}$$
is a rung with antisymmetric kernels 
$\rho_{0202},\ \rho_{1102},\ \rho_{0211},\ \rho_{1111}$, then
$$
E \circ \rho =  - 2\sum_{i_1,i_2,i_3,i_4\in\fX} \Big( 
  \epsilon_{202}
({\sst i_1,i_2,i_3,i_4})\,\psi_{i_1}\psi_{i_2}\xi'_{i_3}\xi'_{i_4}
 +2\,\epsilon_{211}
({\sst i_1,i_2,i_3,i_4})\,\psi_{i_1}\psi_{i_2}\ze'_{i_3}\xi'_{i_4} \Big)
$$
with
$$\eqalign{
\epsilon_{202} & =  e_{202} \circ (C\otimes C) \circ \rho_{0202}
  + e_{211} \circ (C\otimes D+D\otimes C) \circ \rho_{1102} \cr
\epsilon_{211} & =   e_{202} \circ (C\otimes C) \circ \rho_{0211}
  +  e_{211} \circ (C\otimes D+D\otimes C) \circ \rho_{1111} \cr
}$$
}
\prf
i) 
$$\eqalign{
E \circ E'
&= \sum_{i_1,i_2,i_3,i_4\atop \ell_1,\ell_2,\ell_3,\ell_4} \int \hskip-4pt \int \Big[\,    
 e_{202}({\sst i_1,i_2,i_3,i_4})\,\psi_{i_1}\psi_{i_2}\,\lw \xi_{i_3}\xi_{i_4} \rw_C
\,+2e_{211}({\sst i_1,i_2,i_3,i_4})\,\psi_{i_1}\psi_{i_2}\,\ze_{i_3}\xi_{i_4} 
\Big] \cr
& \hskip 1cm \Big[
\,e'_{202}({\sst \ell_1,\ell_2,\ell_3,\ell_4})\,\psi_{\ell_1}\psi_{\ell_2}\,
  \lw \xi_{\ell_3}\xi_{\ell_4} \rw_C
\,+2e'_{211}({\sst \ell_1,\ell_2,\ell_3,\ell_4})\,\psi_{\ell_1}\psi_{\ell_2}\,
  \ze_{\ell_3}\xi_{\ell_4} 
\Big] \,d\mu_C(\xi)\,d\mu_D(\ze)\cr
&\ = -\sum_{i_1,i_2,i_3,i_4\atop \ell_1,\ell_2,\ell_3,\ell_4}
      \psi_{i_1}\psi_{i_2}\psi_{\ell_1}\psi_{\ell_2} \, \Big[
4e_{211}({\sst i_1,i_2,i_3,i_4})\,e'_{211}({\sst\ell_1,\ell_2,\ell_3,\ell_4})\,
  D({\sst\ze_{i_3},\ze_{\ell_3}}) C({\sst\xi_{i_4},\xi_{\ell_4}}) \cr
&\hskip 1.5cm + e_{202}({\sst i_1,i_2,i_3,i_4})\,e'_{202}({\sst \ell_1,\ell_2,\ell_3,\ell_4})
   \big( C({\sst\xi_{i_3},\xi_{\ell_3}})\,C({\sst\xi_{i_4},\xi_{\ell_4}}) 
          - C({\sst\xi_{i_3},\xi_{\ell_4}})\,C({\sst\xi_{i_4},\xi_{\ell_3}}) \big) \Big] \cr
&\ = -\sum_{i_1,i_2,j_1,j_2\atop \ell_1,\ell_2,k_1,k_2}
      \psi_{i_1}\psi_{i_2}\psi_{\ell_1}\psi_{\ell_2} \, \Big[
4e_{211}({\sst i_1,i_2,j_1,j_2})\,e'_{211}({\sst\ell_1,\ell_2,k_1,k_2})\,
  D({\sst\ze_{j_1},\ze_{k_1}}) C({\sst\xi_{j_2},\xi_{k_2}}) \cr
&\hskip 4.5cm + 
  e_{202}({\sst i_1,i_2,j_1,j_2})\,e'_{202}({\sst \ell_1,\ell_2,k_1,k_2})
    C({\sst\xi_{j_1},\xi_{k_1}})\,C({\sst\xi_{j_2},\xi_{k_2}}) \cr 
&\hskip 4.5cm - 
  e_{202}({\sst i_1,i_2,j_1,j_2})\,e'_{202}({\sst \ell_1,\ell_2,k_2,k_1})
    C({\sst\xi_{j_1},\xi_{k_1}})\,C({\sst\xi_{j_2},\xi_{k_2}})     \Big] \cr
&\ = -\sum_{i_1,i_2, \ell_1,\ell_2}
      \psi_{i_1}\psi_{i_2}\psi_{\ell_1}\psi_{\ell_2} \, \Big[
\big(4 e_{211} \circ (C\otimes D) \circ e'_{211} \big)
    ({\sst i_1,i_2, \ell_1,\ell_2}) \cr
& \hskip 4.5cm +2 \big( e_{202} \circ (C\otimes C) \circ e'_{202} \big)
 ({\sst i_1,i_2, \ell_1,\ell_2})  \Big] \cr
&\ = -2 \sum_{i_1,i_2, i_3,i_4}
      \psi_{i_1}\psi_{i_2}\psi_{i_3}\psi_{i_4} \, \Big[
 \big( e_{211} \circ (C\otimes D+D\otimes C) \circ e'_{211} \big)
    ({\sst i_1,i_2, i_3,i_4}) \cr
& \hskip 4.5cm + \big( e_{202} \circ (C\otimes C) \circ e'_{202} \big)
 ({\sst i_1,i_2, i_3,i_4})  \Big] \cr
}$$
For the last two equalities we used the antisymmetry of the kernels.
\Item ii) Similarly,
$$\eqalign{
E\circ \rho
&= \sum_{i_1,i_2,i_3,i_4\atop \ell_1,\ell_2,\ell_3,\ell_4} 
\psi_{i_1}\psi_{i_2}  \int \hskip-4pt \int  \Big[\,    
 e_{202}({\sst i_1,i_2,i_3,i_4})\,\,\lw \xi_{i_3}\xi_{i_4} \rw_C
\,+2e_{211}({\sst i_1,i_2,i_3,i_4})\,\ze_{i_3}\xi_{i_4} 
\Big] \cr
& \hskip 1.2cm \Big[ \Big(
\,\rho_{0202}({\sst \ell_1,\ell_2,\ell_3,\ell_4})\,
\lw \xi_{\ell_1}\xi_{\ell_2}\rw_C \, 
+ 2\,\rho_{1102}({\sst \ell_1,\ell_2,\ell_3,\ell_4})\,
 \ze_{\ell_1}\xi_{\ell_2} \Big) \,\xi'_{\ell_3}\xi'_{\ell_4}  \cr
& \hskip 1cm +\Big(
\,2\,\rho_{0211}({\sst \ell_1,\ell_2,\ell_3,\ell_4})\,
\lw \xi_{\ell_1}\xi_{\ell_2}\rw_C \, 
+ 4\,\rho_{1111}({\sst \ell_1,\ell_2,\ell_3,\ell_4})\,
 \ze_{\ell_1}\xi_{\ell_2} \Big) \,\ze'_{\ell_3}\xi'_{\ell_4}
\Big] \,d\mu_C(\xi)\,d\mu_D(\ze)\cr
&\ = -2\sum_{i_1,i_2, i_3,i_4}
      \psi_{i_1}\psi_{i_2}\xi'_{i_3}\xi'_{i_4} \, \Big[
 \big( e_{211} \circ (C\otimes D+D\otimes C) \circ \rho_{1102} \big)
    ({\sst i_1,i_2, i_3,i_4}) \cr
& \hskip 5cm + \big( e_{202} \circ (C\otimes C) \circ \rho_{0202} \big)
 ({\sst i_1,i_2, i_3,i_4})  \Big] \cr
& -4 \sum_{i_1,i_2, i_3,i_4}
      \psi_{i_1}\psi_{i_2}\ze'_{i_3}\xi'_{i_4} \, \Big[
 \big( e_{211} \circ (C\otimes D+D\otimes C) \circ \rho_{1111} \big)
    ({\sst i_1,i_2, i_3,i_4}) \cr
& \hskip 5cm + \big( e_{202} \circ (C\otimes C) \circ \rho_{0211} \big)
 ({\sst i_1,i_2, i_3,i_4})  \Big] \cr
}$$
\endproof

\lemma{\STM\lemAIII}{
Let $f$ be an antisymmetric four legged kernel and $n\ge 0$. Set $F=Gr(f;\xi)$ and 
$$
E_n(F)=E(F) \circ \rho(F)\circ \rho(F) \cdots \rho(F)
$$
with $n$ copies of $\rho(F)$, where $E(F)$ and $\rho(F)$ were defined in
Definition \defladders.iii. Then
$$
E_n(F)  =   \sum_{i_1,i_2,i_3,i_4\in\fX}  
 f_n ({\sst i_1,i_2,i_3,i_4})
\Big(\,\psi_{i_1}\psi_{i_2}\xi'_{i_3}\xi'_{i_4}
 +2\,\psi_{i_1}\psi_{i_2}\ze'_{i_3}\xi'_{i_4} \Big)
$$ 
with
$$
f_n = \sfrac{(-1)^n}{2} (12)^{n+1}\,f\circ(\cC\circ f)^n
$$

}

\prf
Observe that
$$\eqalign{
E(F) &= {4\choose 2} \sum_{i_1,i_2,i_3,i_4} f({\sst i_1,i_2,i_3,i_4})
  \big[ \psi_{i_1}\psi_{i_2}\xi'_{i_3}\xi'_{i_4}
 +2\,\psi_{i_1}\psi_{i_2}\ze'_{i_3}\xi'_{i_4} \big] \cr
\rho(F) &= {4\choose 2} \sum_{i_1,i_2,i_3,i_4} f({\sst i_1,i_2,i_3,i_4})
\big[ \xi_{i_1}\xi_{i_2}\xi'_{i_3}\xi'_{i_4}
    +2\,\ze_{i_1}\xi_{i_2}\xi'_{i_3}\xi'_{i_4} 
    +2\,\xi_{i_1}\xi_{i_2}\ze'_{i_3}\xi'_{i_4}
    +4\,\ze_{i_1}\xi_{i_2}\ze'_{i_3}\xi'_{i_4} \big]
}\EQN\eqAI$$
Since $E(F)=E_0(F)$, this proves the case $n=0$. We now perform induction 
on $n$. By Lemma \lemAII\ and the induction hypothesis
$$
E_{n+1}(F)= \sum_{i_1,i_2,i_3,i_4\in\fX} \Big( 
  \epsilon_{202}
({\sst i_1,i_2,i_3,i_4})\,\psi_{i_1}\psi_{i_2}\xi'_{i_3}\xi'_{i_4}
 +2\,\epsilon_{211}
({\sst i_1,i_2,i_3,i_4})\,\psi_{i_1}\psi_{i_2}\ze'_{i_3}\xi'_{i_4} \Big)
$$
with
$$\eqalign{
\epsilon_{202} = \epsilon_{211} 
&=  - 2{4 \choose 2}\big[ f_n \circ (C\otimes C) \circ f
  + f_n \circ (C\otimes D+D\otimes C) \circ f \big] \cr
& = -12 \,f_n \circ \cC \circ f = f_{n+1}
}$$
\endproof

\proposition{\STM\propAIV}{
Let $f$ be an antisymmetric four legged kernel. Set $F=Gr(f;\xi)$
and let $L_r(F) = E(F) \circ \rho(F)\circ\cdots\circ\rho(F)\circ E(F)$ the ladder of length $r\ge 1$ with $r-1$ rungs $\rho(F)$. Then
$$
L_r(F) = \sfrac{(-1)^r}{2} (12)^{r+1}\,Gr\big(\psi;\ f\circ(\cC\circ f)^r\big)
$$
}

\prf
By Lemma \lemAII, Lemma \lemAIII\  and (\eqAI)
$$\eqalign{
L_r(F) &=  E_{r-1}(F) \circ E(F) \cr
    & = -2 {4\choose 2}\, Gr\big(\psi;\ f_{r-1} \circ (C\otimes C) \circ f 
            + f_{r-1}\circ (C\otimes D +D\otimes C)\circ f \big) \cr
    & = -12\,Gr\big(\psi;\ f_{r-1} \circ \cC \circ f\big) \cr
    & = \sfrac{(-1)^r}{2}\,(12)^{r+1}\,Gr\big(\psi;\ f\circ(\cC\circ f)^r\big)\cr
}$$
\endproof

\vfill\eject

\titlea{ References}\PG\pgRIIref

\item{[BS]} F. Berezin, M. Shubin: {\it The Schr\"odinger Equation}. Kluwer 1991. 
Supplement 3: D.Le\u ites, {\bf Quantization and supermanifolds}.
\smallskip%
\item{[FKLT1]} J. Feldman, H. Kn\"orrer, D. Lehmann, E. Trubowitz, {\bf Fermi 
Liquids in Two-Space Dimensions}, in {\it Constructive Physics} V. Rivasseau ed. 
Springer Lecture Notes in Physics 446, 267-300 (1995).
\smallskip%
\item{[FKLT2]} J. Feldman, H. Kn\"orrer, D. Lehmann, E. Trubowitz, 
{\bf Are There Two Dimensional Fermi Liquids?}, in {\it Proceedings of the XIth
International Congress of Mathematical Physics}, D. Iagolnitzer ed., 
440-444 (1995).
\smallskip%
\item{[FKT1]} J. Feldman, H. Kn\"orrer, E. Trubowitz, 
{\bf A Representation for Fermionic Correlation Functions},
Communications in Mathematical Physics {\bf 195},  465--493 (1998).
\item{[FKTf1]} J. Feldman, H. Kn\"orrer, E. Trubowitz, 
{\bf A Two Dimensional Fermi Liquid, Part 1: Overview}, preprint.
\smallskip%
\item{[FKTo3]} J. Feldman, H. Kn\"orrer, E. Trubowitz, 
{\bf Single Scale Analysis of Many Fermion Systems, Part 3: Sectorized Norms}, preprint.
\smallskip%
\item{[FMRT]} J. Feldman, J. Magnen, V. Rivasseau, E. Trubowitz, 
{\bf An Infinite Volume Expansion for Many Fermion Green's Functions},
Helvetica Physica Acta {\bf 65}, 679-721 (1992).
\smallskip%
\item{[FST1]} J. Feldman, M. Salmhofer, E. Trubowitz, 
{\bf Perturbation Theory around Non-nested Fermi Surfaces, I: Keeping the Fermi 
Surface Fixed}, Journal of Statistical Physics {\bf 84}, 1209-1336 (1996).
\smallskip%
\item{[FST2]} J. Feldman, M. Salmhofer, E. Trubowitz, 
{\bf Regularity
of the Moving Fermi Surface: RPA Contributions}, Communications
on Pure and Applied Mathematics,  {\bf LI} 1133-1246 (1998).
\smallskip%
\item{[FST3]}  J. Feldman, M. Salmhofer, E. Trubowitz,
{\bf Regularity of Interacting Nonspherical Fermi Surfaces: The Full 
Self--Energy}, Communications on Pure and Applied Mathematics {\bf LII},
 273-324  (1999).
\smallskip%
\item{[FST4]} 
 J. Feldman, M. Salmhofer, and E. Trubowitz,
{\bf An inversion theorem in Fermi surface theory}, 
Communications on Pure and Applied Mathematics, to appear.

\vfill\eject

\hoffset=-0.1in
\titlea{Notation}\PG\pgRIInot

\vfill

\centerline{
\vbox{\offinterlineskip
\hrule
\halign{\vrule#&
         \strut\hskip0.05in\hfil#\hfil&
         \hskip0.05in\vrule#\hskip0.05in&
          #\hfil\hfil&
         \hskip0.05in\vrule#\hskip0.05in&
          #\hfil\hfil&
           \hskip0.05in\vrule#\cr
height2pt&\omit&&\omit&&\omit&\cr
&Not'n&&Description&&Reference&\cr
height2pt&\omit&&\omit&&\omit&\cr
\noalign{\hrule}
height2pt&\omit&&\omit&&\omit&\cr
&$Z(f)$&&degree zero component of $f$&&Definition \defsuperalgebra.iii&\cr
height2pt&\omit&&\omit&&\omit&\cr
&$\bigwedge V$&&Grassmann algebra over $V$&&Example \exsuper&\cr
height2pt&\omit&&\omit&&\omit&\cr
&$\bigwedge\nolimits_A V $&&Grassmann algebra over $V$ with coefficients
in $A$&&Example \exsuper&\cr
height4pt&\omit&&\omit&&\omit&\cr
&$A_m[n_1,\cdots,n_r]$&&partially antisymmetric elements of $A_m\otimes V^{\otimes (n_1+\cdots+n_r)}$
&&Definition \defAmnonetonr&\cr
height4pt&\omit&&\omit&&\omit&\cr
&$\susywedge E,\ \susywedge\nolimits_A E$&&enlarged algebra&&
Definition \defenlalg&\cr
height2pt&\omit&&\omit&&\omit&\cr
&${\rm Ev}$&&evaluation map&&
Definition \defenlalg&\cr
height4pt&\omit&&\omit&&\omit&\cr
&$\int e^{\Sigma\, \xi_i\ze_i}\, d\mu_C(\xi)$
&&$=e^{-1/2\,\Sigma\, \ze_i C_{ij} \ze_j}$
Grassmann Gaussian  integral&&before Definition \defrengroup&\cr
height4pt&\omit&&\omit&&\omit&\cr
&$\Om_C(W)(\psi)$&&$=\log \sfrac{1}{Z}\int e^{W(\psi +\xi)} d\mu_C(\xi)$
renormalization group map&&Definitions \defrengroup, \defrengroupInfDim&\cr
height4pt&\omit&&\omit&&\omit&\cr
&$\cS(f)$&&$=\sfrac{1}{Z(U,C)}\, \int f(\xi)\, e^{U(\xi)}\, d\mu_C(\xi)$
Schwinger functional&&before Remark \remrenschw&\cr
height4pt&\omit&&\omit&&\omit&\cr
&$R$&&$R$--operator&&before Theorem \theoremIII&\cr
height2pt&\omit&&\omit&&\omit&\cr
&$R_C(K_1,\cdots,K_\ell)$&&$\ell^{\rm th}$ Taylor coefficient of $R$&&
(\eqRCdef)&\cr
height2pt&\omit&&\omit&&\omit&\cr
&$\cR_{K,C}(f)$&&
$\lww  \int\hskip-4pt\int \lW  e^{\lw K(\xi,\xi',\eta)\rw_{\xi'}} -1 \rW_\eta\,f(\eta)\,
d\mu_C(\xi')\,d\mu_C(\eta) \rww_\xi$&&Definition \defcalR&\cr
height2pt&\omit&&\omit&&\omit&\cr
&$\tilde R_C(K_2,\cdots,K_\ell)(f)$&&
$\int \hskip-4pt\int \lW \big( \smprod\limits_{i=2}^\ell \lw K_i(\xi,\xi'+\xi^{\prime\prime},\eta')\rw_{\xi^{\prime\prime}} \big) \rW_{\eta'} \,f(\eta+\eta')\,d\mu_{C}(\xi^{\prime\prime})\,d\mu_{C}(\eta')$&&
(\eqndefrctilde)&\cr
height4pt&\omit&&\omit&&\omit&\cr
&$\lw e^{\Sigma\, \xi_i\ze_i}\rw_{\xi,C}$&&$=e^{1/2\,\Sigma\, \ze_i C_{ij} \ze_j}\,e^{\Sigma\, \xi_i\ze_i}$
Wick ordering&&after Remark \remnormal&\cr
height2pt&\omit&&\omit&&\omit&\cr
&$\Cont{i}{j}{C},\ \cont{\xi}{\xi'}{C}$&&contractions&&Definitions \defContract,
\defGrasscontract&\cr
height2pt&\omit&&\omit&&\omit&\cr
&$\fN_d$&&norm domain&&Definition \defFancynormdomain&\cr
height2pt&\omit&&\omit&&\omit&\cr
&$\cb$&&contraction bound&&Definition \defcontractintbound.i&\cr
height2pt&\omit&&\omit&&\omit&\cr
&$\ib$&&integral bound&&Definition \defcontractintbound.ii&\cr
height2pt&\omit&&\omit&&\omit&\cr
&$\cb,\ib,J$&&improved integration constants&&Definition \defimprnorm&\cr
height2pt&\omit&&\omit&&\omit&\cr
&$N(f\cl \al)$&&$\sfrac{1}{\ib^2}\,\cb\!\sum_{m,n_1,\cdots, n_r\ge 0}\,
\al^{|n|}\,\ib^{|n|} \,\|f_{m; n_1,\cdots, n_r}\|$&&Definition \deffunctnorm&\cr
height2pt&\omit&&\omit&&\omit&\cr
&$\rho(F)$&&rung&&Definition \defladders&\cr
height2pt&\omit&&\omit&&\omit&\cr
&$E(F)$&&end&&Definition \defladders&\cr
height2pt&\omit&&\omit&&\omit&\cr
&$L_r(F)$&&ladder of length $r$&&Definition \defladders&\cr
height2pt&\omit&&\omit&&\omit&\cr
&$\|\,\cdot\,\|_p$&&configuration of seminorms&&Definition \defimprconf&\cr
height2pt&\omit&&\omit&&\omit&\cr
&$P$&&projects $\lw f(\psi;\xi)\rw_\xi$ to $f_{4,0}(\psi,0) + f_{2,0}(\psi,0)$&&Definition \defP&\cr
height2pt&\omit&&\omit&&\omit&\cr
&$Q(\bar K_1,\cdots,\bar K_\ell)$&&
enlarged algebra analog of $R_C(K_1,\cdots,K_\ell)$&&Definition \defQ&\cr
height2pt&\omit&&\omit&&\omit&\cr
&$Gr(\xi;f)$&&Grassmann function&&Definition \defAI.ii&\cr
height2pt&\omit&&\omit&&\omit&\cr
&$\circ$&&convolution&&Definition \defAI.iv&\cr
height2pt&\omit&&\omit&&\omit&\cr
}\hrule}}

\end